\documentclass[12pt]{article}
\usepackage{a4wide, amsmath, latexsym, epsfig,
  graphicx, rotating, fancyhdr, tabularx, afterpage}
\usepackage{axodraw}
\usepackage{subfigure}

\newcommand\ba{\begin{eqnarray}}
\newcommand\ea{\end{eqnarray}}

\def\order#1{{\mathcal O}\left(#1\right)}

\begin{document}
\begin{titlepage}

\begin{flushright}
{ IFJPAN-IV-2012-14 \hskip 1 cm CERN-PH-TH/2012-354}
\end{flushright}

\vspace{0.1cm}
\begin{center}
{\Huge \bf QED Bremsstrahlung in decays of electroweak  bosons}
\end{center}

\vspace*{1mm}

\begin{center}
   {\bf  A.B. Arbuzov$^{a}$, R.R. Sadykov$^{a}$ and Z. W\c{a}s$^{b,c}$.}\\
{\em $^a$ Joint Institute for Nuclear Research, Joliot-Curie str. 6, Dubna, 141980, Russia}\\
       {\em $^b$  Institute of Nuclear Physics, PAN,
        Krak\'ow, ul. Radzikowskiego 152, Poland}\\
{\em $^c$ CERN PH-TH, CH-1211 Geneva 23, Switzerland }
\end{center}
\vspace{.1 cm}
\begin{center}
{\bf   ABSTRACT  }
\end{center}
Isolated lepton momenta, in particular their directions are the most precisely measured quantities 
in $pp$ collisions at LHC.
This offers opportunities for multitude of precision measurements.
It is of practical importance to verify if precision measurements with leptons in the final state  require  {\it all}
theoretical effects evaluated simultaneously or if QED bremsstrahlung in the final state can be separated 
without unwanted precision loss.

Results for final state bremsstrahlung in the decays of narrow resonances
are obtained from the Feynman rules of QED in an unambiguous way and can be controlled  with a very high precision. 
Also for resonances of non-negligible width, if 
calculations  are 
appropriately performed, such separation from the remaining electroweak effects
 can be expected.

Our paper is devoted to validation that final state QED bremsstrahlung can
indeed  be separated from the
rest of QCD and electroweak effects, in  the  production and decay of
$Z$ and $W$ bosons, and to estimation
of the resulting systematic error. 
The  quantitative discussion is based on
 Monte Carlo programs
{\tt PHOTOS} and {\tt SANC}, as well as on  {\tt KKMC} which is used
for benchmark results. 
We show, that for a large classes of $W$ and $Z$ boson observables as used at LHC, theoretical error
on photonic bremsstrahlung   is 0.1 or 0.2\%, depending on  the program options used.
An overall theoretical error on QED final state radiation, {\it i.e.} taking into account missing corrections
due to pair emission and interference with initial state radiation is estimated respectively at 0.2\% or 0.3\%
again depending on the program option used.


\vfill %
\vspace{0.1 cm}
\begin{flushleft}
{  IFJPAN-IV-2012-14  \hskip 1 cm  CERN-PH-TH/2012-354\\
 December, 2012}
\end{flushleft}

\vspace*{1mm}
\bigskip
\footnoterule
\noindent
{\footnotesize \noindent  $^{\dag}$
This project is financed in part from funds of Polish National Science
Centre under decisions DEC-2011/03/B/ST2/00220, DEC-2012/04/M/ST2/00240 and by Russian Foundation for Basic Research,
 grant No 10-02-01030-a.
}
\end{titlepage}

\section {Introduction}

Several of the most important measurements at LHC experiments, such as Higgs boson
searches \cite{:2012gk}, precision measurements of the $W$ boson 
mass \cite{Besson:2009zzb,Aaltonen:2009aa,Aaltonen:2012bp} or measurements 
of electroweak boson pair production \cite{Aaltonen:2009aa} rely on a precise
reconstruction of momenta for the final state leptons \cite{Aad:2011mk}. 
A substantial effort of the experimental community was devoted to 
optimize detector design and understand detector responses.
Precision of 0.1\%  (even 0.01\% for lepton directions) is of no exception. 
For  more details, see {\it e.g.}  reviews \cite{Aad:2008zzm,Aad:2011mk,ATLAS:2011yga}.

The  QED effects of the final state radiation (FSR) play an important role in such experimental studies.
Final state bremsstrahlung
is included in all simulation chains and 
indeed should be studied together with the detector  response to leptons. It can not be separated,
because of infrared singularities of QED.
Different approaches based on different theoretical simplifications are at present in use. 
At the level of the collinear approximation, 
expressions for higher order FSR corrections
are not only well defined but  are in fact process independent. In general, QED calculations are process dependent
but methods for obtaining results with ${\cal O}(\alpha^2)$ corrections and resummation of higher order effects
are well established as well as techniques for  evaluating theoretical errors. There is no need
for introducing effective models. Situation is different  if incoming or outgoing particles are unstable, an effort
like documented in \cite{Bardin:1999ak} is then needed. 
In case 
of $Z$ and $W$ decays if the narrow width approximation is used, the theoretical framework for  
QED FSR
bremsstrahlung effects is  unambiguous. In case of the $Z$ decay (in fact for hard processes 
mediated by s-channel $Z/\gamma^*$ exchange) the  framework in which the  QED final state bremsstrahlung is
separated from other contributions is defined unambiguously as well. This 
 was explored in $e^+e^-$ collisions at LEP and  high precision solutions were proposed.

For LEP I experiments Monte Carlo simulation programs based 
on exclusive exponentiation and featuring second order matrix elements were developed \cite{Jadach:1999vf}. 
Considerable  theoretical effort was invested in the reordering of the perturbative expansion. This 
opened a scheme
for proper resummation of vacuum polarization diagrams into terms contributing to  $Z$ width and remaining vacuum 
polarization corrections \cite{Bohm:1986rj}. The definition of final state bremsstrahlung was affected in 
a minimal way. The by-product of this effort was separation of the electroweak corrections into 
QED parts, corrections to the $Z$ boson propagator (which have to be resummed to all orders) and remaining weak corrections
which, with the proper choice of the calculation schemes, were small.

In the case  of LEP II this theoretical approach had to be reconsidered, because of
the new $W^+W^-$ pair production processes. The gauge 
cancellation between diagrams of electroweak bosons exchanged in $s$ and $t$ channels, had to be carefully
respected. The
resummation of the dominant contribution of the vacuum polarization had to be restricted to the case of the constant width.
It was not necessary however  to reopen discussion on details of scheme for final state bremsstrahlung 
calculations because of the relatively small available statistics of  $W$-pair 
samples \cite{Altarelli:1996gh,Altarelli:1996ww} and thus 
limited interest in high precision calculations for QED FSR bremsstrahlung\footnote{ Consequences of
resummation of parts of the electroweak effects are complex but will  not be covered in this paper. 
Let us point to another  
theoretical constraint restricting naive resummation. It could be observed that even in the case of initial 
state QED bremsstrahlung for the process $e^+e^- \to \nu_e \bar \nu_e$, previously performed step of
resummation had to be  partially revisited, because of amplitude featuring $t$-channel $W$ exchange 
  \cite{Bardin:2001vt}. For the complete second order matrix element of the two hard photon emission,
the diagrams of  charged Higgs boson exchange had to be taken into account \cite{Was:2004ig}.}.

It is important to stress that the QED final state effects 
in processes where leptons are produced through decays of $W$ or $Z/\gamma^*$ can be 
calculated and simulated with an essentially arbitrary high precision.  
They form a separate class of Feynman diagrams and developed already techniques should be explored at LHC, 
especially as the
attractiveness of such an approach was confirmed  \cite{Group:2012gb} in the context 
of $W$ mass measurement of CDF and D0 collaborations. With the ever increasing precision,
effects beyond photonic final state bremsstrahlung have to be of course considered as part of FSR effects as well, in particular 
those due to emission of extra lepton pairs. Also interference effects, such as 
initial-final state bremsstrahlung interference, has to be taken into account.

The interference become an issue  for separating  QED FSR radiation from the rest of the
electroweak corrections, at the level of cross section.  That is why,
 the interference of photon emission from the initial state quarks and the final 
state lepton  has to be discussed carefully.
It is of  practical importance to verify if the suppression of the interference
due to boson's lifetime survives experimental cuts.
If it is the case, then  separation of effects due to QED final state  bremsstrahlung 
and remaining parts of the electroweak and strong interaction corrections
can be conveniently explored in the experimental studies. 

For this  paper we 
assume, that in practical applications, all other corrections than  QED final 
state interactions, that is remaining  electroweak, and initial state hadronic 
interactions 
are expected  to be measured
with the properly defined observables.
Thanks to this  approach we will be able to achieve very competitive   precision of theoretical predictions on the
class of corrections directly affecting lepton momenta measurements.
This represents a  complementary approach to the one used in
\cite{Richardson:2010gz}. There,
emphasis was put on the use of electroweak calculations together with all 
hadronic initial state interactions necessary for complete predictions for observables.

Our paper is organized as follows.
In Section 2 we describe two programs {\tt PHOTOS} and {\tt SANC}, and their theoretical base.
  Section 3 is devoted to tests of the first order QED calculations.
This is of importance in itself but also represent  consistency
 checks of definitions of this part of electroweak 
corrections which will be considered at the amplitude level as QED final state bremsstrahlung (FSR). 
Definition of observables and calculation schemes used all over the paper are also given in this section.
 Section 4 is devoted 
to discussion of results for multiphoton emission and theoretical uncertainties in $\phi^*_\eta$ measurements. 
Section~5 is devoted 
to discussion for other than photonic bremsstrahlung effects which contribute to theoretical error of QED FSR simulated 
using { \tt PHOTOS} or { \tt SANC}. In particular,
comments 
on emissions of pairs and interferences between photon emission from final state and other sources 
are given here. 
Section 6, Summary, closes the paper.

Our discussion of theoretical error is 
limited to systematic error of QED FSR only, but is performed in the context of full event generation. 
Other effects, like orientation of spin state for the intermediate $W$'s or $Z$'s, affecting input for calculation  of  QED FSR spin  amplitudes
are addressed in Section 5.3. 
Precision tag for the QED FSR calculation implemented in { \tt PHOTOS} or { \tt SANC} is finally given for 
a broad class of observables: first for 
the photonic bremsstrahlung and then for complete FSR corrections\footnote{
In this paper we use the name {\it photonic bremsstrahlung} whenever we want to stress that only diagrams resulting 
from supplementing  Born level amplitudes with photon lines are considered. The  name {\it final state radiation} 
is used when we stress presence of additional pairs  and {\it final state interaction} when we want to discuss
separation with remaining parts of electroweak interactions.
}.
\section {Description of the programs}

Usually when discussing  phenomenological processes at LHC in the context of 
the detector response one concentrates on description of the  hard process and  initial state  QCD
effects embedded {\it e.g.} in general purpose Monte Carlo generators featuring parton shower, 
models of underlying events  and finally  QCD NLO or NNLO corrections 
to the hard scattering process itself are taken into account,
 see {\it e.g.} \cite{Buckley:2011ms} for review.

In this paper we  concentrate on the QED FSR effects. This determine choice of programs
which will  be used by us for  simulations. The final state interactions, 
consist of  QED bremsstrahlung in decays of electroweak bosons (including cases of
substantial virtualities) and to some degree on the other parts of weak 
corrections as well. Let us recall the massive effort of years 1980-2000 for establishing definition 
of calculation scheme at LEP \cite{Altarelli:1989hv,Altarelli:1989hw,Altarelli:1989hx} where theoretically 
sophisticated and numerically essential
separation of electroweak corrections into initial state, final state, vacuum polarization 
(including definition of the  width) and interferences was established.

In this context discussion of theoretical errors of all parts necessary for predictions is  important as effects of 
new physics and their  interfering genuine weak backgrounds is not straightforward to disentangle.
This represents however further separate  work on weak corrections. In the present paper we 
concentrate on final state radiation, especially on the final state photonic bremsstrahlung.

 For the purpose of this studies two programs featuring QED Final State Radiation for LHC applications
 will be first  described and later used.
We will start with presentation of the {\tt SANC} system~\cite{Andonov:2004hi}, 
because 
it features complete electroweak corrections as well. Description of {\tt PHOTOS} 
\cite{Barberio:1990ms,Barberio:1993qi,Golonka:2005pn,Davidson:2010ew} implementing QED FSR photonic 
bremsstrahlung  only,
will follow.

It might be useful to note that abbreviations LO and NLO in {\tt SANC} and {\tt PHOTOS} have somewhat
different meaning. In the case of {\tt SANC}, ``LO'' (the Leading Order) means just the tree-level
Born cross section, while an exclusive-exponentiation-like notation is adopted in {\tt PHOTOS}, where ``LO'' supposes
exponentiation/resummation of the terms responsible after partial phase space integration, 
for leading logarithmic terms of photonic bremsstrahlung.
Full coverage of multiphoton phase-space is assured. 
The same concerns ``NLO'':
in {\tt SANC} it means the one-loop approximation (Born plus ${\mathcal O}(\alpha)$ EW corrections),
while in {\tt PHOTOS} exponentiation of the ${\mathcal O}(\alpha)$ result is assumed.

\subsection{{ \tt SANC}}

{\tt SANC} is a computer system for Support of Analytic and Numeric 
calculations for experiments at Colliders~\cite{Andonov:2004hi}.
It can be accessed through the Internet at 
{\tt http://sanc.jinr.ru/} and  at {\tt http://pcphsanc.cern.ch/}.
The {\tt SANC} system is suited for calculations of one-loop QED, EW, and QCD
radiative corrections to various SM processes.
Automatized analytic calculations in {\tt SANC} provide
FORM and FORTRAN modules~\cite{Andonov:2008ga}, which can be used
as building blocks in computer codes for particular applications. 

For Drell-Yan-like processes within the {\tt SANC} project there are  implemented:\\[.1cm]
--- complete one-loop EW RC in CC~\cite{Arbuzov:2005dd} and NC~\cite{Arbuzov:2007db} 
    processes;\\[.1cm]
--- photon induced DY processes~\cite{Arbuzov:2007kp};\\[.1cm]
--- higher order photonic FSR in the collinear leading logarithmic approximation;\\[.1cm]
--- higher order photonic and pair FSR in the QED leading logarithmic 
    approximation~\cite{Kuraev:1985hb,Skrzypek:1992vk,Arbuzov:1999cq};\\[.1cm]
--- complete NLO QCD corrections~\cite{Andonov:2007zz,Andonov:2009nn};\\[.1cm]
--- Monte Carlo integrators~\cite{Bardin:2012jk} and event generators;\\[.1cm]
--- interface to parton showers in {\tt PYTHIA} and {\tt HERWIG} based on the standard 
    Les Houches Accord format.

Tuned comparisons with results of {\tt HORACE} \cite{CarloniCalame:2007cd} and {\tt Z(W)GRADE} \cite{Baur:2004ig} for EW RC to CC and NC DY 
were performed within the scope of {\it Les Houches} '05, '07 and {\it TEV4LHC} '06 workshops.
A good agreement achieved in these comparisons confirms correctness of the 
implementation of the complete one-loop EW corrections in all these programs.

An important feature of the {\tt SANC} approach is the possibility to 
control and directly access different contributions to the observables
being under consideration. In particular, {\tt SANC} code allows to separate
effects due to the final state radiation, the interference of 
initial and final radiation, the so called pure weak contributions, {\it etc.} 

Separation of the FSR contribution in the case of neutral current DY processes
is straightforward, it naturally appears at the level of Feynman diagrams.
But the corresponding separation in the case of the charged current DY processes
is not so trivial. In general it is even not gauge invariant. 
For the sake of {\it tuned comparison} with {\tt PHOTOS}, a special prescription for this
separation\footnote{ This prescription  should be respected also in electroweak, non-QED FSR calculations
 used together with {\tt PHOTOS} in practical applications.
}
was introduced into {\tt SANC}. 

Let us consider a formal separation of the pure weak (PW)
and QED contributions $\delta^{PW}$ and $\delta^{QED}$ 
to the total $ W^+ \to u + \bar{d} $ decay width
\begin{equation} \label{gamma_w}
\Gamma_W^{PW+QED} = \Gamma^{LO}(\delta^{PW} + \delta^{QED}).
\end{equation}
This process is described by 6 QED-like diagrams with virtual photon line and 3 other ones with real photon
emission which together 
lead to the  formula
\begin{equation}
\delta^{QED} = \frac{\alpha}{\pi}\left[ Q_W^2 \left( \frac{11}{6} - \frac{\pi^2}{3}\right)
+ (Q_u^2 + Q_d^2) \left( \frac{11}{8} - \frac{3}{4}\log{\frac{M_W^2}{\mu^2_{PW}}} \right) \right],
\nonumber
\end{equation}
where parameter $\mu_{PW}$ is the 't~Hooft scale introduced for separation of QED and PW contributions.
In order to separate the FSR QED contribution, we choose  $ \mu_{PW} = M_W \exp(-\frac{11}{12}) $.
This value of the 't~Hooft scale makes the total QED contribution to the $W$ boson decay
being equal to zero. This is in agreement with the corresponding treatment in {\tt PHOTOS}, where
by construction the effect of FSR to a process does not change 
the normalization of the cross section.

\subsection{{ \tt PHOTOS}}

Already in the era of data analysis of LEP experiments simulation of bremsstrahlung in decays of resonances and particles required 
specialized tools. In parallel to programs oriented toward highest possible overall precision for the whole
 processes in $e^+e^-$ collisions
such as { \tt KKMC} \cite{Jadach:1999vf} or { \tt KORALZ} \cite{Jadach:1993yv}, programs dealing with decays only, gradually 
became of a broad use. The { \tt PHOTOS} Monte Carlo was one of such applications
\cite{Barberio:1990ms,Barberio:1993qi}. Naturally comparisons with these high precision generators  became parts 
of test-beds for { \tt PHOTOS} package. 

The principle of { \tt PHOTOS} algorithm is to replace, on the basis 
of well defined rules, the decay vertex embedded in the event record such as { \tt HEPEVT} \cite{Altarelli:1989wu} or HepMC
\cite{Dobbs:2001ck} with the new one, where additional photons are added.  Such solution, initially
 not aimed for high precision simulations, turned out to be very effective and precise as well. Phase space parameterization was 
carefully documented in \cite{Nanava:2006vv}. Gradually for selected decays 
\cite{Nanava:2006vv,Golonka:2006tw,Nanava:2009vg,Xu:2012px}, also exact  matrix elements were 
implemented and could be activated in place of universal kernels\footnote{Prior to introduction of the {\tt C++} interface 
matrix element kernels were available for our test only. They require more detailed information from the event record which was
available from {\tt PHOTOS} interface  in {\tt FORTRAN}.
}.
 Originally \cite{Barberio:1990ms}, only 
single photon radiation was possible and 
approximations in the universal kernel were present even in the soft photon region. With time,
multiphoton radiation was introduced \cite{Golonka:2005pn} and then installation of exact first order matrix elements
 in $W$ and $Z$ decays became available with C++ implementation of { \tt PHOTOS} \cite{Davidson:2010ew}.
The algorithm of { \tt PHOTOS} is constructed in such a way, that the same function, but with different input kinematical variables,
 is used if the single photon 
emission   or  full multiphoton emission is requested. Such an arrangement enables tests 
in a rigorous first order emission environment. For multiphoton emission, the same kernel is used
iteratively, thanks to the factorization 
properties. Technical checks are thus spared. Optimal solution for the iteration was chosen
and verified with alternative calculations \cite{RichterWas:1993ta,RichterWas:1994ep} based on the second order matrix element. 
It was later extended to the multiphoton
 case for  $Z$ decays in Ref.~\cite{Golonka:2006tw}. Numerical tests of that paper, 
for distributions  of generic kinematical observables pointed to the theoretical precision for the simulation 
of photon bremsstrahlung
of the 0.1\% level. 

When presenting numerical results from {\tt PHOTOS} we always refer to its {\tt C++} version \cite{Davidson:2010ew} with matrix elements for $W$ and $Z$ decays switched on. If the LO level is explicitly  mentioned, matrix elements 
are replaced by universal kernels and algorithm as of {\tt FORTRAN} version 2.14 \cite{Golonka:2005pn}, or higher is used.
At present version \cite{Davidson:2010ew} represents the up-to-date version of {\tt PHOTOS}; it is available, 
with few technical updates, from LCG library\cite{LCG} as well.

\section {Definitions and results of the first order calculations }

As a first step of our tests we have compared numerical results obtained from { \tt PHOTOS} and from { \tt SANC} 
programs in case of the single photon emission. These tests cross-check conventions used and numerical stability of the two calculations. They also verify
 the proper choice of parameters in {\tt PYTHIA8} generator, which is used to produce electroweak Born level events,
on which {\tt PHOTOS} is activated.
We have monitored distributions for the
following observables:
pseudorapidity $\eta$ of $\ell^-$, transverse mass $M_T$ of $\ell^-\bar{\nu}_\ell$ pair and transverse momentum  $p_T$ of
$\ell^-$ in the case of charged current and pseudorapidity $\eta$ of $\ell^-$, invariant mass $M$ of
$\ell^+\ell^-$ pair and transverse momentum $p_T$ of $\ell^-$ in the case of neutral current.

The following  set of input parameters was used:
\begin{equation}
\begin{split}
&G_\mu = 1.6637 \times 10^{-5} \; \mathrm{GeV}^{-2},\\
&M_W = 80.403 \; \mathrm{GeV}, \quad \Gamma_W = 2.091 \; \mathrm{GeV},\\
&M_Z = 91.1876 \; \mathrm{GeV}, \quad \Gamma_Z = 2.4952 \; \mathrm{GeV},\\
&V_{ud} = 0.9738, \quad V_{us} = 0.2272, \quad V_{cd} = 0.2271, \quad V_{cs} = 0.9730,\\
&m_e = 0.511 \; \mathrm{MeV}, \quad m_\mu = 0.10566 \; \mathrm{GeV},
\end{split}
\end{equation}

and the following experiment motivated cuts were applied on momenta of the final state leptons:
\begin{equation}
\begin{split}
\text{NC:} \quad &|\eta(\ell^{+})| < 10, \quad |\eta(\ell^{-})| < 10, \quad
p_\perp(\ell^{+}) > 15 \; \mathrm{GeV}, \quad p_\perp(\ell^{+}) > 15 \; \mathrm{GeV},\\
& 70 \; \mathrm{GeV} < M(\ell^{+}\ell^{-}) < 110 \; \mathrm{GeV};\\
\text{CC:} \quad &|\eta(\ell^{-})| < 10, \quad
p_\perp(\ell^{-}) > 0.1\; \mathrm{GeV}, \quad p_\perp(\bar{\nu}_\ell) > 0.1\; \mathrm{GeV}.
\end{split}
\end{equation}

We work in the running width scheme for $W$ and $Z$ boson propagators and fix  value of the weak mixing angle:
$ \cos{\theta_W} = M_W/M_Z$, $\sin^2{\theta_W} = 1 - \cos^2{\theta_W} $. The value of the electromagnetic coupling $\alpha$
is evaluated in the $G_\mu$-scheme using the Fermi constant $G_\mu$: the effective coupling is defined by
$\alpha_{G_\mu} = \sqrt{2}G_\mu M_W^2\sin^2{\theta_W}/\pi$.

To compute the hadronic cross section we have used CTEQ6L1 set of parton distribution functions with running factorization scale
$\mu_r^2 = \hat{s}$, where $\hat{s}$ is squared total energy of the colliding partons in their center-of-mass system.

Comparison is performed for muon and electron final states. For the electron final states and
for each observable the {\it bare} and {\it calo} results are provided. In the {\it calo} case the four-momenta of the final
electron and photon are combined into effective four-momentum of the electron when the separation
$\Delta R = \sqrt{(\Delta\eta(e,\gamma))^2 + (\Delta\phi(e,\gamma))^2} < 0.1$.

\begin{figure}[htp!]
\begin{tabular}{ccc}
\subfigure{
\includegraphics[%
  width=0.40\columnwidth]{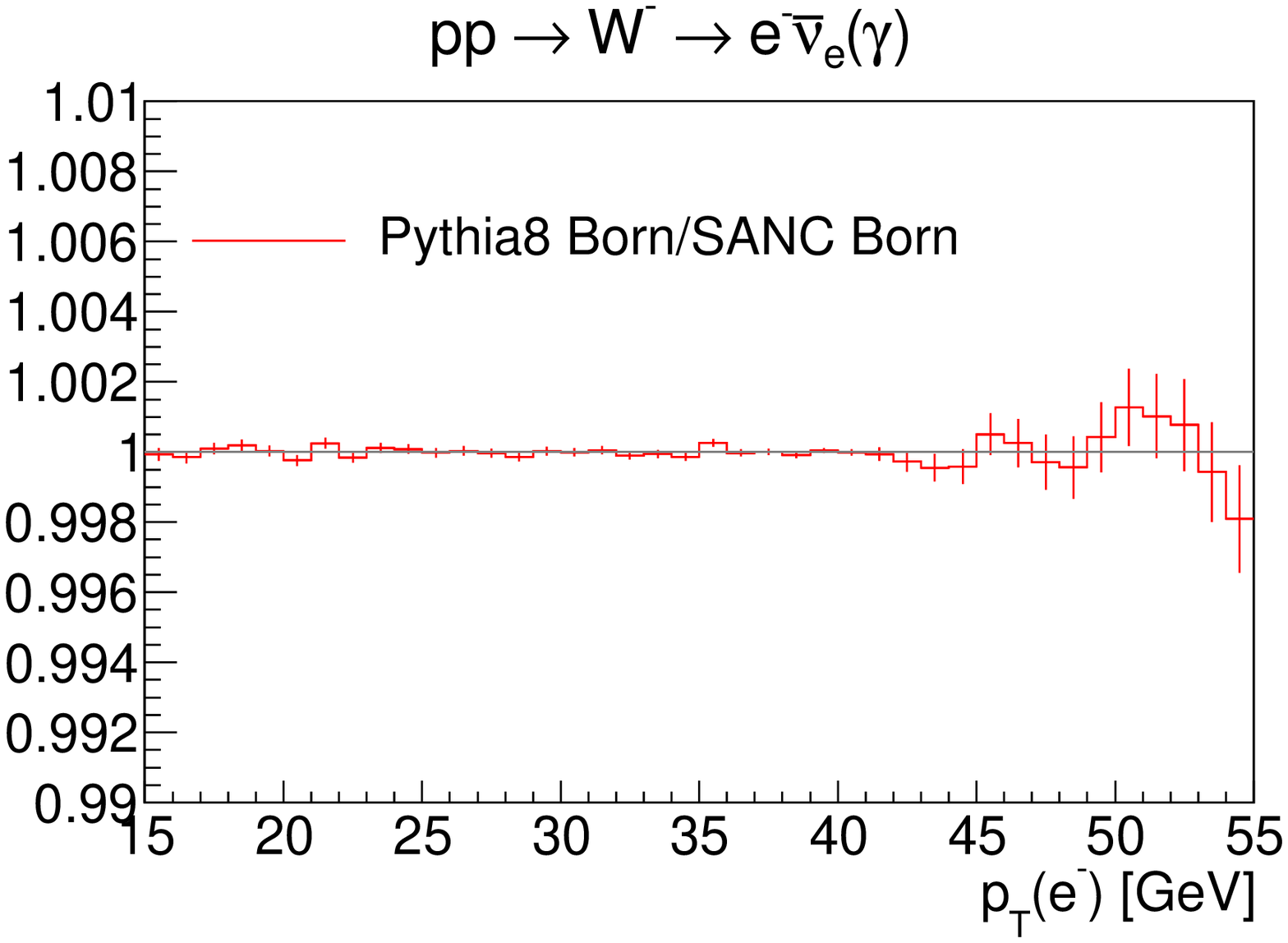}} & \subfigure{\includegraphics[%
  width=0.40\columnwidth]{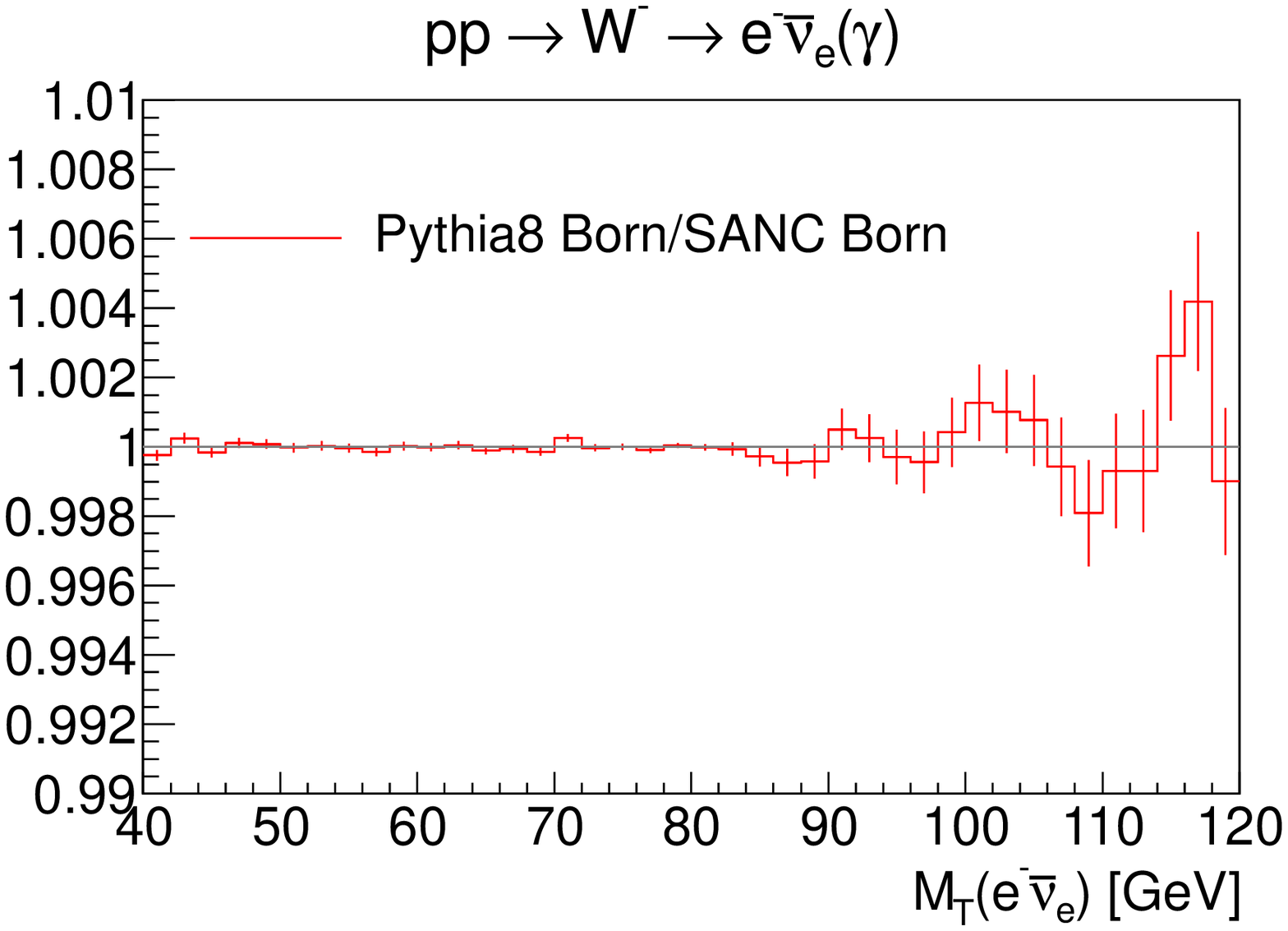}} \\
\subfigure{  \includegraphics[%
  width=0.40\columnwidth]{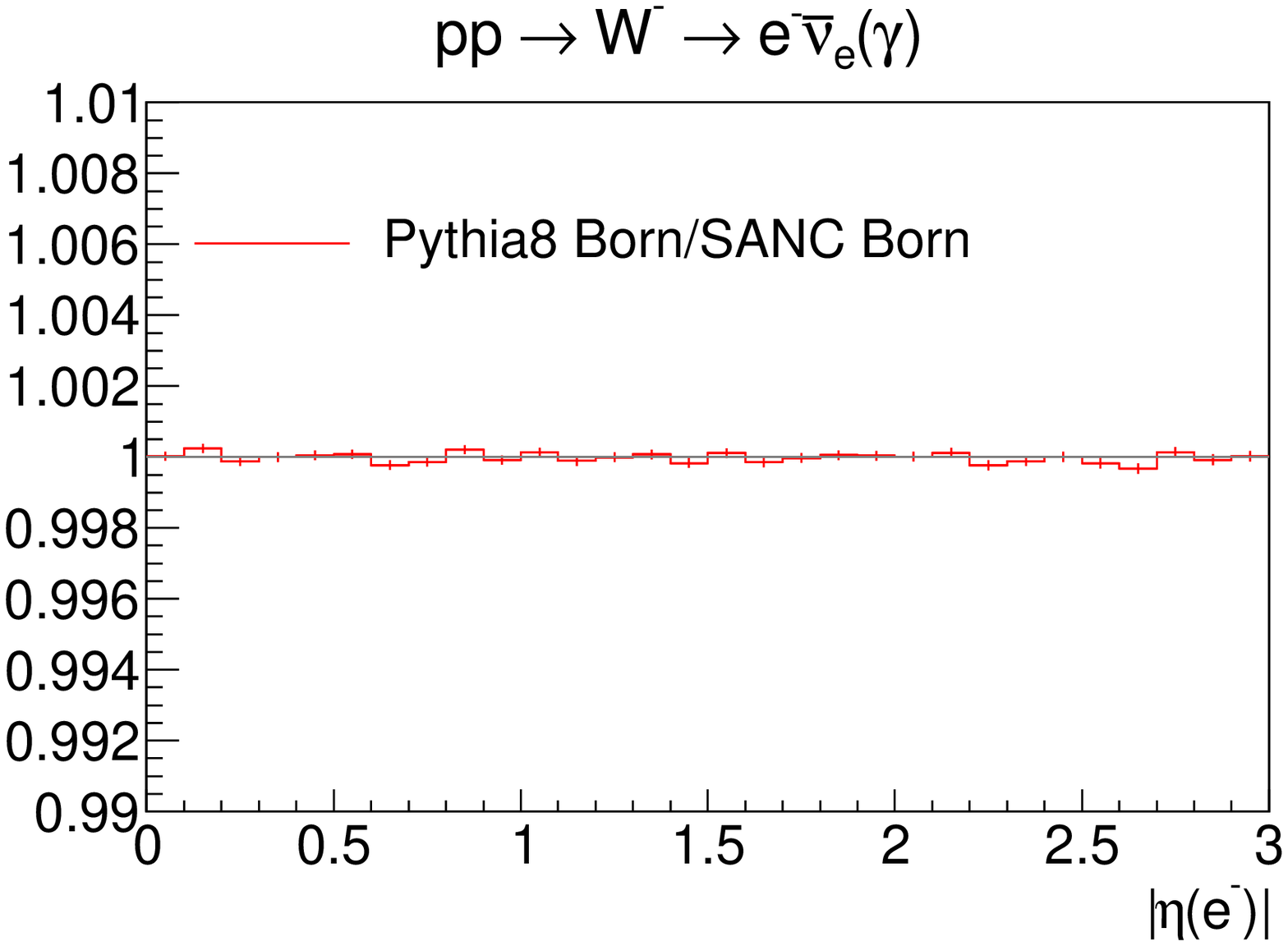}}
\end{tabular}
\caption{Ratios for Born-level distributions in $W\to e\nu$ decay.  
 \label{WBornRatio}}
\end{figure}

\begin{figure}[htp!]
\begin{tabular}{ccc}
\subfigure{
\includegraphics[%
  width=0.40\columnwidth]{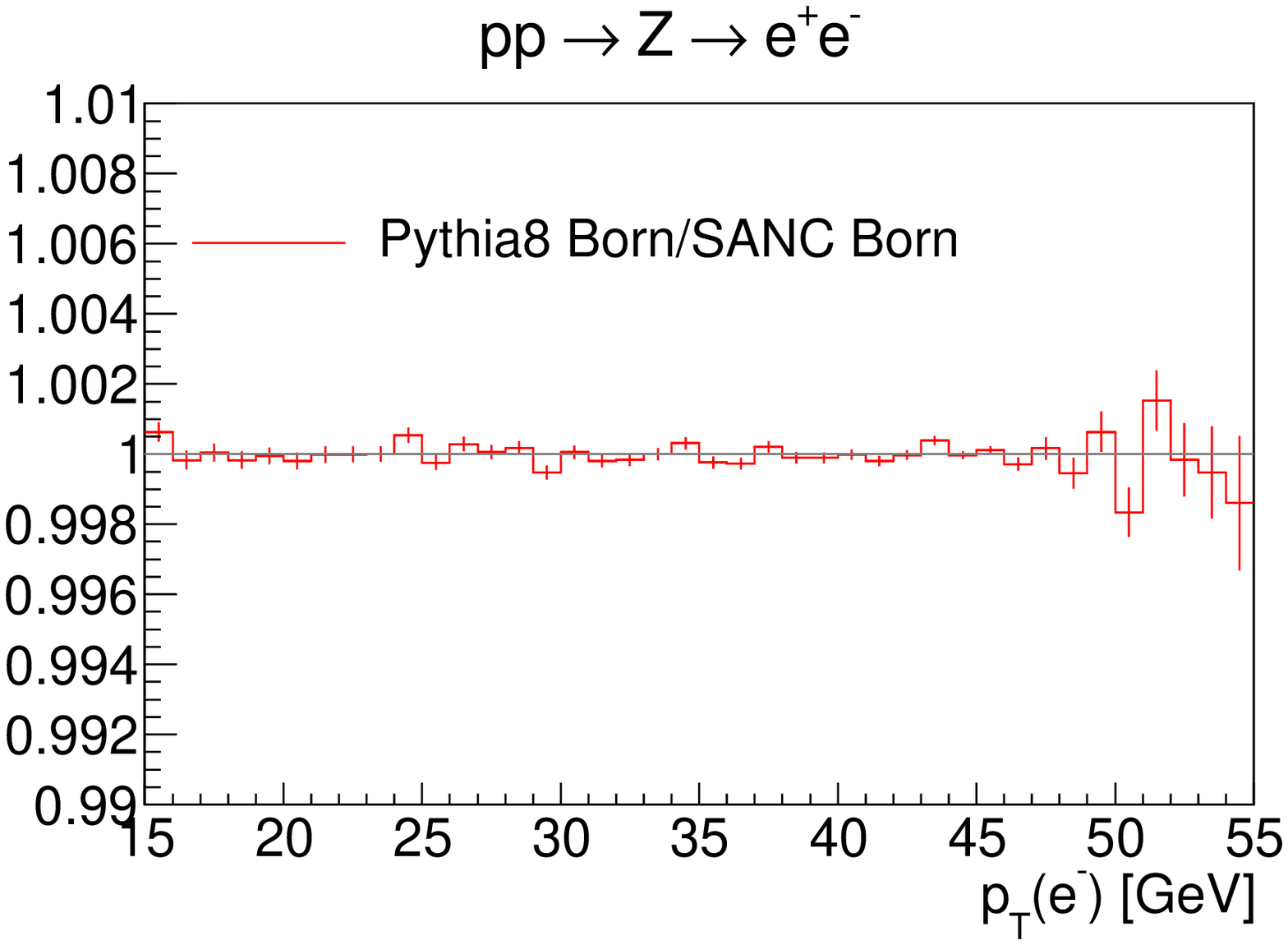}} & \subfigure{\includegraphics[%
  width=0.40\columnwidth]{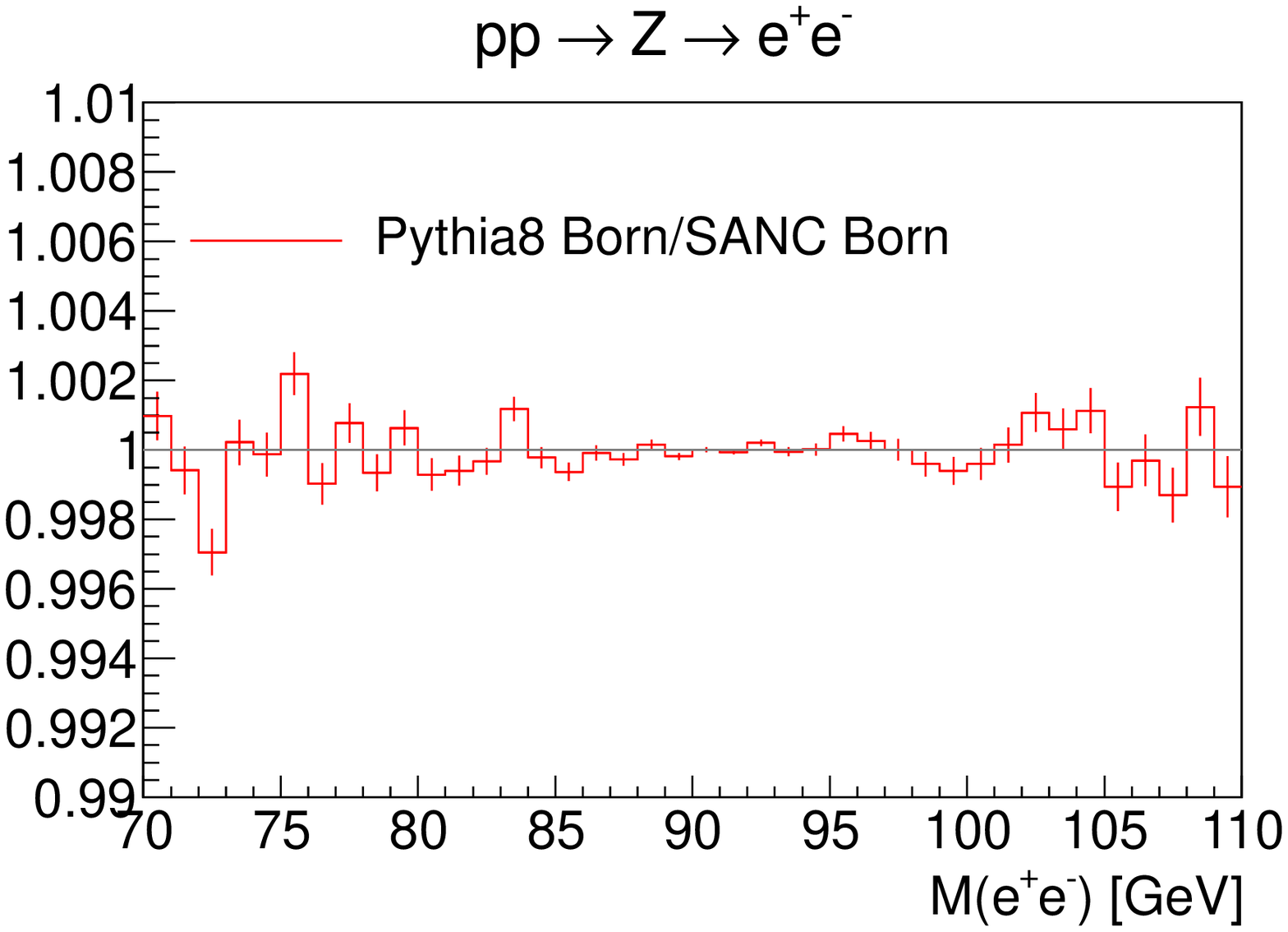}} \\
\subfigure{  \includegraphics[%
  width=0.40\columnwidth]{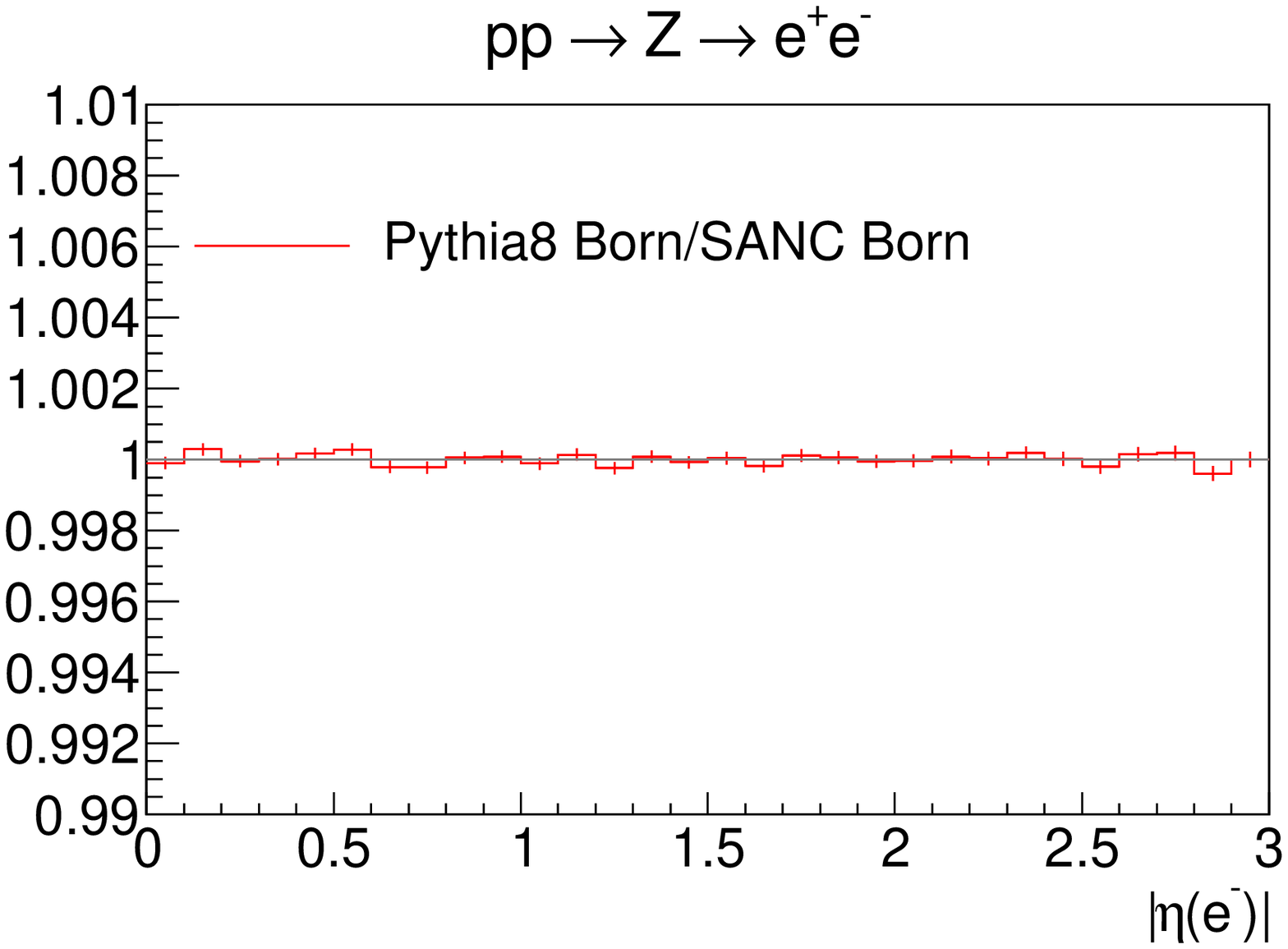}}
\end{tabular}
\caption{Ratios for Born-level distributions in $Z\to ee$ decay.
 \label{ZBornRatio}}
\end{figure}

To check that the normalization of Born cross sections is properly adjusted between simulations using {\tt SANC}
and {\tt PHOTOS}, we have completed tests at a
sub-permille precision level. The corresponding results for electrons and  ratios of differential cross sections 
are shown in Fig. \ref{WBornRatio} for CC and in Fig. \ref{ZBornRatio} for NC.

\begin{figure}[htp!]
\begin{tabular}{ccc}
\subfigure{
\includegraphics[%
  width=0.40\columnwidth]{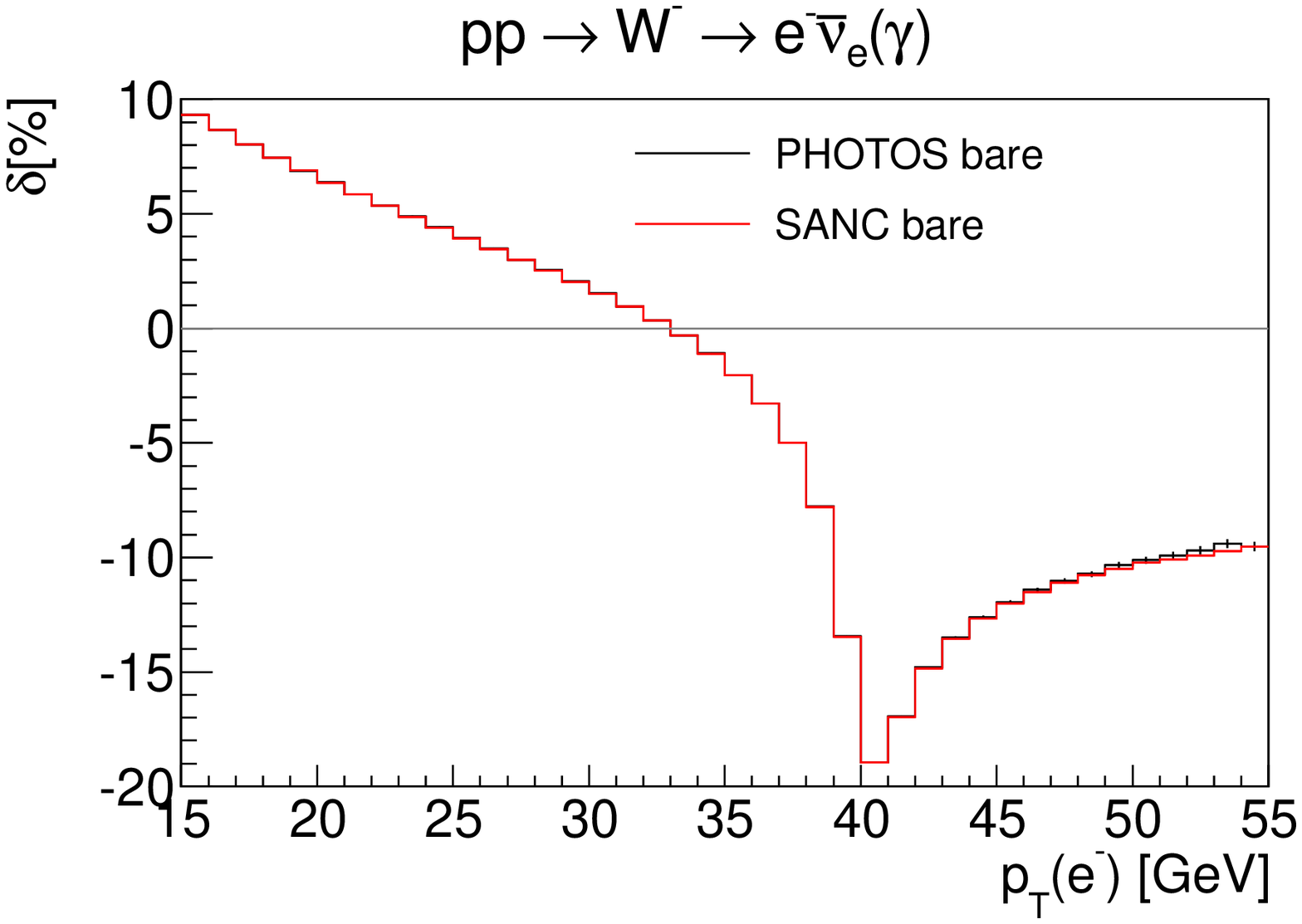}} & \subfigure{\includegraphics[%
  width=0.40\columnwidth]{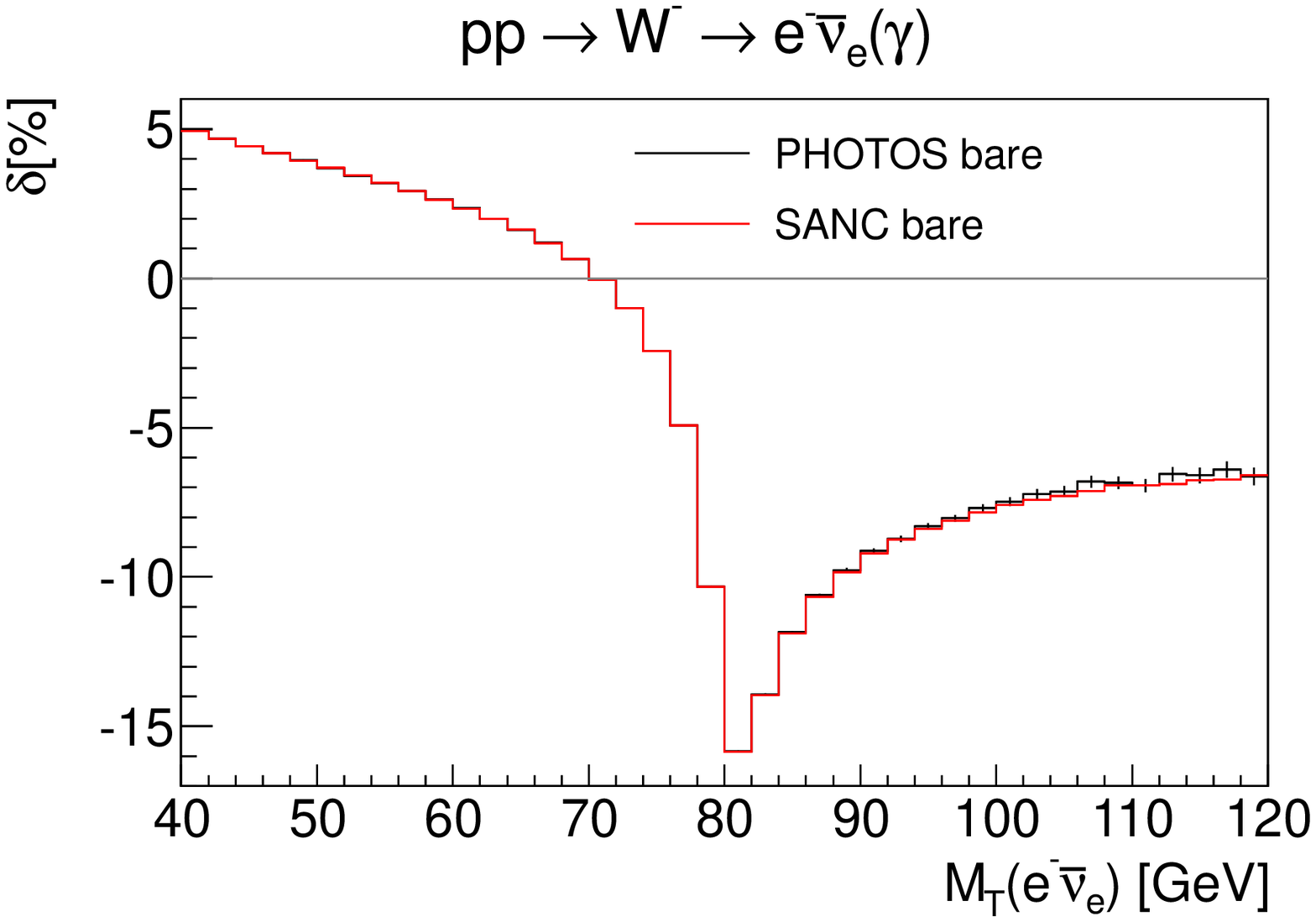}} \\
\subfigure{  \includegraphics[%
  width=0.40\columnwidth]{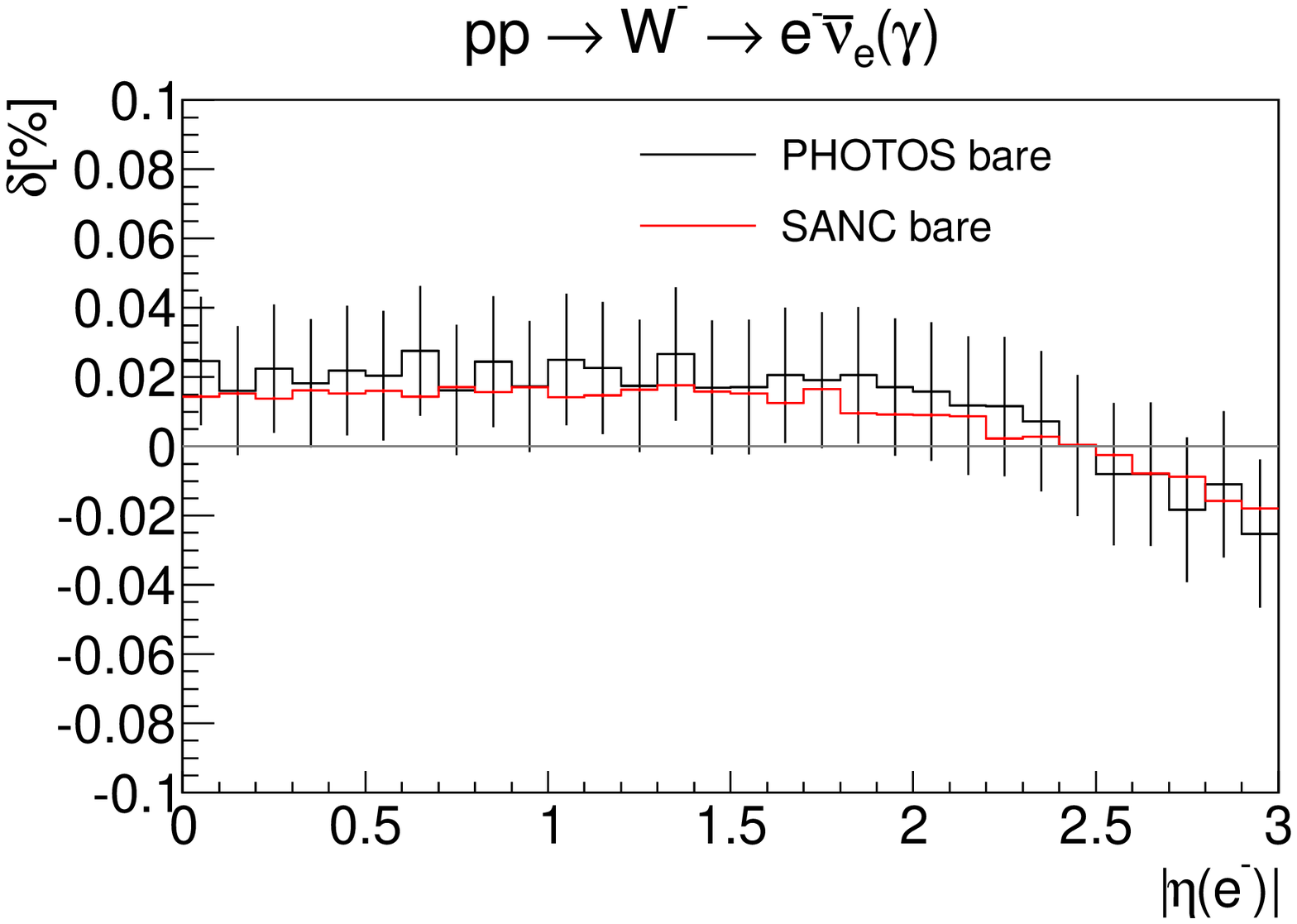}}
\end{tabular}
\caption{$\mathcal{O}(\alpha)$ corrections for  basic  kinematical distributions from { \tt PYTHIA+PHOTOS} and { \tt SANC} in 
$W\to e \nu$ decay.
 \label{WfirstELbare}}
\end{figure}

\begin{figure}[htp!]
\begin{tabular}{ccc}
\subfigure{
\includegraphics[%
  width=0.40\columnwidth]{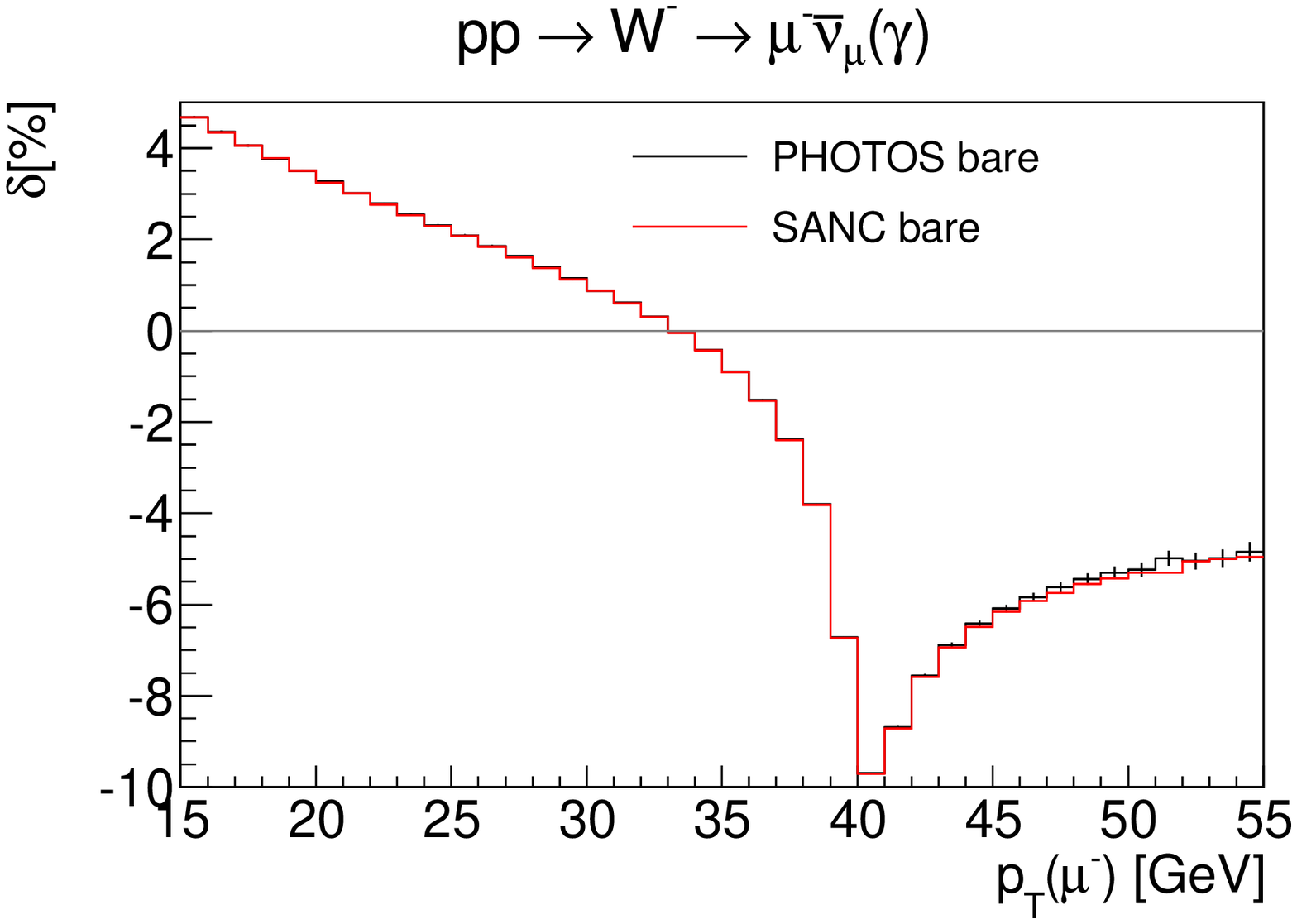}} & \subfigure{\includegraphics[%
  width=0.40\columnwidth]{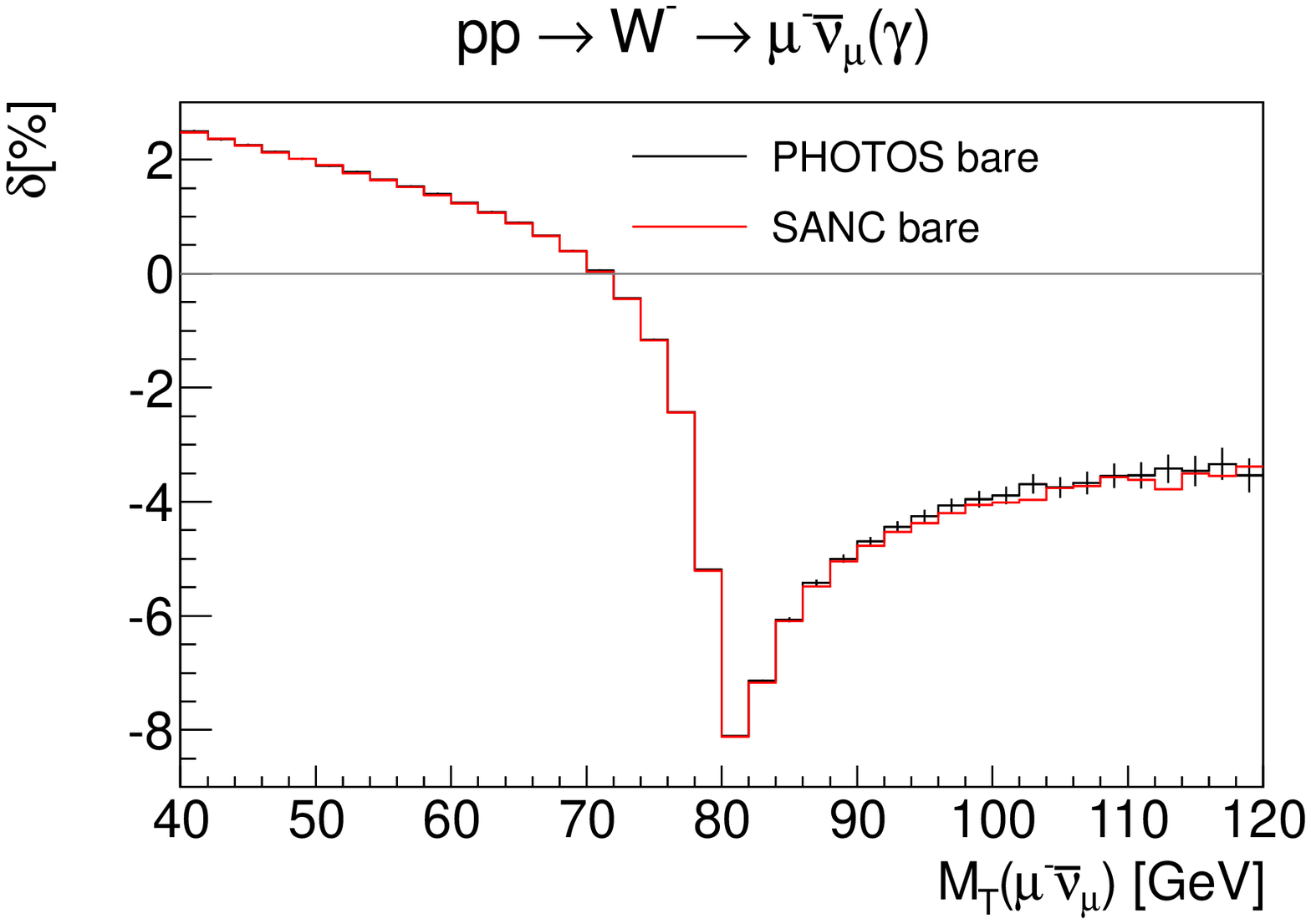}} \\
\subfigure{  \includegraphics[%
  width=0.40\columnwidth]{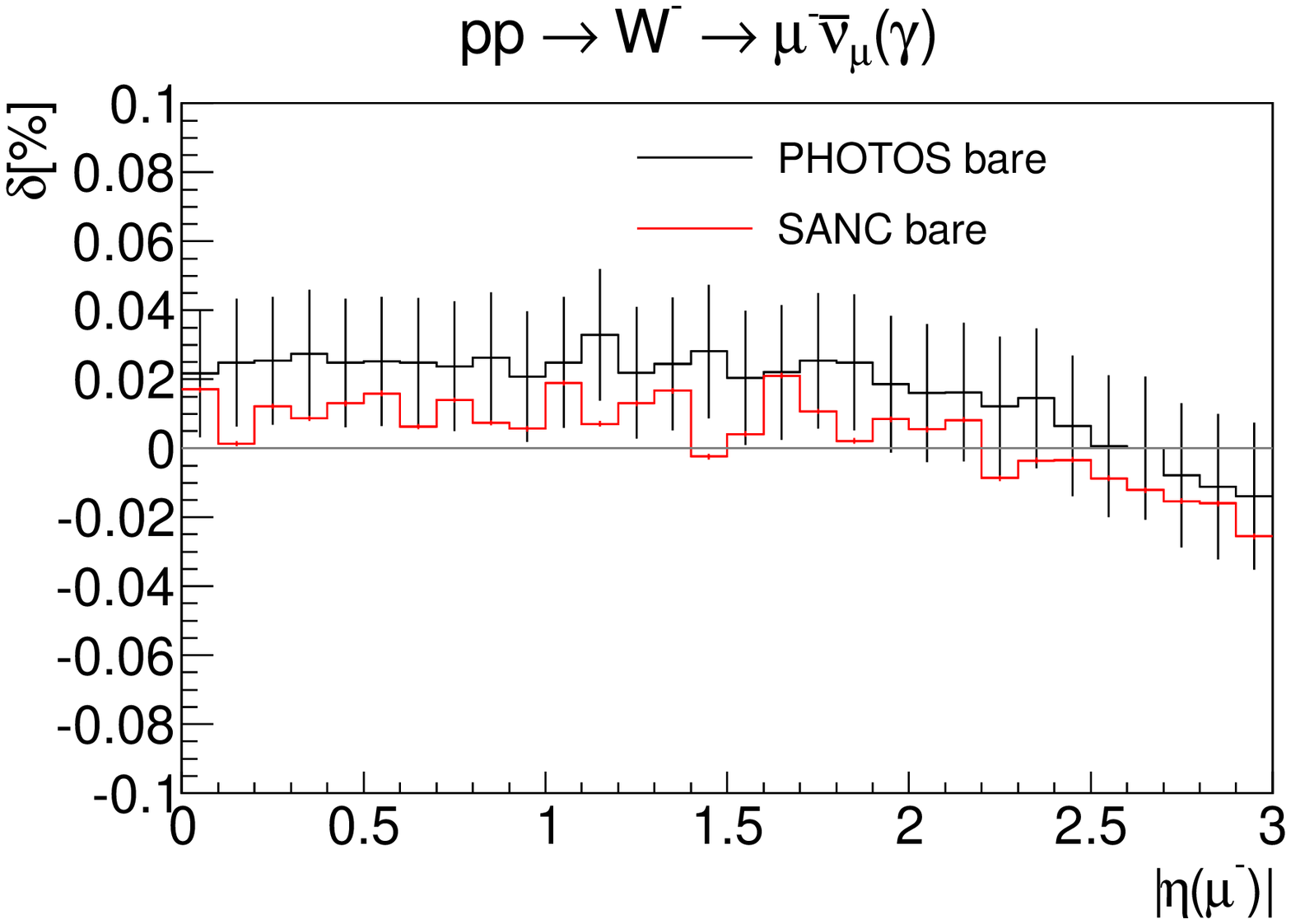}}
\end{tabular}
\caption{$\mathcal{O}(\alpha)$ corrections for  basic kinematical distributions from { \tt PYTHIA+PHOTOS} and { \tt SANC} in $W\to \mu \nu$ decay.
 \label{WfirstMUbare}}
\end{figure}

\begin{figure}[htp!]
\begin{tabular}{ccc}
\subfigure{
\includegraphics[%
  width=0.40\columnwidth]{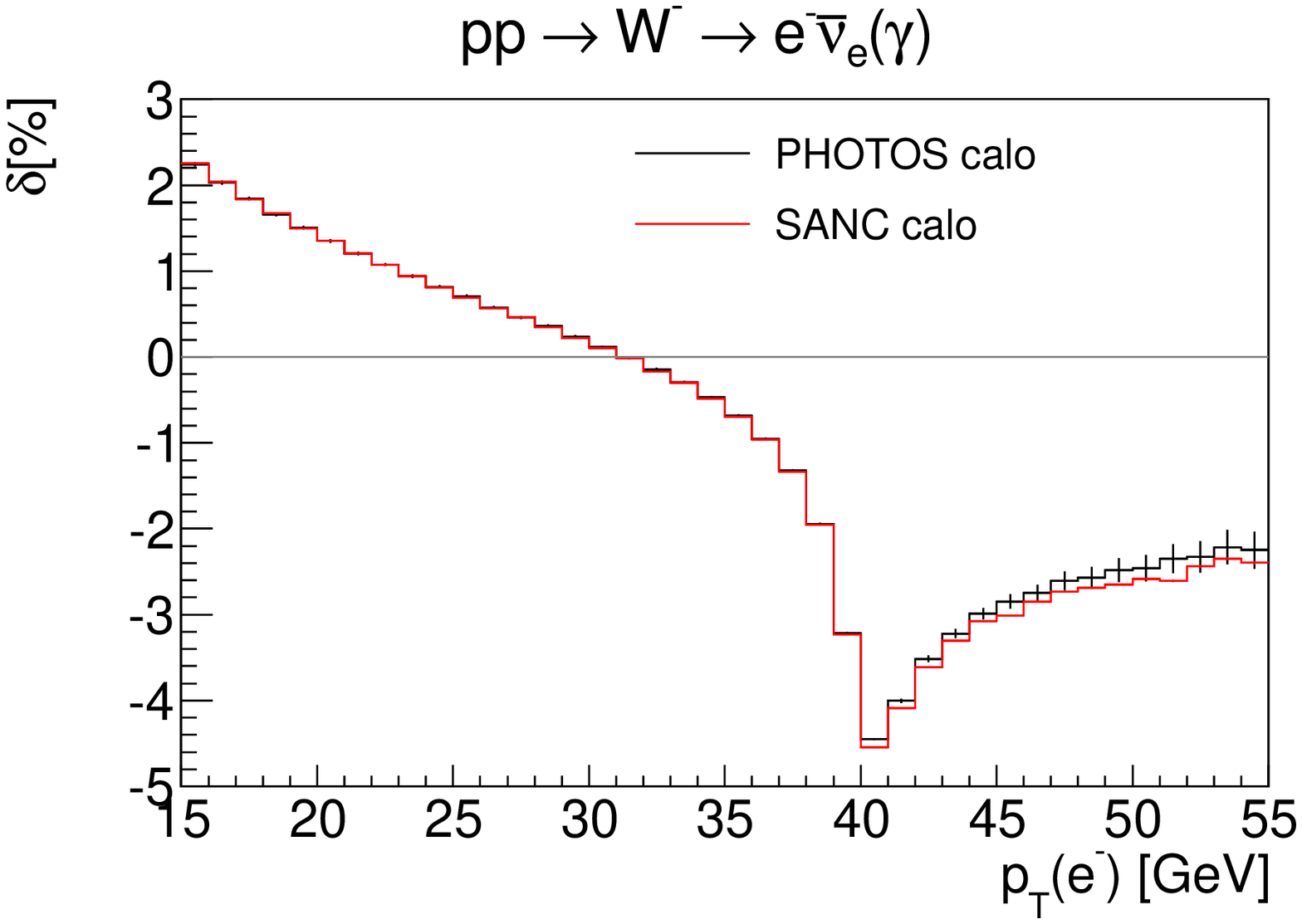}} & \subfigure{\includegraphics[%
  width=0.40\columnwidth]{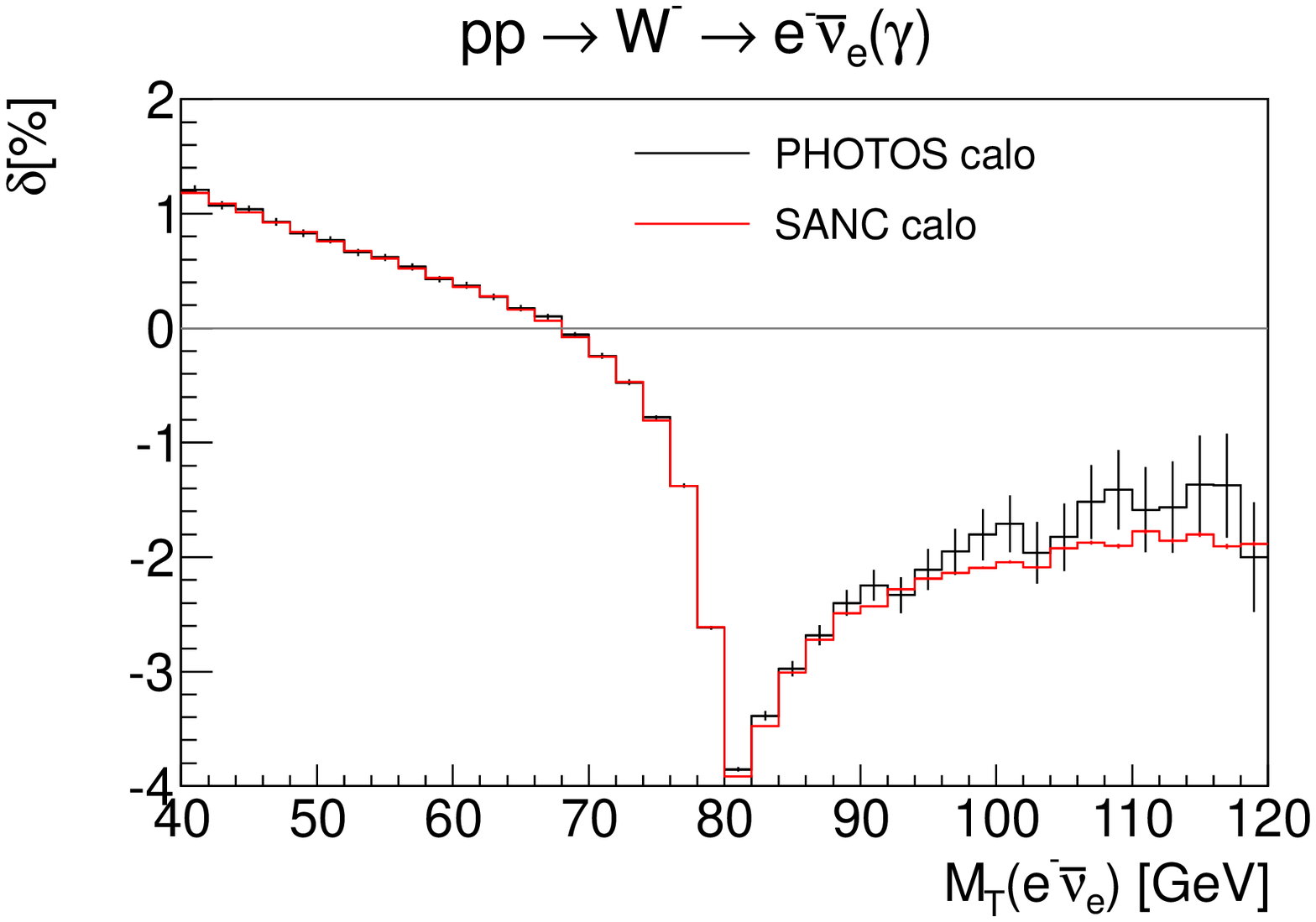}} \\
\subfigure{  \includegraphics[%
  width=0.40\columnwidth]{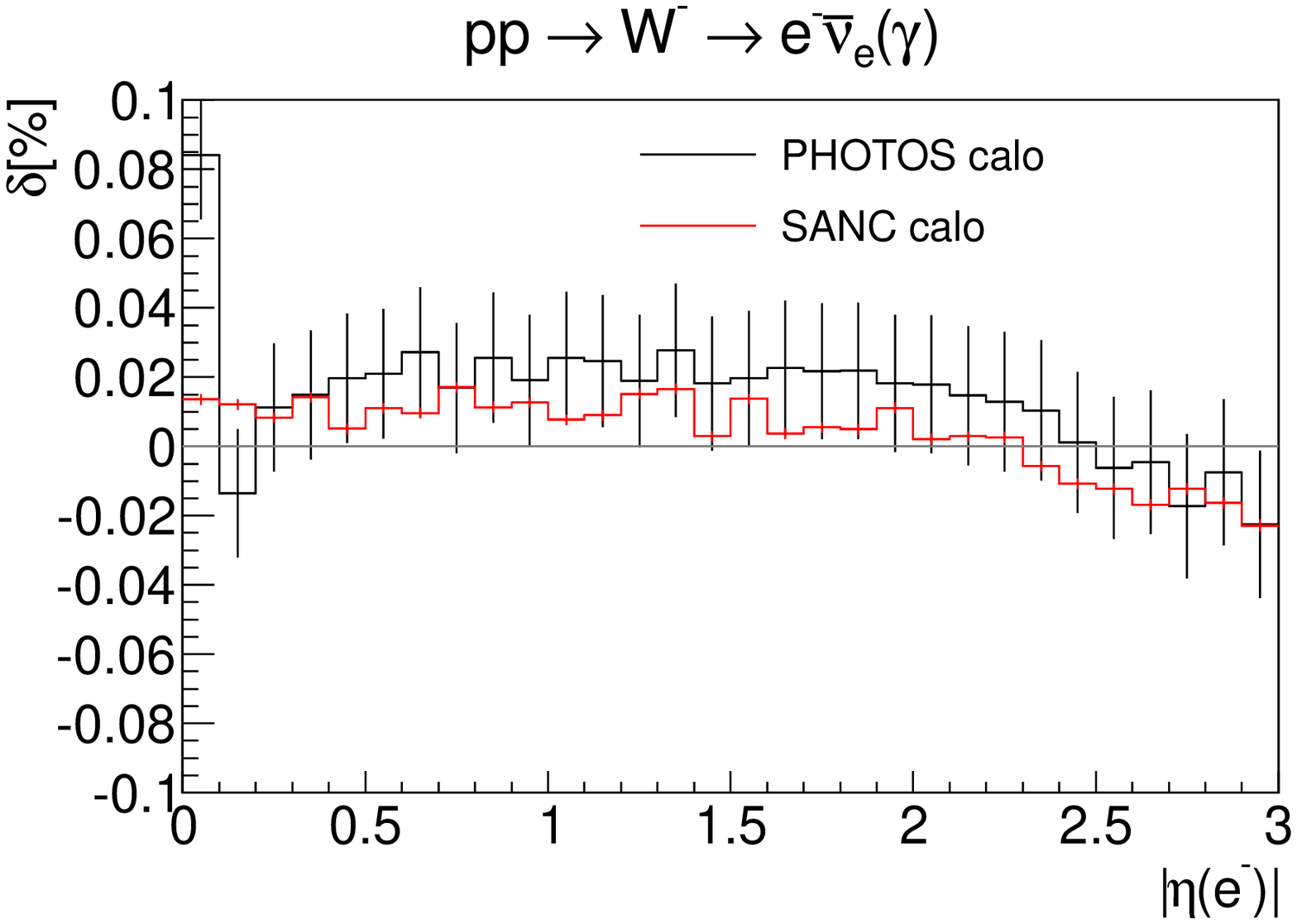}}
\end{tabular}
\caption{$\mathcal{O}(\alpha)$ corrections for  basic kinematical distributions from { \tt PYTHIA+PHOTOS} and { \tt SANC} in $W\to e \nu$ decay  ({\it calo} electrons).
 \label{WfirstELcalo}}
\end{figure}

\begin{figure}[htp!]
\begin{tabular}{ccc}
\subfigure{
\includegraphics[%
  width=0.40\columnwidth]{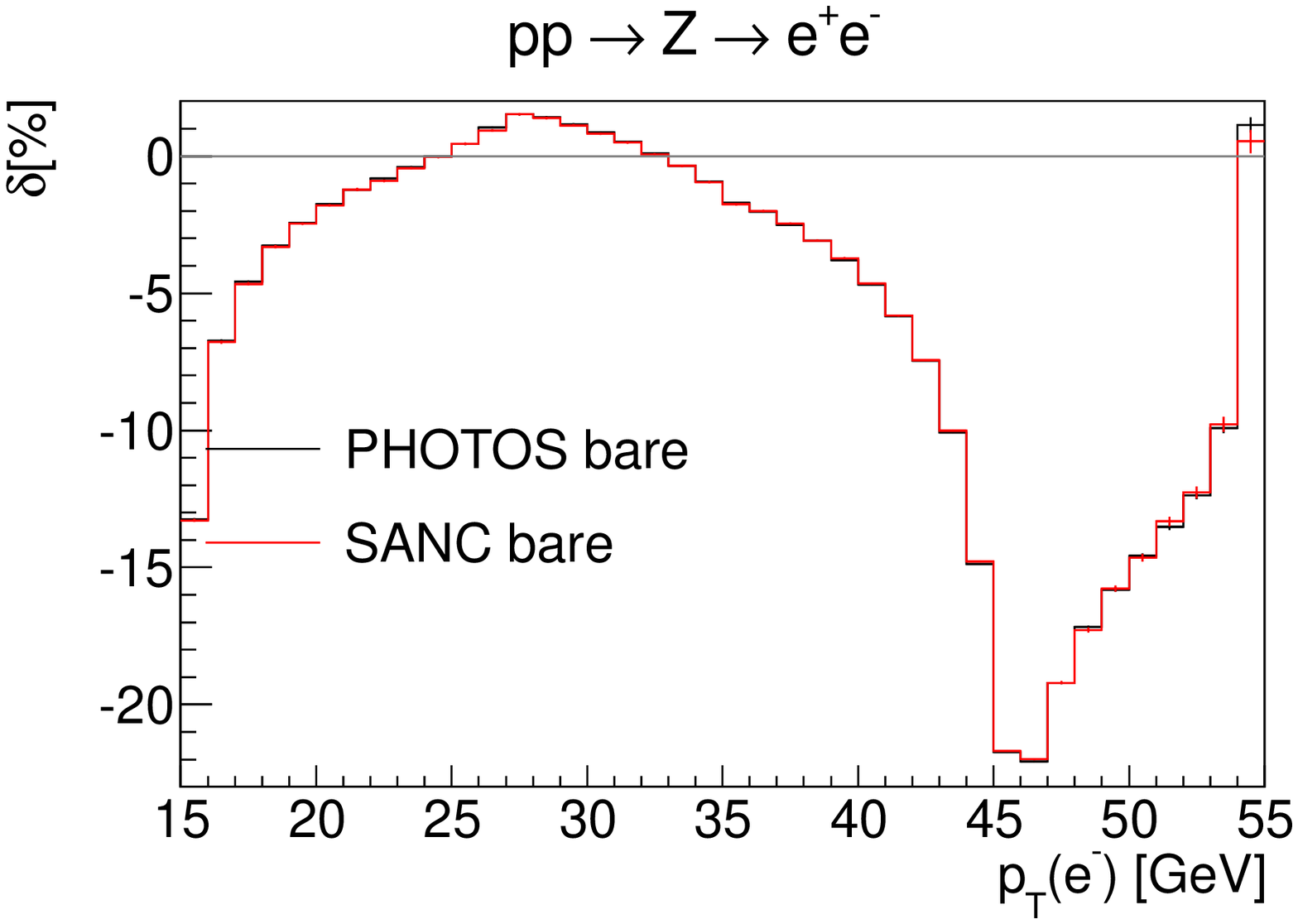}} & \subfigure{\includegraphics[%
  width=0.40\columnwidth]{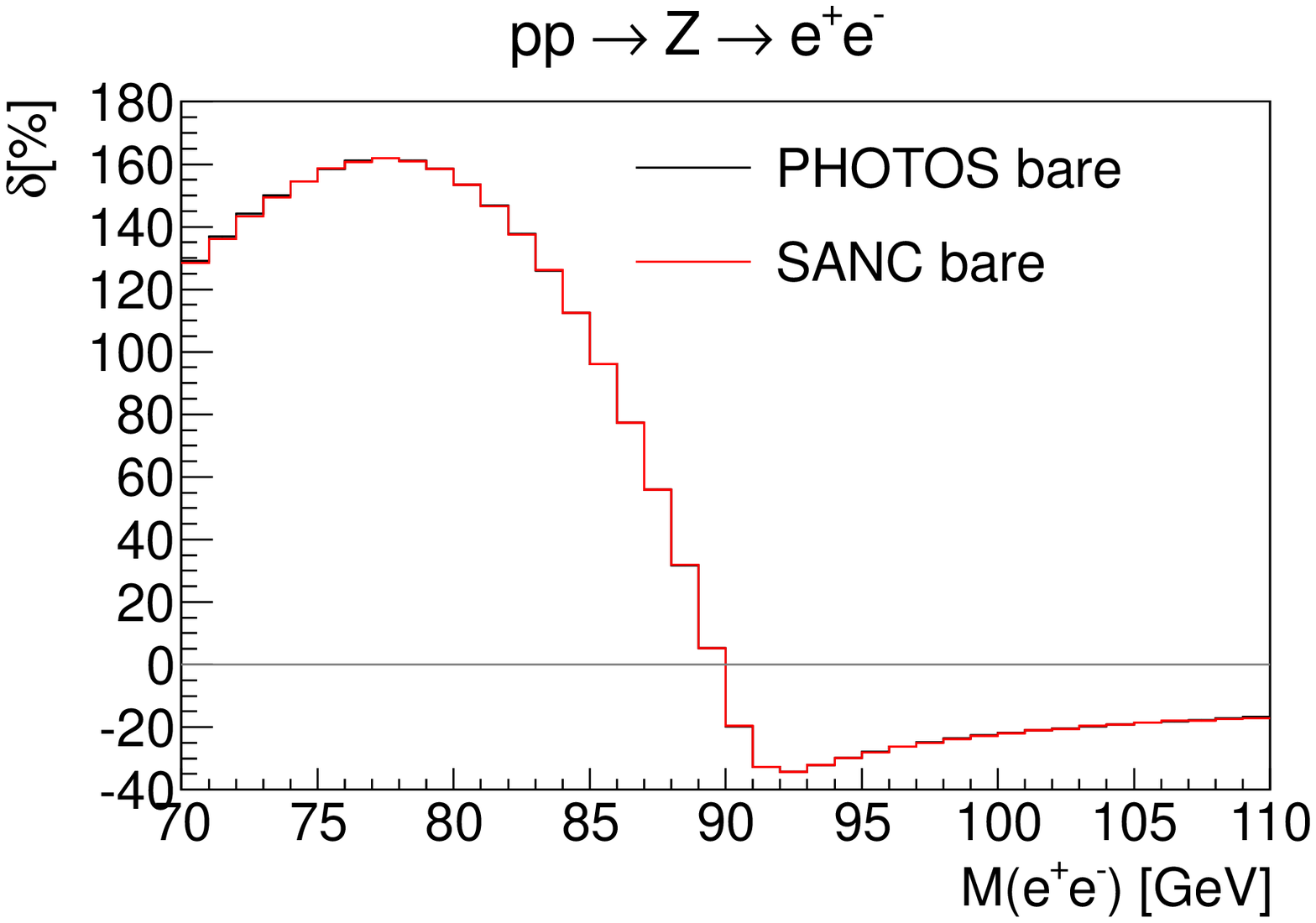}} \\
\subfigure{  \includegraphics[%
  width=0.40\columnwidth]{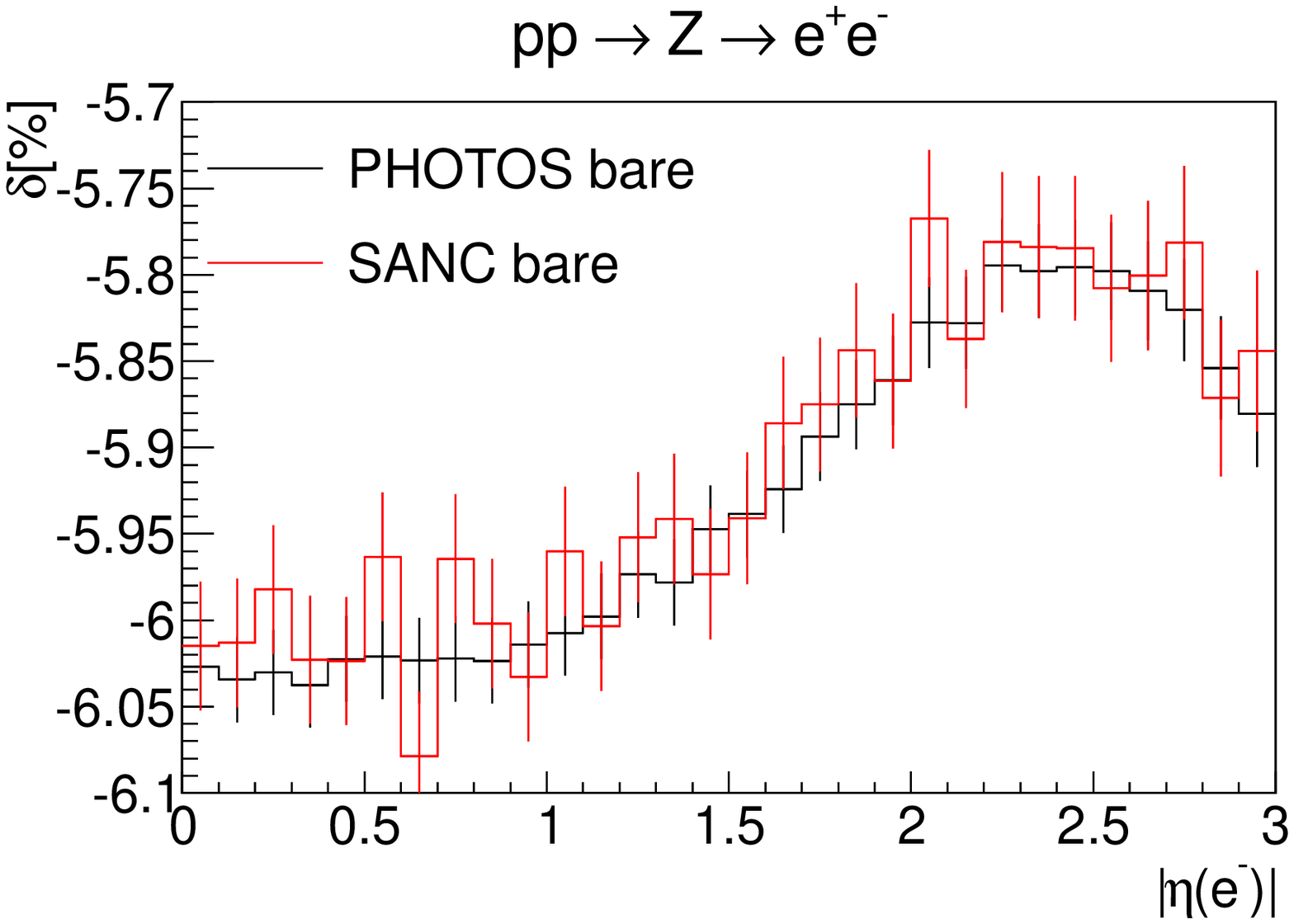}}
\end{tabular}
\caption{$\mathcal{O}(\alpha)$ corrections for  basic kinematical distributions from { \tt PYTHIA+PHOTOS} and { \tt SANC} in $Z\to ee$ decay.
 \label{ZfirstELbare}}
\end{figure}

\begin{figure}[htp!]
\begin{tabular}{ccc}
\subfigure{
\includegraphics[%
  width=0.40\columnwidth]{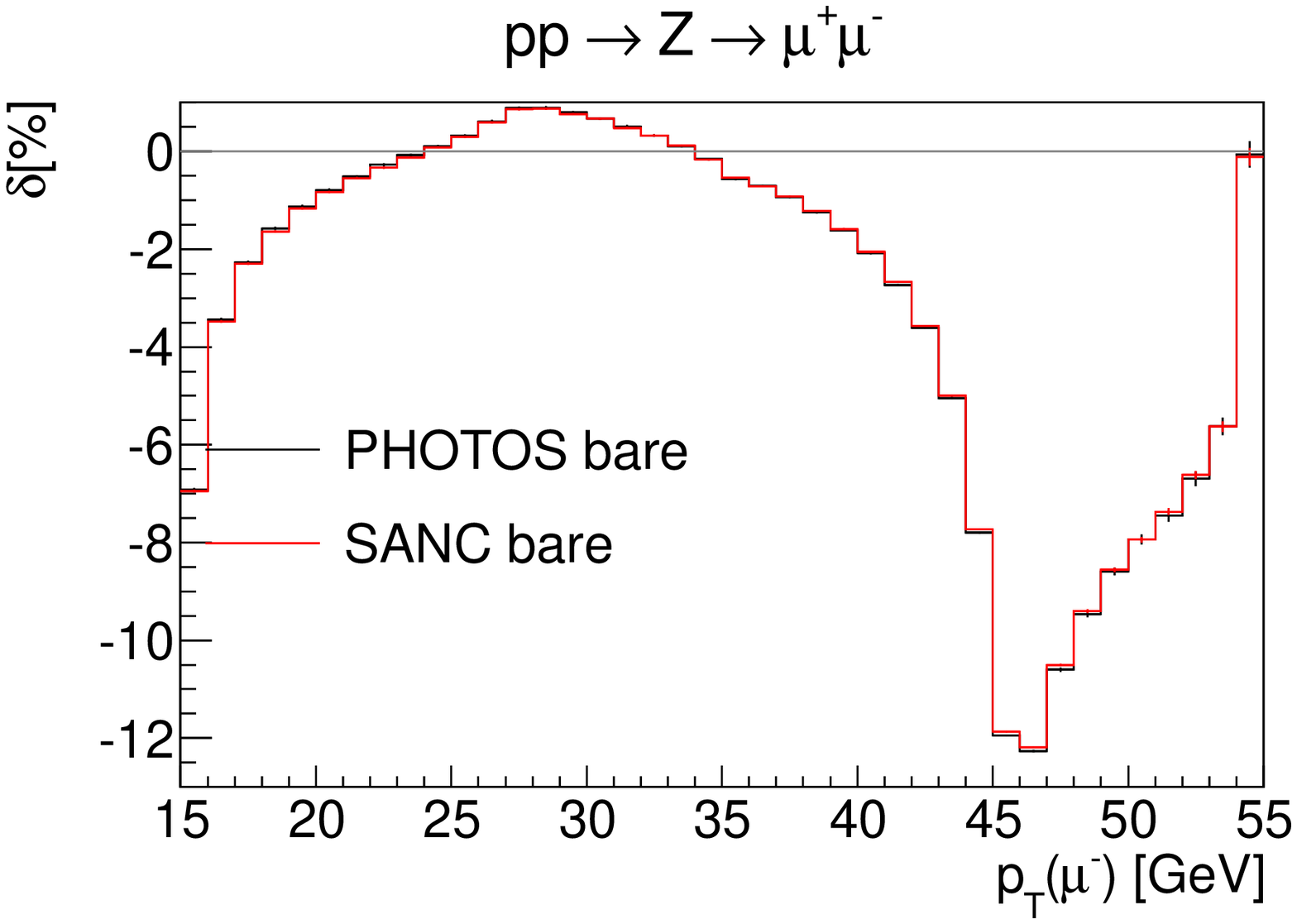}} & \subfigure{\includegraphics[%
  width=0.40\columnwidth]{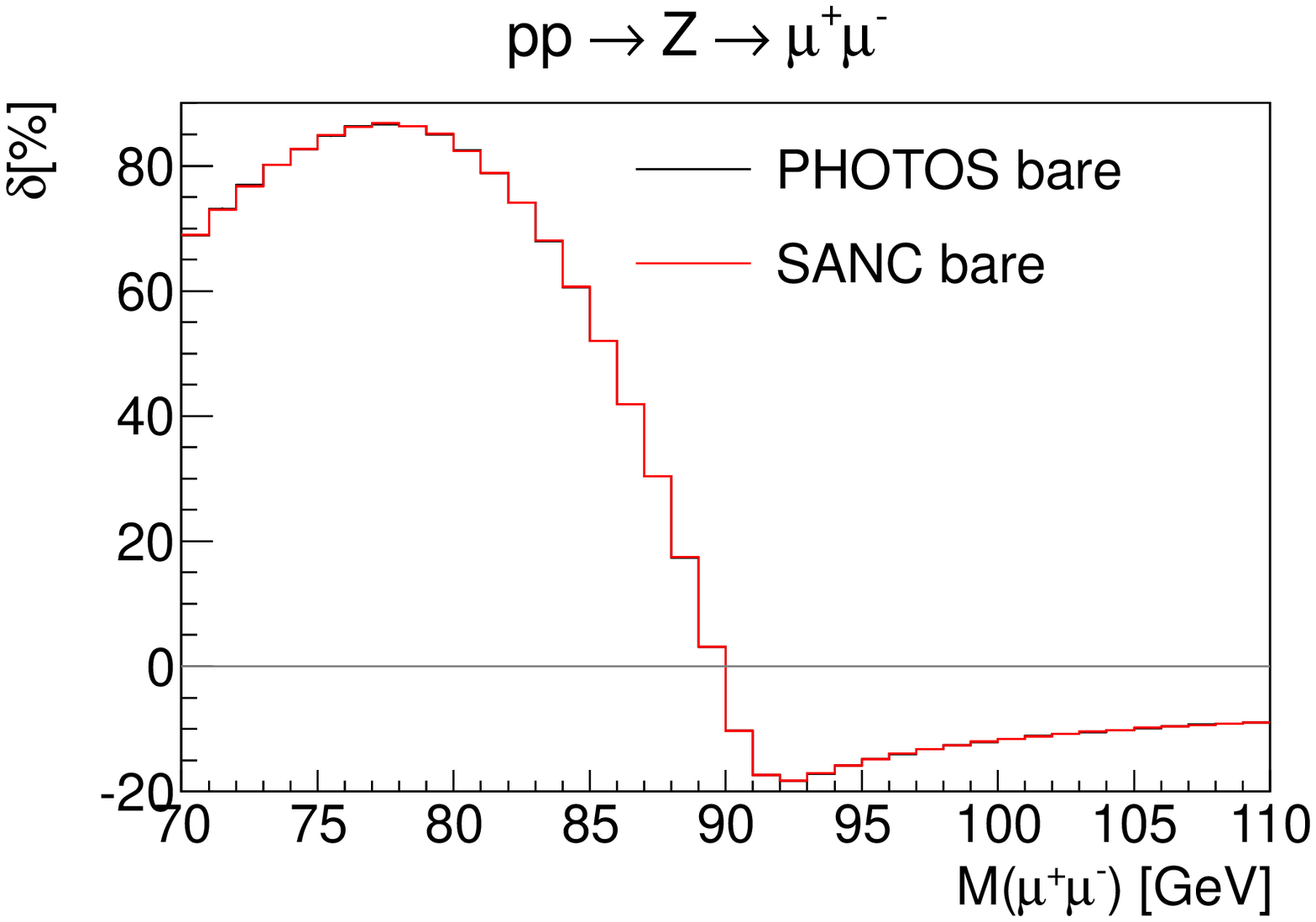}} \\
\subfigure{  \includegraphics[%
  width=0.40\columnwidth]{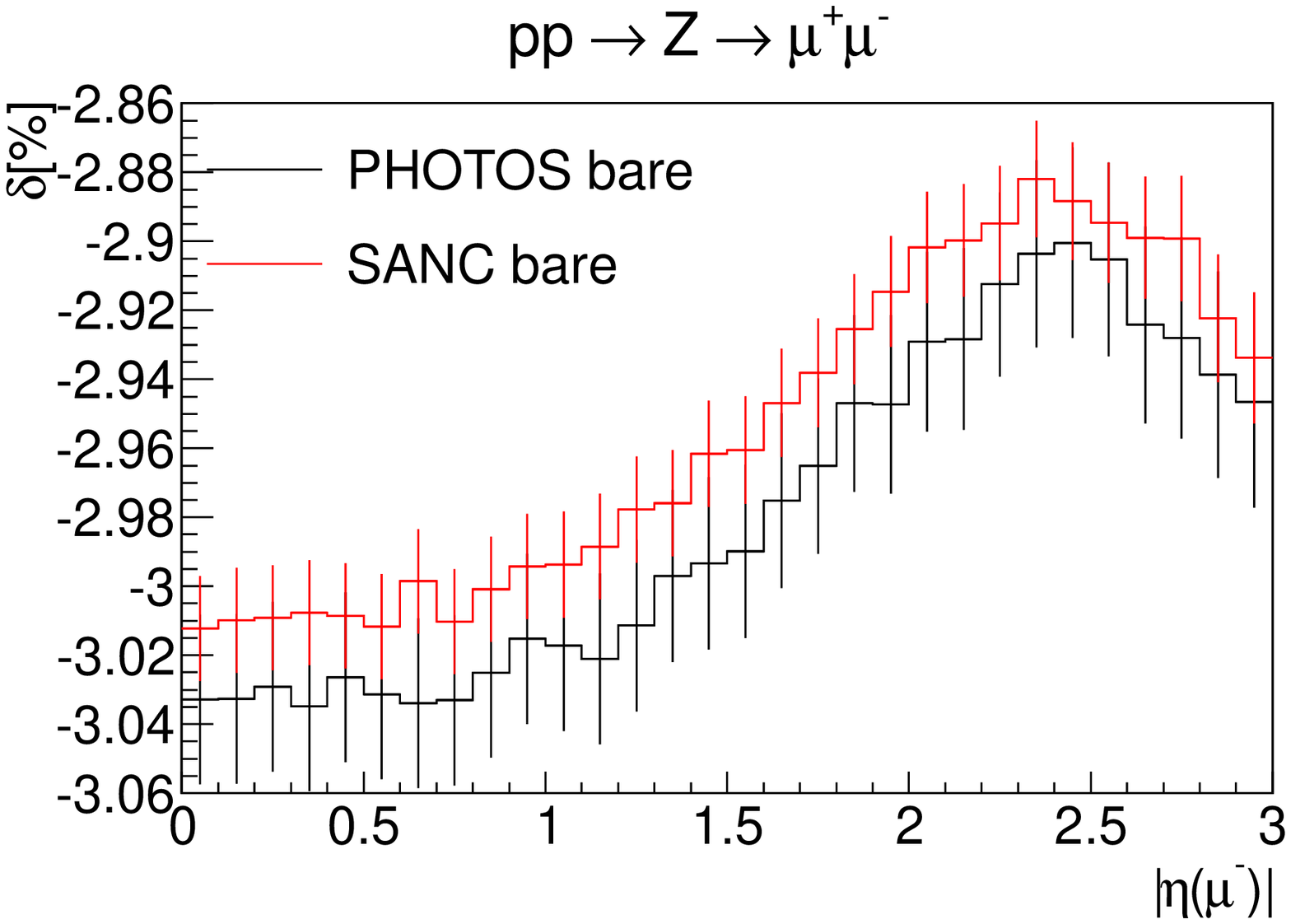}}
\end{tabular}
\caption{$\mathcal{O}(\alpha)$ corrections for  basic kinematical distributions from { \tt PYTHIA+PHOTOS} and { \tt SANC} in $Z\to \mu\mu$ decay.
 \label{ZfirstMUbare}}
\end{figure}

\begin{figure}[htp!]
\begin{tabular}{ccc}
\subfigure{
\includegraphics[%
  width=0.40\columnwidth]{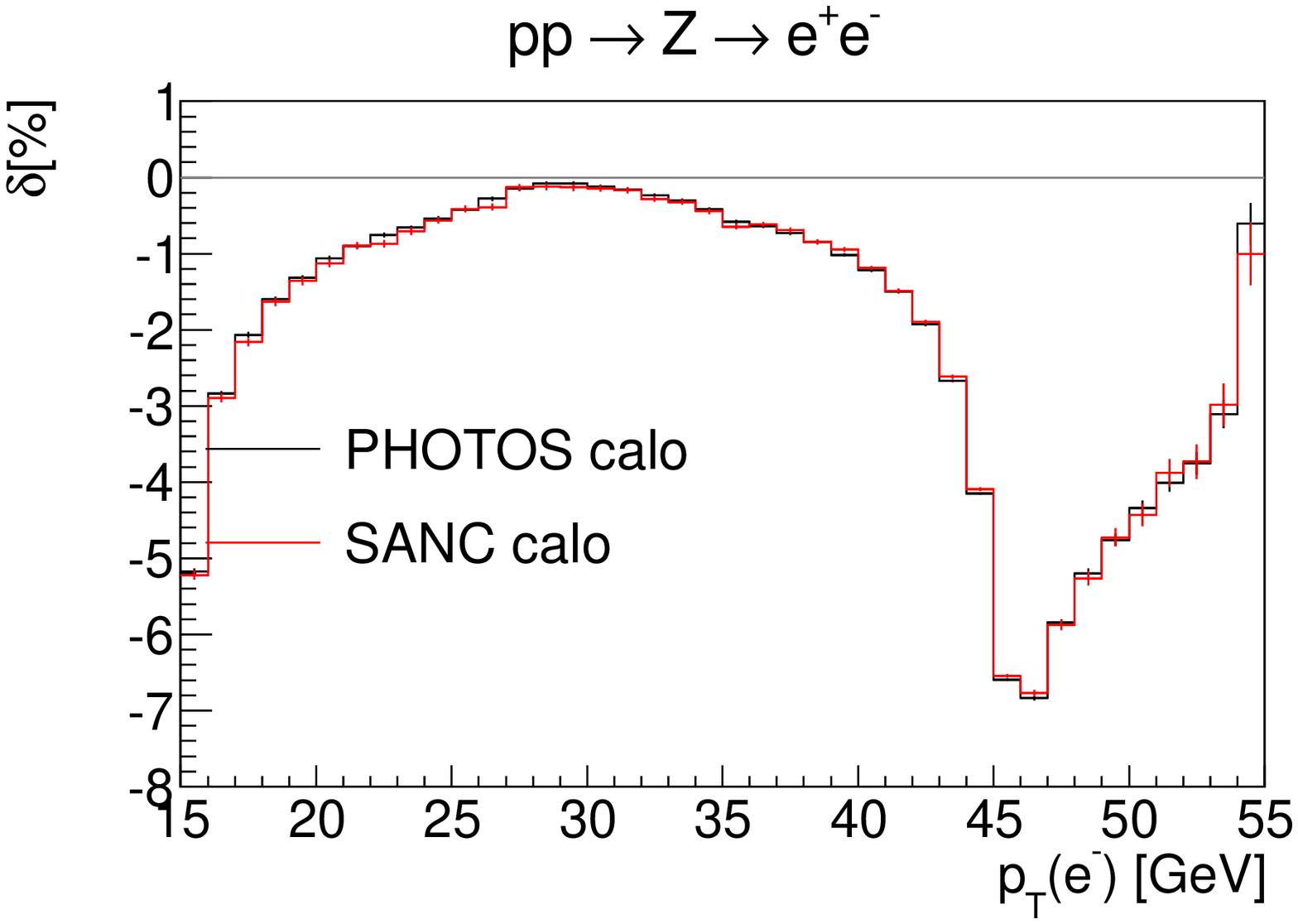}} & \subfigure{\includegraphics[%
  width=0.40\columnwidth]{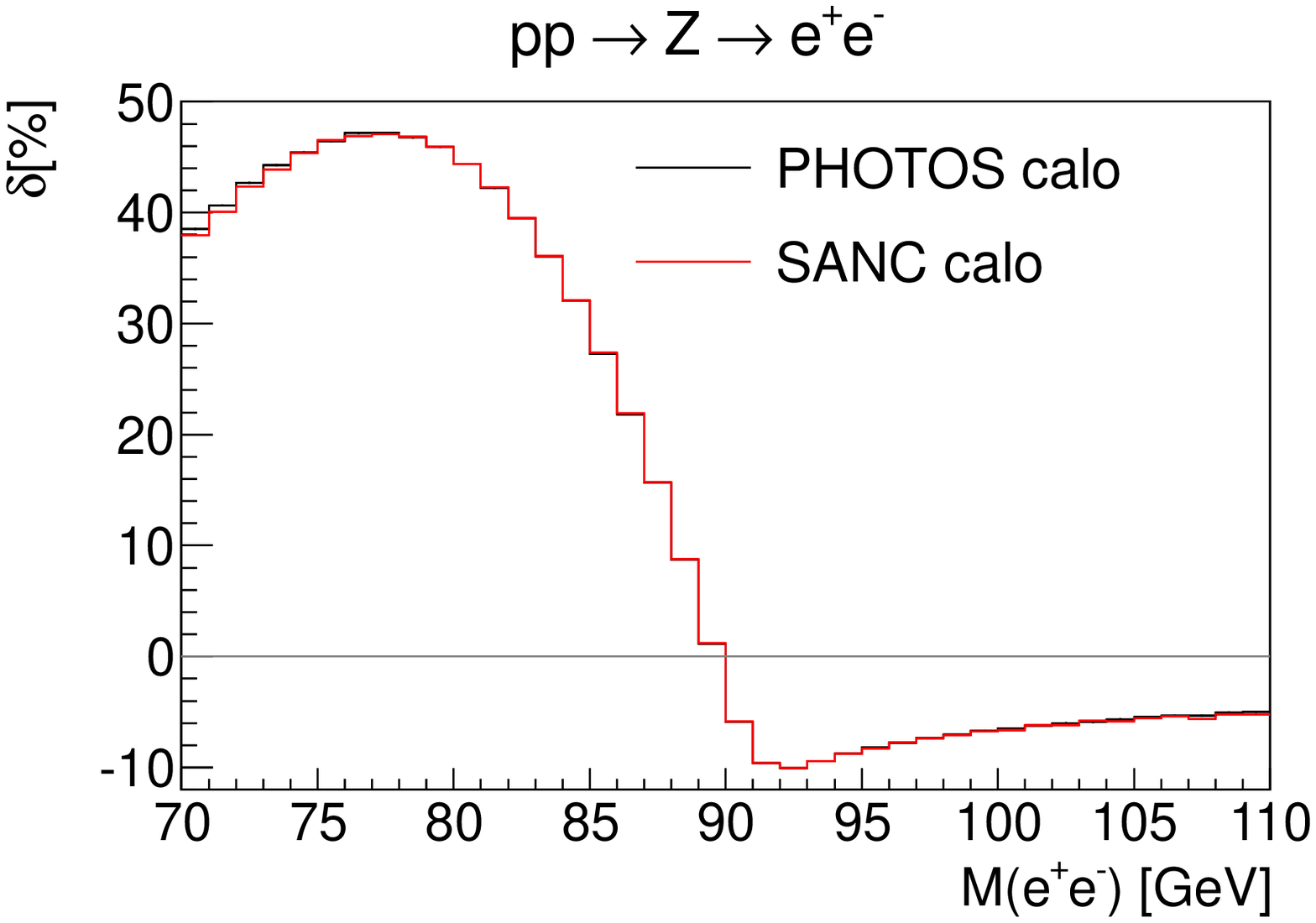}} \\
\subfigure{  \includegraphics[%
  width=0.40\columnwidth]{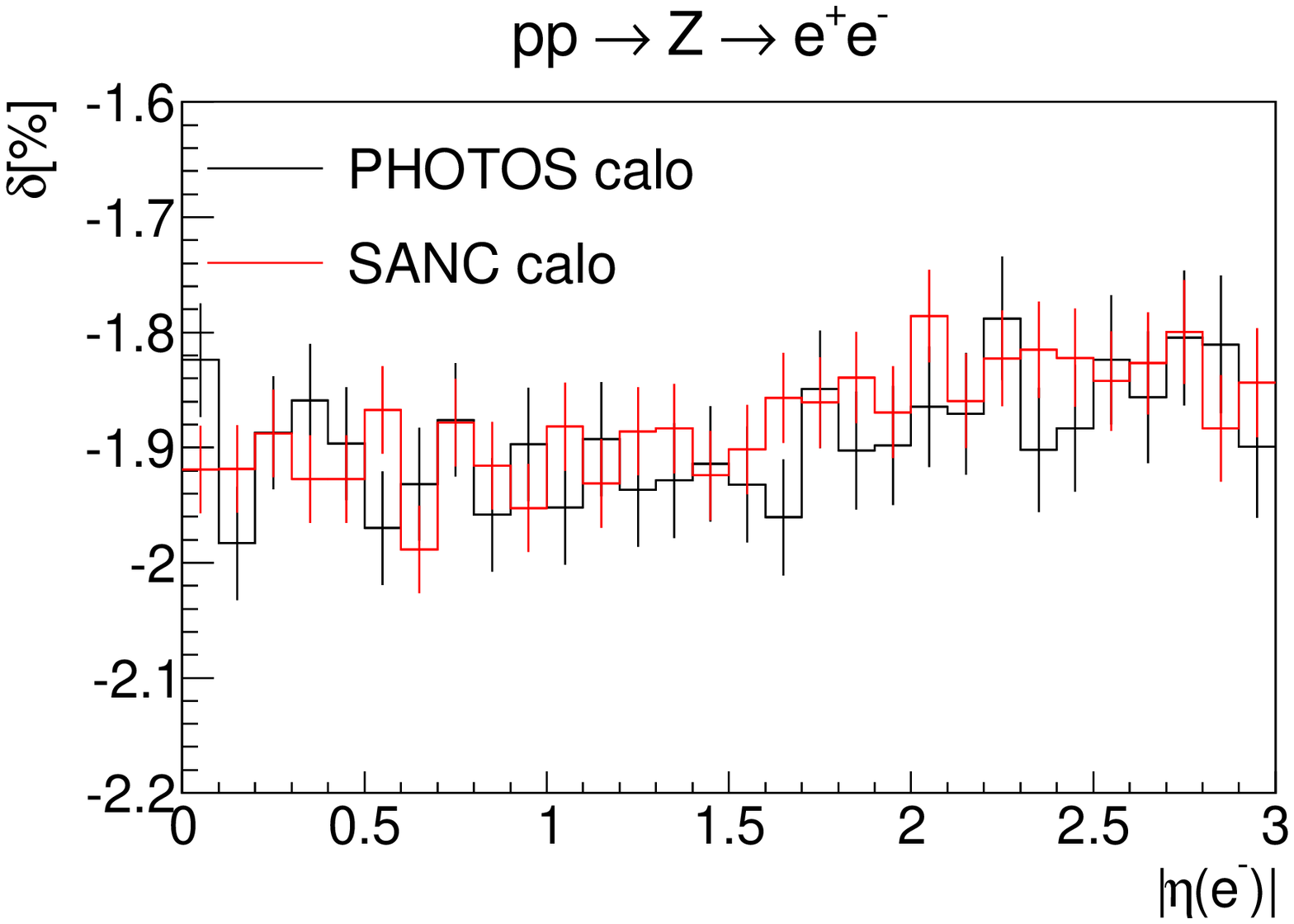}}
\end{tabular}
\caption{$\mathcal{O}(\alpha)$ corrections for  basic kinematical distributions from { \tt PYTHIA+PHOTOS} and { \tt SANC} 
in $Z\to ee$ decay ({\it calo} electrons).
 \label{ZfirstELcalo}}
\end{figure}

For {\tt SANC} and {\tt PYTHIA+PHOTOS} cases,
the results for $\mathcal{O}(\alpha)$ corrections which are defined by
$\delta = (\sigma^{\mathcal{O}(\alpha)}-\sigma^{Born})/\sigma^{Born}$ are presented on Figs. \ref{WfirstELbare}-\ref{ZfirstELcalo}.
As we can see from these figures agreement at the one-loop level was found to be
excellent, at the level of 0.01\% for both $Z$ and $W$ decays, once biases due to technical parameter separating hard and soft photon emission was properly tuned between the two 
calculations.

\subsection {Dependence on technical parameters}

\begin{figure}[htp!]
\includegraphics[width=1.0\textwidth]{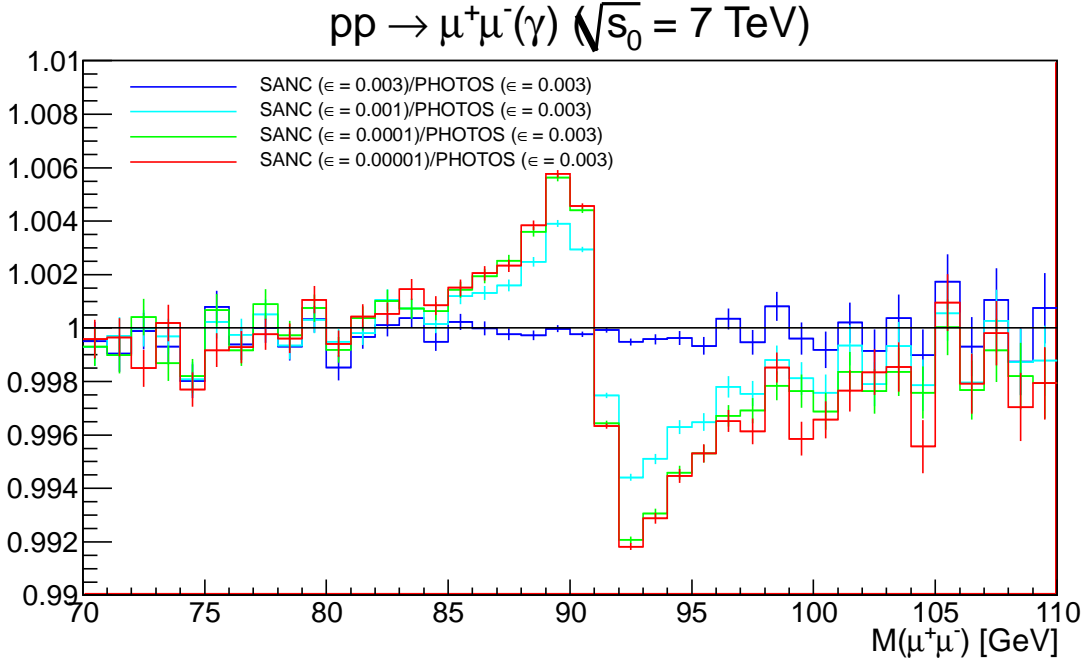}
\caption{Ratio of invariant mass distributions from {\tt SANC} and {\tt PHOTOS} in $Z$ 
decay as function of $k_0=\epsilon=2E_{\gamma,min}/\sqrt{s}$.}
\label{EpsiDep}
\end{figure}

While performing tests we had to address well known technical problem of the so called  ``$k_0$ bias''. 
In case of fixed order correction implemented into Monte Carlo algorithm,  a threshold on energy for emitted photon, typically in the rest-frame
of the decaying particle, has to be introduced. It regularizes infrared singularity. Below this threshold photons are simply integrated out and resulting contribution is summed up
with virtual corrections to cancel the infrared singularity. Unfortunately $k_0\to 0$ limit can not be reached, unless one accept working with negative event weights. The dependence on the technical parameter $k_0$ is however small
for inclusive quantities such as the total cross section. The effect becomes enhanced on differential distributions near the resonance peaks.
Such an effect can be observed  in Fig.~\ref{EpsiDep} for invariant mass $m_{\mu\mu}$ in $Z\to \mu\mu(\gamma)$ decay. 
Generally this dependence is nowadays of no interest\footnote{It is used however when { \tt PHOTOS} is combined with {\tt POWHEG} 
for soft photon emissions only.} as in practical application options of the programs 
with multiple photon emission
should be used. This example however may be instructive for studies 
of ambiguities in implementation like in \cite{Barze:2012tt} where { \tt PHOTOS} is used for soft photon emission only, 
while the hard photon phase space 
is populated with the help of genuine {\tt POWHEG}  simulation for the final and initial state emissions simultaneously.

\section {Multiple photon emissions}  

\begin{figure}[htp!]
\begin{tabular}{ccc}
\subfigure{
\includegraphics[%
  width=0.40\columnwidth]{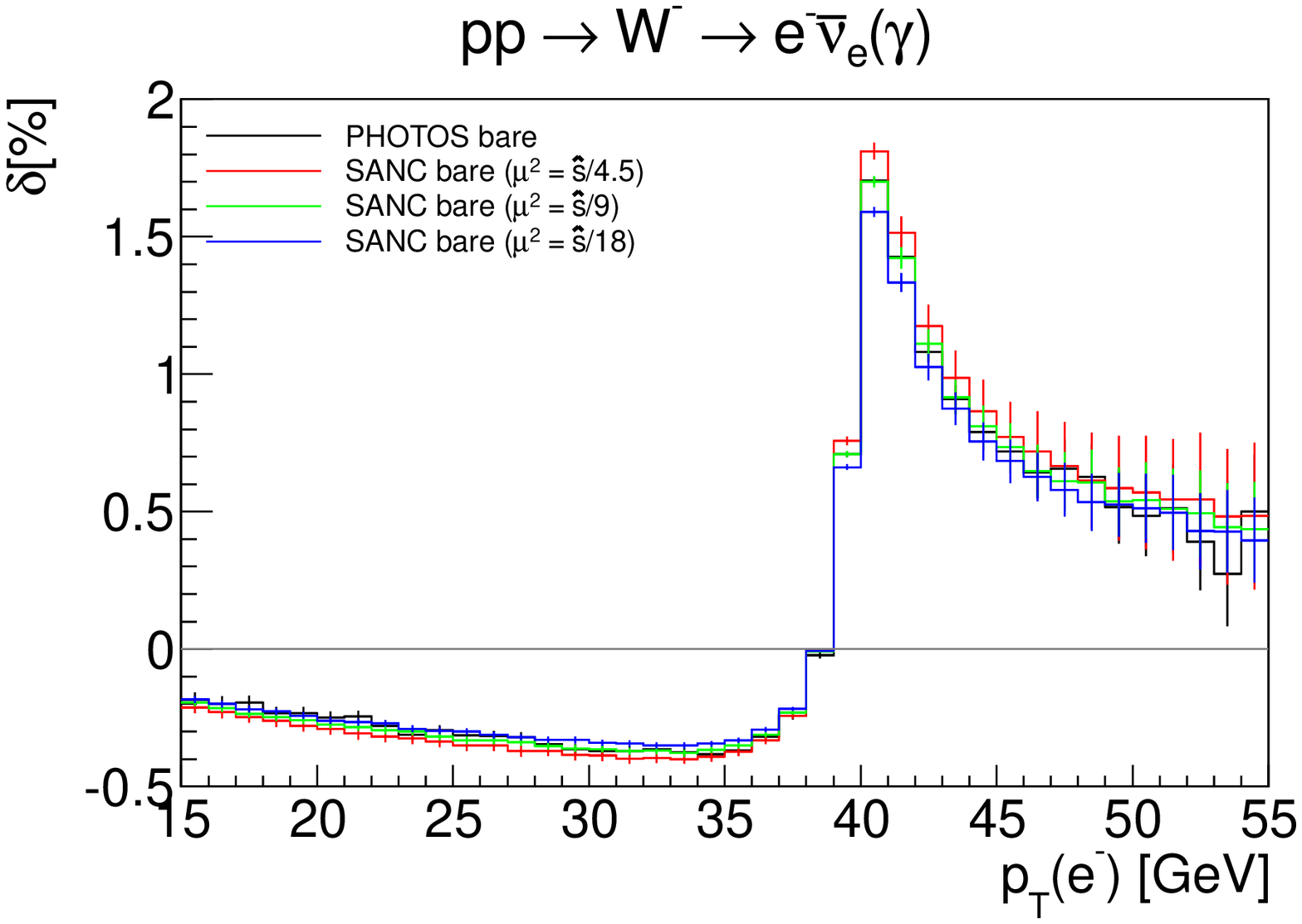}} & \subfigure{\includegraphics[%
  width=0.40\columnwidth]{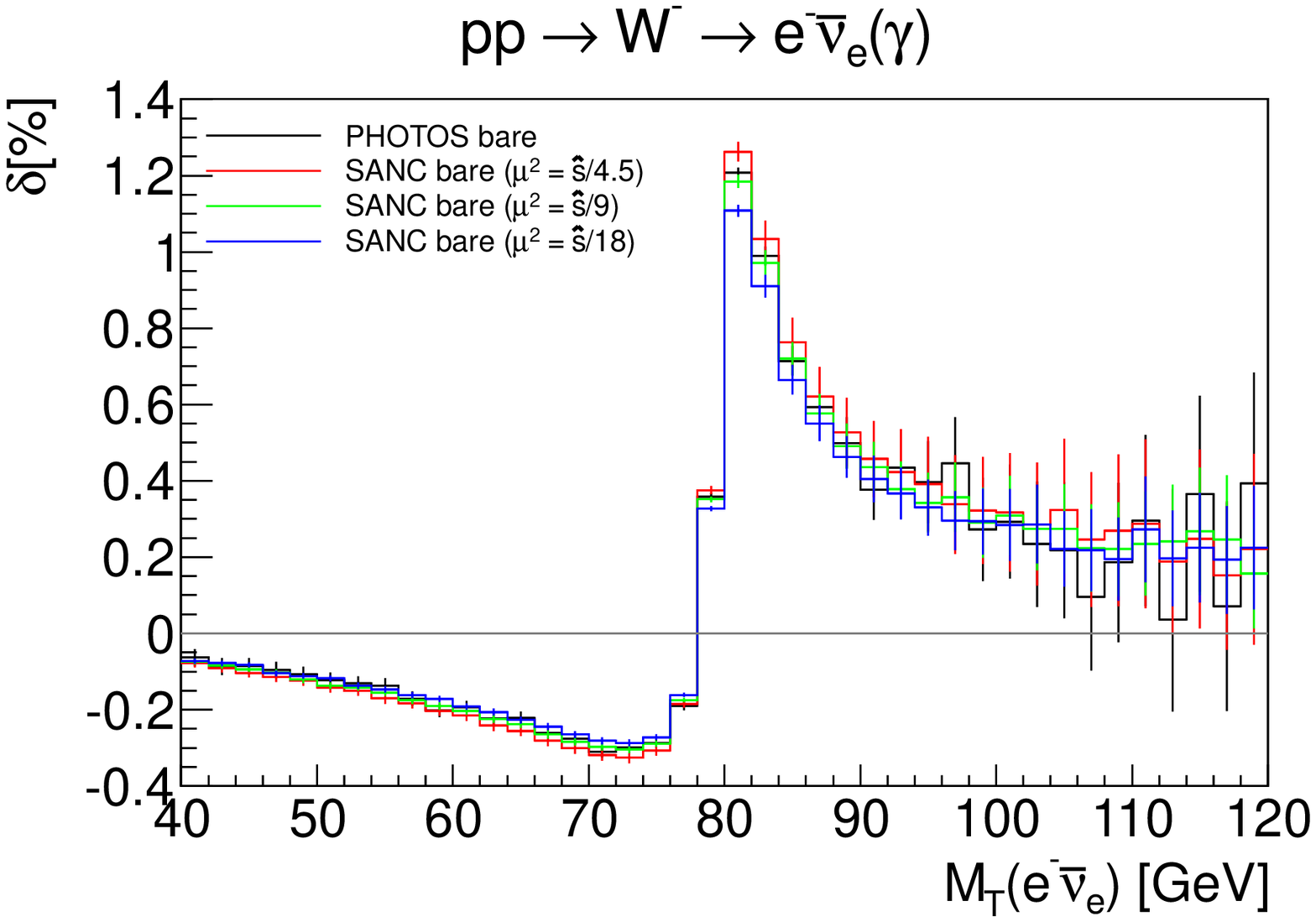}} \\
\subfigure{  \includegraphics[%
  width=0.40\columnwidth]{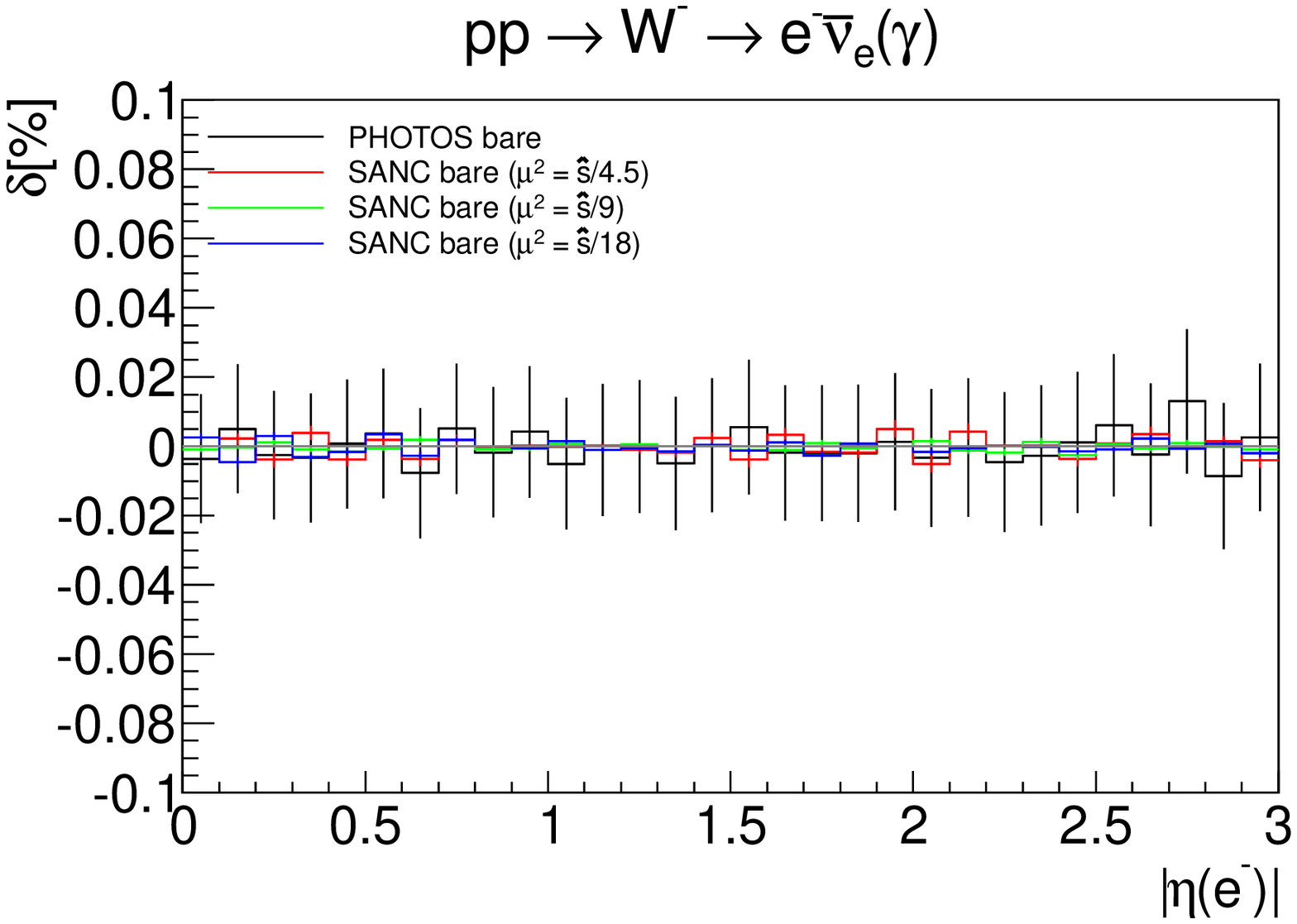}}
\end{tabular}
\caption{\small Higher order corrections for  basic kinematical  distributions from { \tt PYTHIA+PHOTOS} and { \tt SANC} in $W\to e \nu$ decay.
 \label{WhoELbare}}
\end{figure}

\begin{figure}[htp!]
\begin{tabular}{ccc}
\subfigure{
\includegraphics[%
  width=0.40\columnwidth]{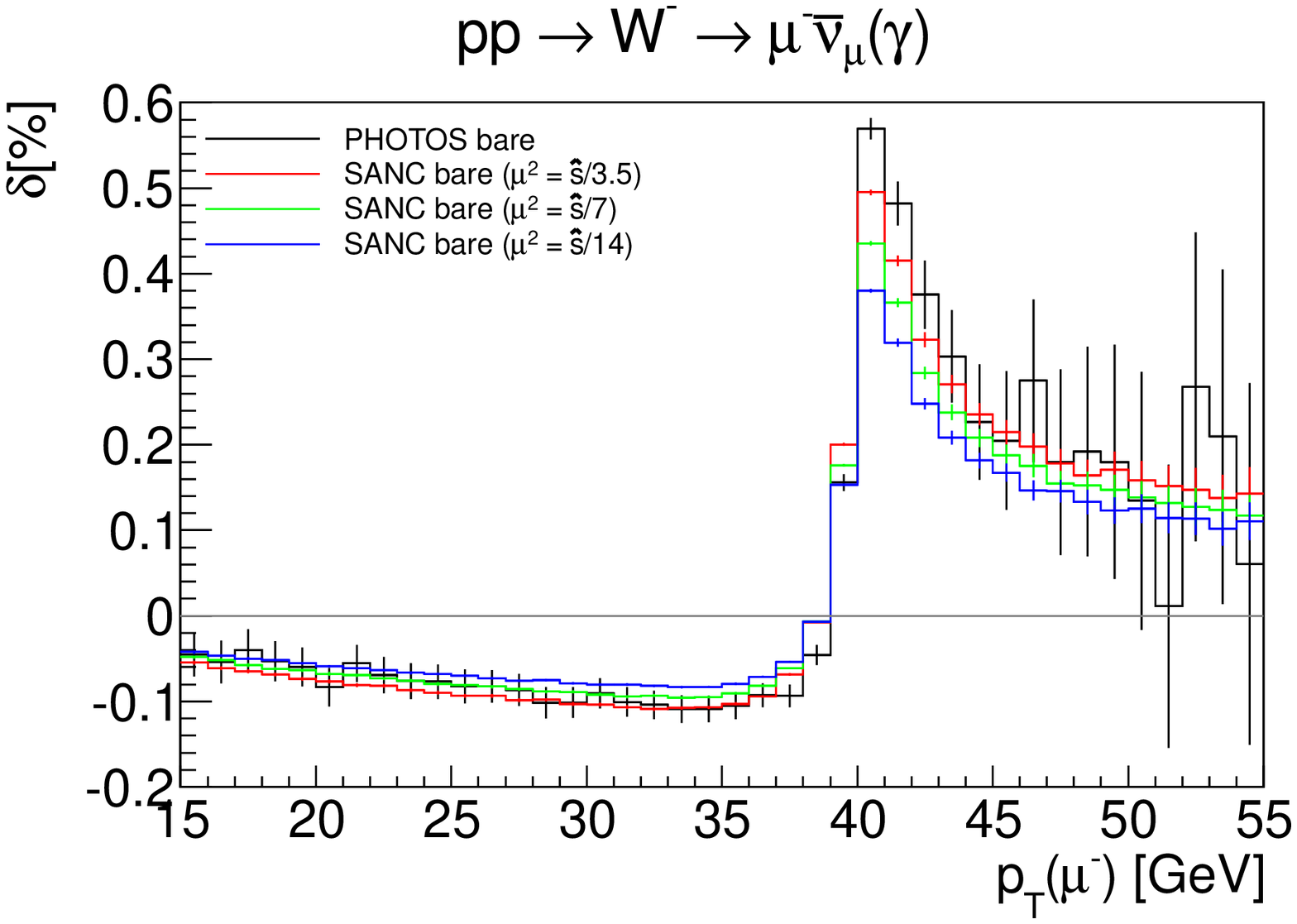}} & \subfigure{\includegraphics[%
  width=0.40\columnwidth]{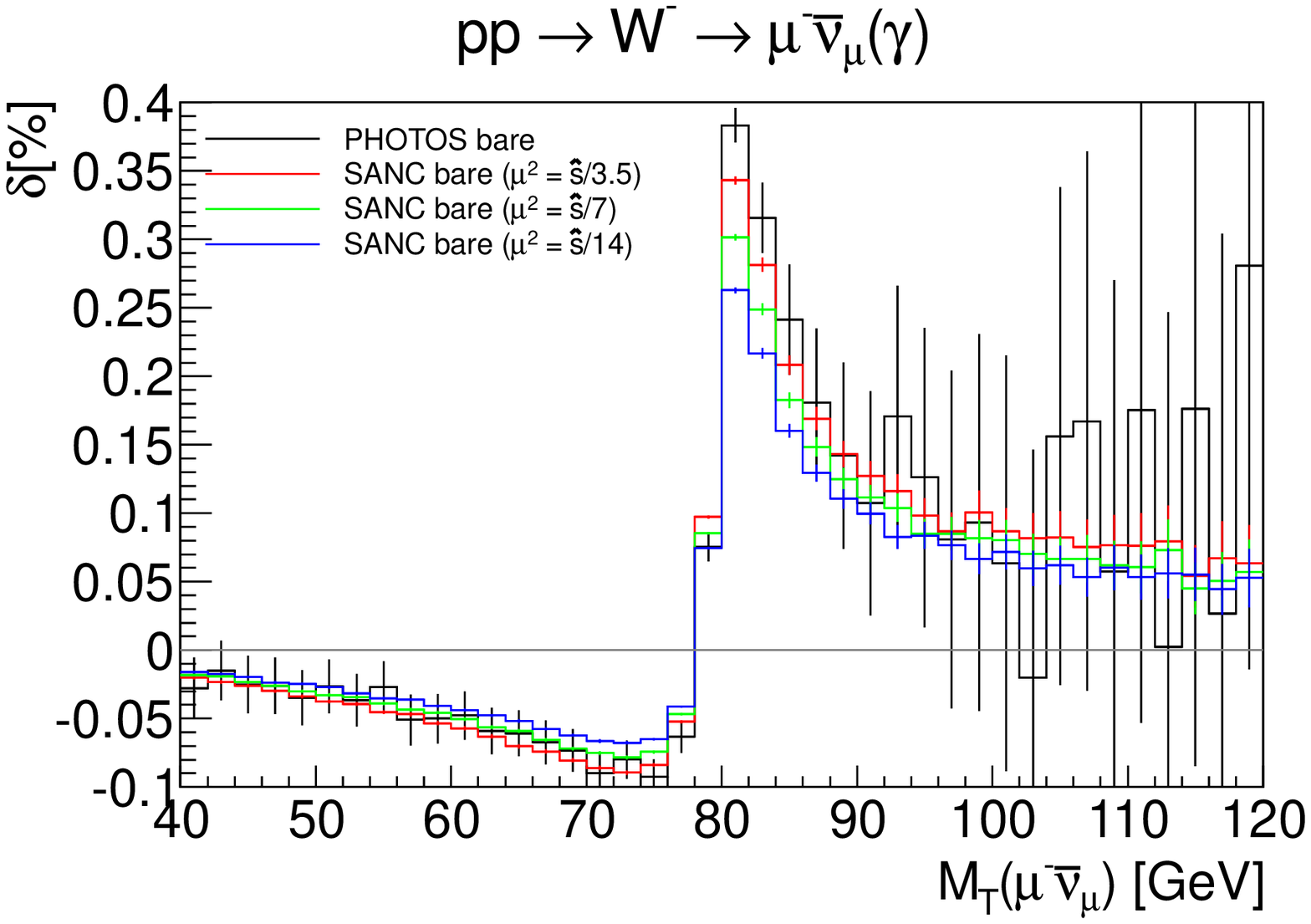}} \\
\subfigure{  \includegraphics[%
  width=0.40\columnwidth]{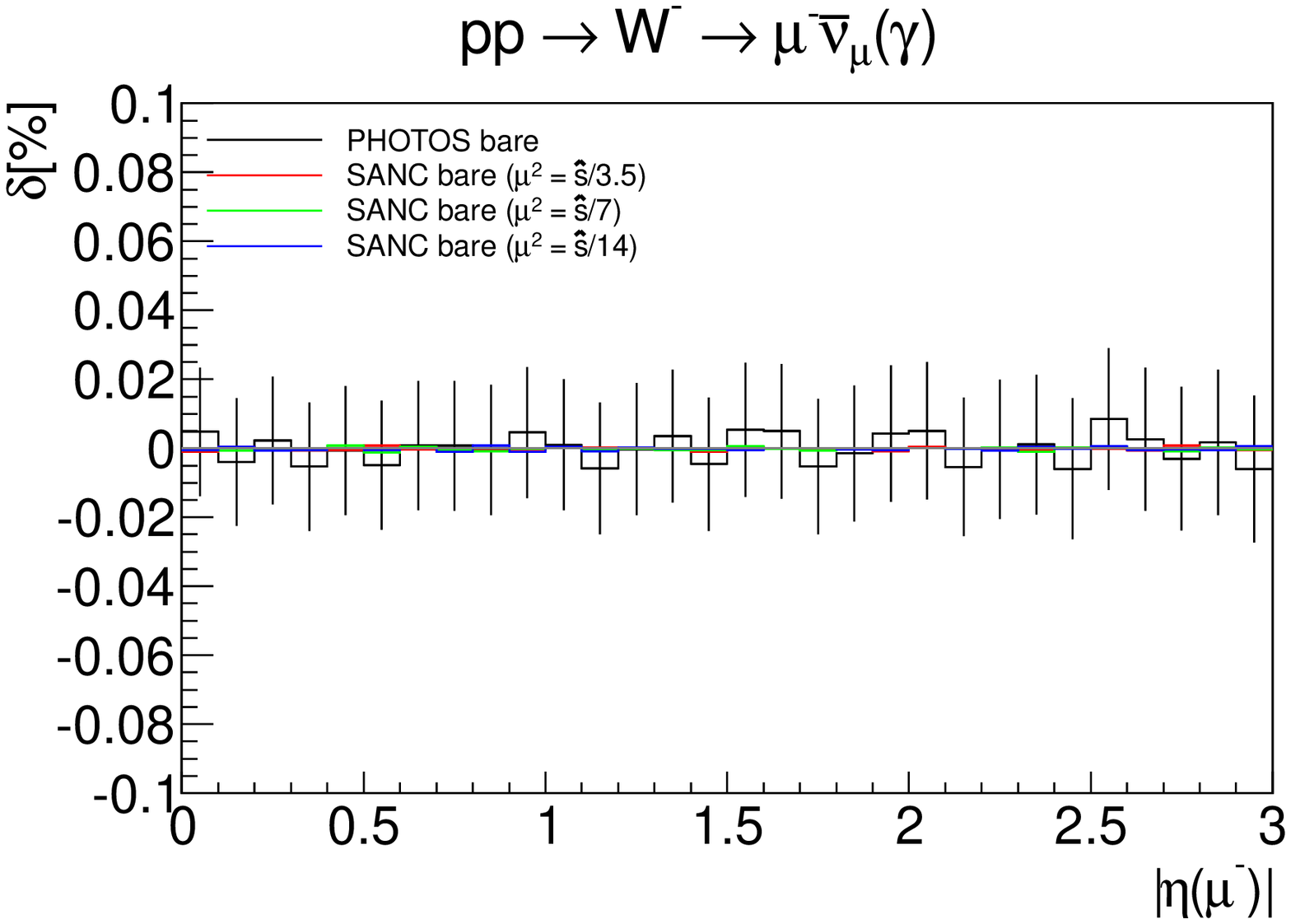}}
\end{tabular}
\caption{\small Higher order corrections for  basic kinematical distributions from { \tt PYTHIA+PHOTOS} and { \tt SANC} in $W\to \mu \nu$ decay.
 \label{WhoMUbare}}
\end{figure}

\begin{figure}[htp!]
\begin{tabular}{ccc}
\subfigure{
\includegraphics[%
  width=0.40\columnwidth]{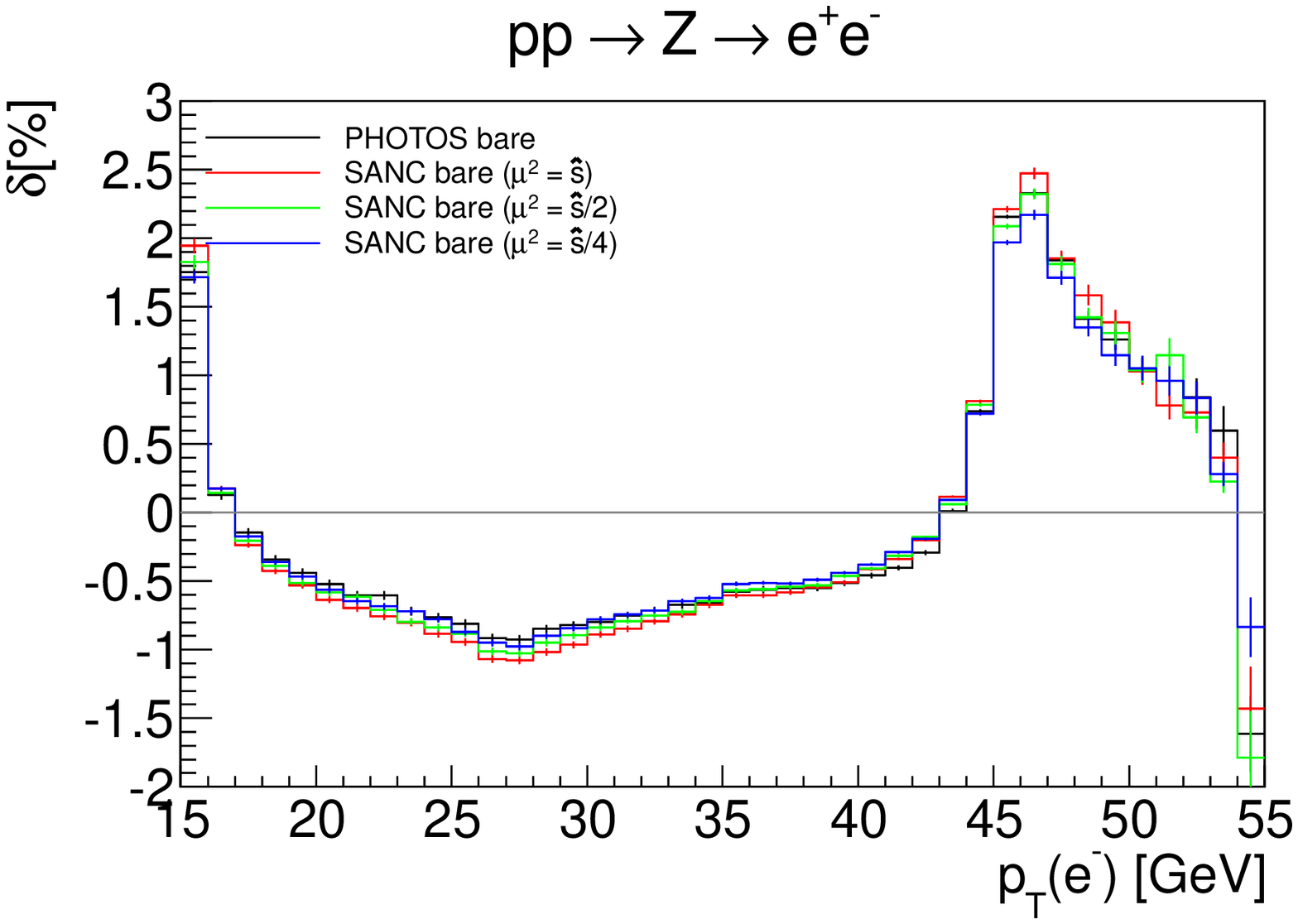}} & \subfigure{\includegraphics[%
  width=0.40\columnwidth]{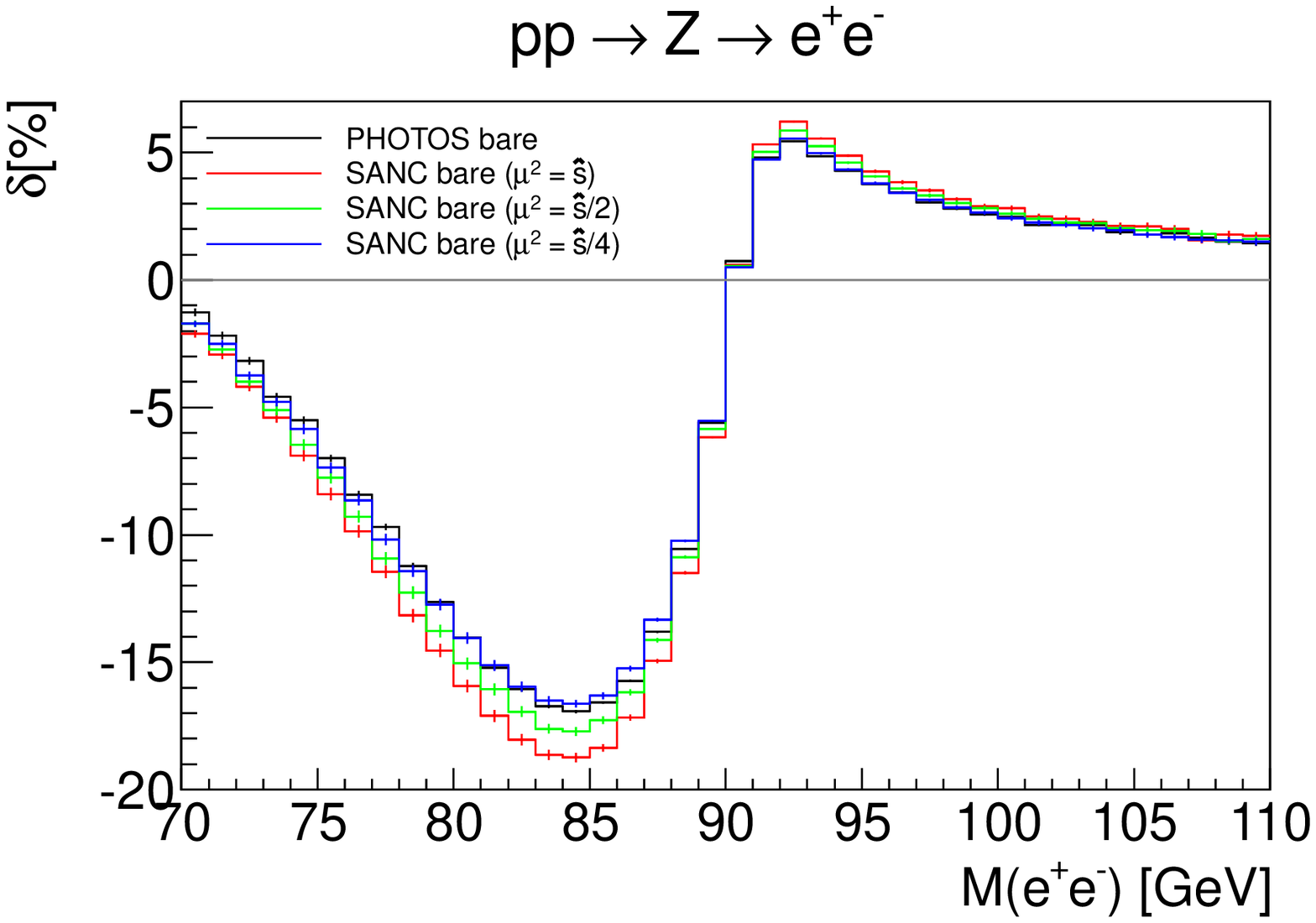}} \\
\subfigure{  \includegraphics[%
  width=0.40\columnwidth]{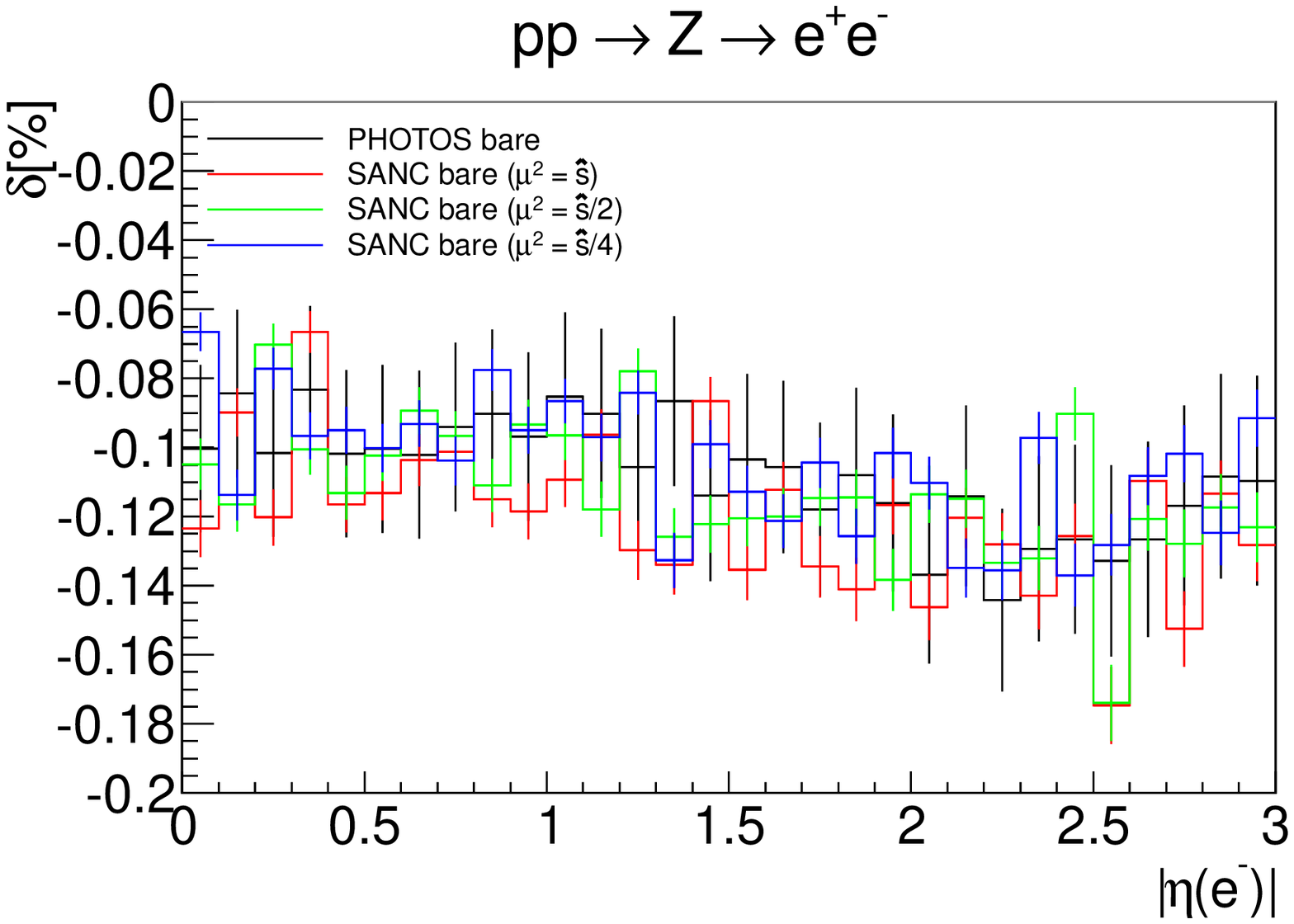}}
\end{tabular}
\caption{\small Higher order corrections for  basic kinematical distributions from { \tt PYTHIA+PHOTOS} and { \tt SANC} in $Z\to ee$ decay.
 \label{ZhoELbare}}
\end{figure}

\begin{figure}[htp!]
\begin{tabular}{ccc}
\subfigure{
\includegraphics[%
  width=0.40\columnwidth]{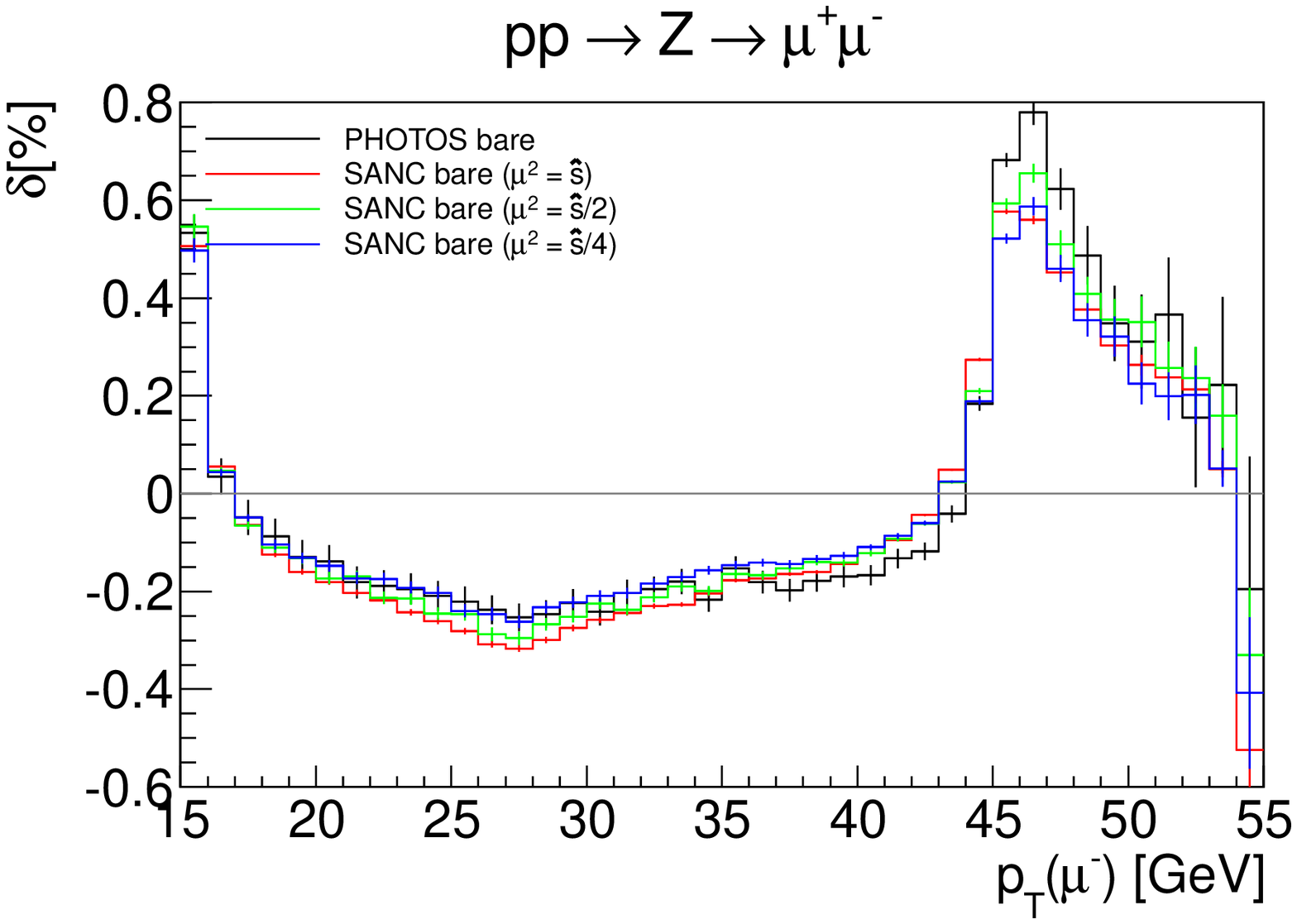}} & \subfigure{\includegraphics[%
  width=0.40\columnwidth]{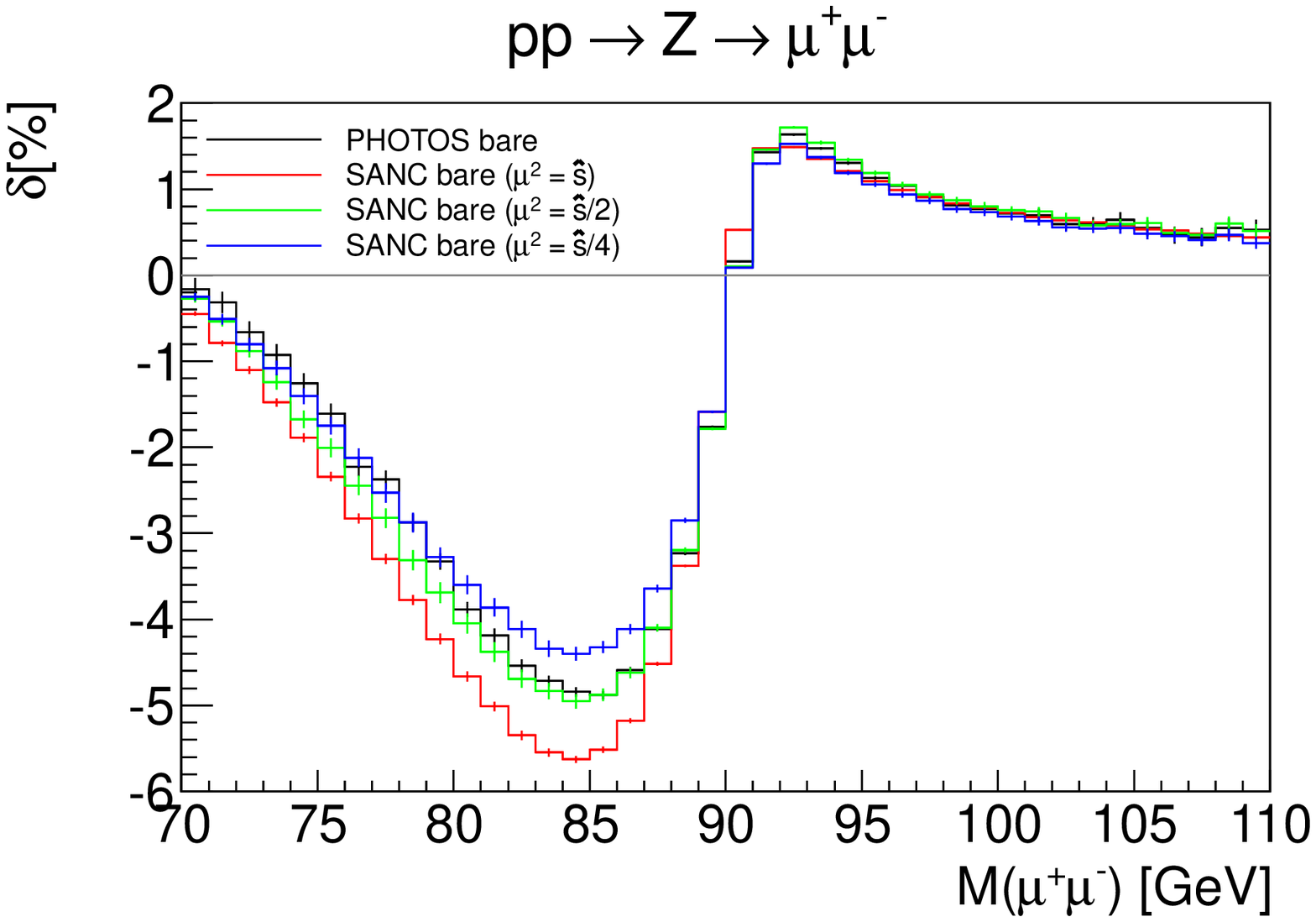}} \\
\subfigure{  \includegraphics[%
  width=0.40\columnwidth]{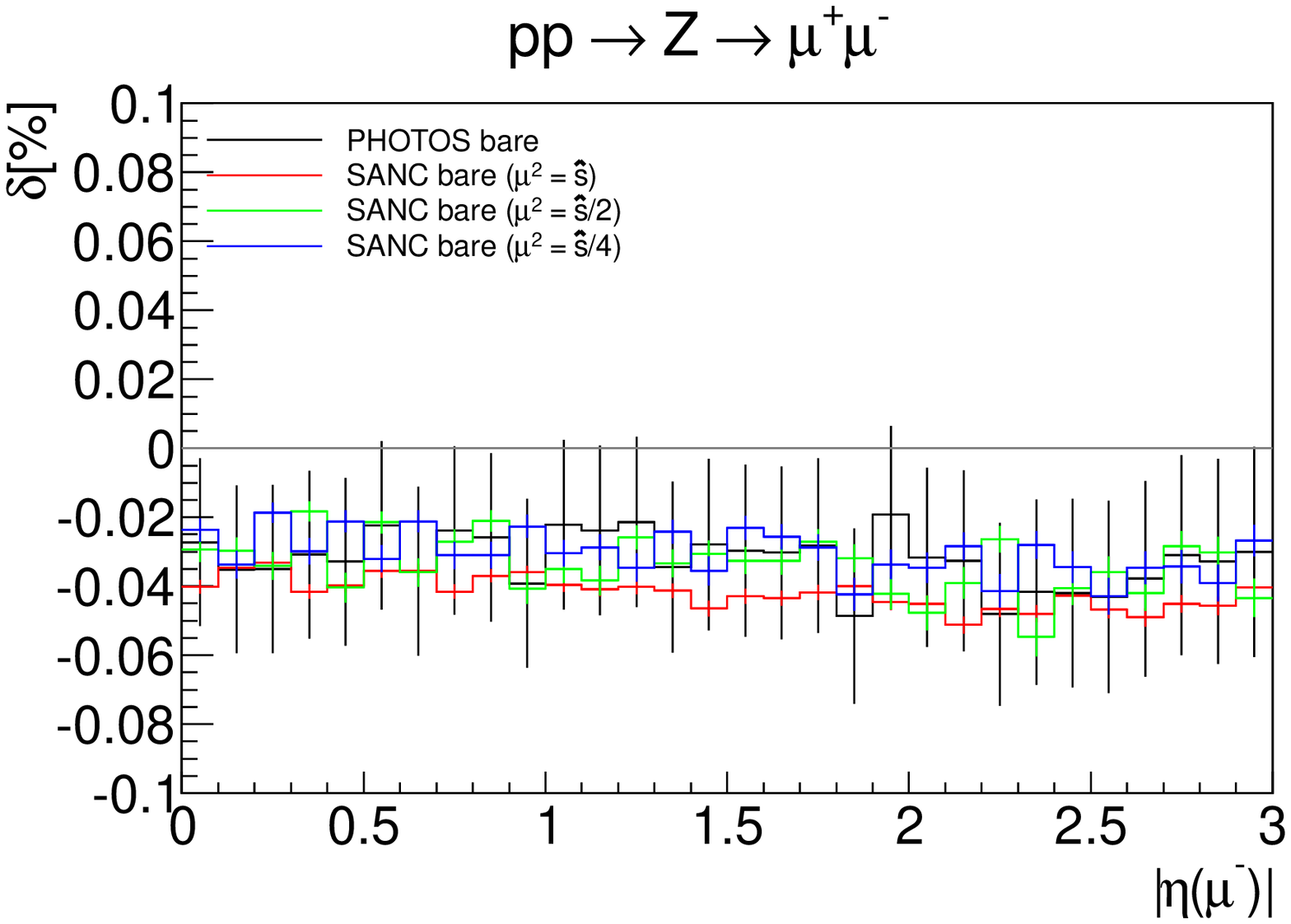}}
\end{tabular}
\caption{\small Higher order corrections for  basic kinematical distributions from { \tt PYTHIA+PHOTOS} and { \tt SANC} in $Z\to \mu\mu$ decay.
 \label{ZhoMUbare}}
\end{figure}

Let us now turn to the same observables, but calculated with the multiple photon emission option 
of { \tt SANC} and { \tt PHOTOS}, suitable for the actual comparisons with the data.
One can see from Figs \ref{WhoELbare}-\ref{ZhoMUbare}, that agreement is a bit worse 
than for single photon case, but well within expected theoretical precision. One should stress
that these results represent quantitative comparisons of different
approximations used in the {\tt SANC} and {\tt PHOTOS} as well. In fact, the approximations used, 
contrary to the single photon case, are not identical. 
Results of {\tt PHOTOS} represents the NLO  calculations with exponentiation and resummation of the collinear terms
of the first order photon emission matrix element. Results of  {\tt SANC} use the
collinear leading logarithmic approximation which introduces by construction numerical
dependence on the corresponding QED factorization scale $\mu^2$.
Differences due to non-optimal choice of the scale $\mu^2$ in {\tt SANC} are below several permille for differential distributions and below
0.1\% otherwise, see Figs.  \ref{WhoELbare}-\ref{ZhoMUbare}.
Additional effort and care in estimation of the size of 
seemingly minor effects may be required, if further improvements on  theoretical precision, beyond 0.1\% are needed.

\subsection{Comparisons of  {\tt PHOTOS} with {\tt KKMC}}
\label{sub:KKMC-PHOTOS}
In Ref.~\cite{Golonka:2006tw} we have demonstrated physics reasons behind very good, 0.1\% level, agreement between 
{\tt PHOTOS} and {\tt KKMC} results for final state photonic bremsstrahlung. Still until now,
only in case of $Z/\gamma^*$ intermediate state rigorous tests with second order QED matrix element Monte Carlo are available at 
present, and only in restricted condition of
no hadronic activities in the  initial state. 
For the reference simulations of $q \bar q \to Z/\gamma^* \to l^+ l^- (n \gamma)$ 
processes { \tt KKMC} Monte Carlo \cite{Jadach:1999vf} 
was used\footnote{This program, at present can not be used with simulation of the whole 
processes at LHC. An effort in this 
direction should be mentioned, see ref.~\cite{Yost:2012az}
but the corresponding results are not available for us at this moment.}. Also the Born level 
$q \bar q \to Z/\gamma^* \to l^+ l^- $ events from {\tt KKMC} were used for simulations with {\tt PHOTOS}. 
For both cases the monochromatic series of events
with fixed virtuality of intermediate $Z/\gamma^*$ state has been generated. 
This provides source of particularly valuable 
benchmarks as {\tt KKMC}  is the only program which features exclusive exponentiation 
combined with spin amplitudes for double photon emissions. 
As the numerical results of such quite extensive tests, more than 1000 figures are collected
on our web page \cite{webTestsPhotos}: plots for $Z\to \mu^+\mu^- (n\gamma)$
and $Z\to e^+e^- (n\gamma)$ are presented there.

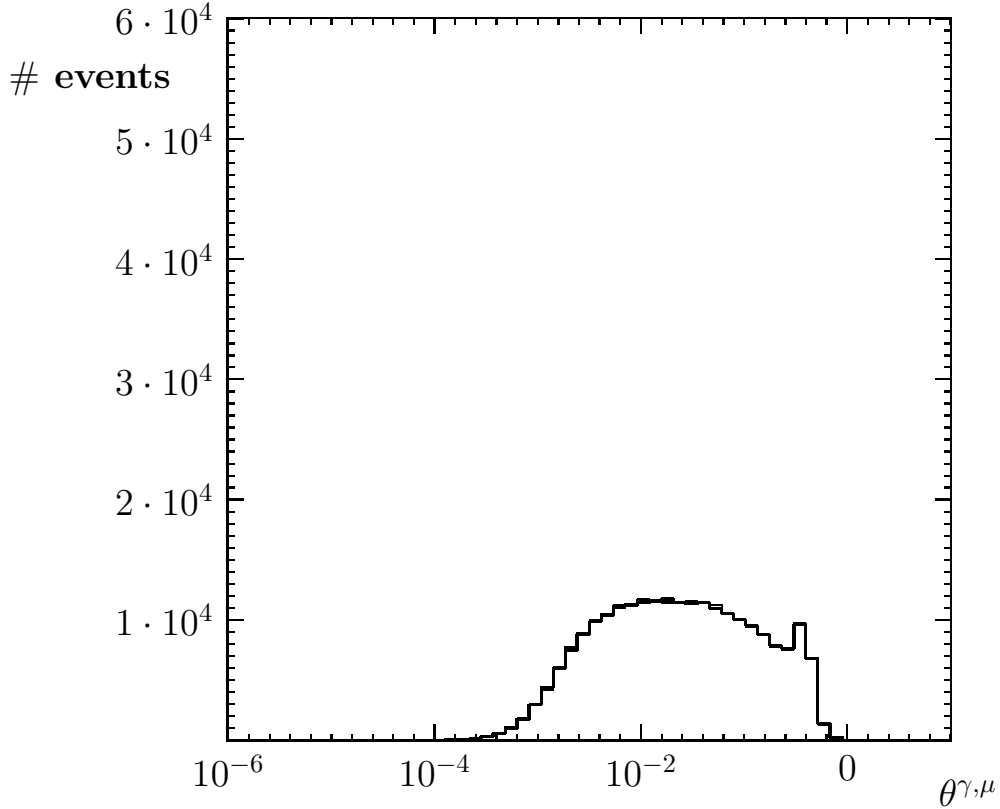
\begin{figure}[!ht]
\centering
\setlength{\unitlength}{0.08mm}
\begin{picture}(1600,1500)
\put(300,250){\begin{picture}( 1200,1200)
\put(0,0){\framebox( 1200,1200){ }}
\multiput(    0.00,0)(  342.86,0){   4}{\line(0,1){25}}
\multiput(    0.00,0)(   34.29,0){  36}{\line(0,1){10}}
\multiput(    0.00,1200)(  342.86,0){   4}{\line(0,-1){25}}
\multiput(    0.00,1200)(   34.29,0){  36}{\line(0,-1){10}}
\put(   0,-25){\makebox(0,0)[t]{\large $   10^{-6} $}}
\put( 343,-25){\makebox(0,0)[t]{\large $   10^{-4} $}}
\put( 686,-25){\makebox(0,0)[t]{\large $   10^{-2} $}}
\put(1029,-25){\makebox(0,0)[t]{\large $    0 $}}
\put(1229,-65){\makebox(0,0)[t]{\large $    \theta^{\gamma,\mu} $}}
\multiput(0,  199.98)(0,  200.00){   6}{\line(1,0){25}}
\multiput(0,   19.98)(0,   20.00){  60}{\line(1,0){10}}
\multiput(1200,  199.98)(0,  200.00){   6}{\line(-1,0){25}}
\multiput(1200,   19.98)(0,   20.00){  60}{\line(-1,0){10}}
\put(-25, 200){\makebox(0,0)[r]{\large $    1\cdot 10^{   4} $}}
\put(-25, 400){\makebox(0,0)[r]{\large $    2\cdot 10^{   4} $}}
\put(-25, 600){\makebox(0,0)[r]{\large $    3\cdot 10^{   4} $}}
\put(-25, 800){\makebox(0,0)[r]{\large $    4\cdot 10^{   4} $}}
\put(-25,1000){\makebox(0,0)[r]{\large $    5\cdot 10^{   4} $}}
\put(-25,1200){\makebox(0,0)[r]{\large $    6\cdot 10^{   4} $}}
\put(-95,1100){\makebox(0,0)[r]{\large {\bf $\#$ events}}}
\end{picture}}
\put(300,250){\begin{picture}( 1200,1200)
\thinlines 
\newcommand{\x}[3]{\put(#1,#2){\line(1,0){#3}}}
\newcommand{\y}[3]{\put(#1,#2){\line(0,1){#3}}}
\newcommand{\z}[3]{\put(#1,#2){\line(0,-1){#3}}}
\newcommand{\e}[3]{\put(#1,#2){\line(0,1){#3}}}
\y{   0}{   0}{   0}\x{   0}{   0}{  20}
\y{  20}{   0}{   0}\x{  20}{   0}{  20}
\y{  40}{   0}{   0}\x{  40}{   0}{  20}
\y{  60}{   0}{   0}\x{  60}{   0}{  20}
\y{  80}{   0}{   0}\x{  80}{   0}{  20}
\y{ 100}{   0}{   0}\x{ 100}{   0}{  20}
\y{ 120}{   0}{   0}\x{ 120}{   0}{  20}
\y{ 140}{   0}{   0}\x{ 140}{   0}{  20}
\y{ 160}{   0}{   0}\x{ 160}{   0}{  20}
\y{ 180}{   0}{   0}\x{ 180}{   0}{  20}
\y{ 200}{   0}{   0}\x{ 200}{   0}{  20}
\y{ 220}{   0}{   0}\x{ 220}{   0}{  20}
\y{ 240}{   0}{   0}\x{ 240}{   0}{  20}
\y{ 260}{   0}{   0}\x{ 260}{   0}{  20}
\y{ 280}{   0}{   0}\x{ 280}{   0}{  20}
\y{ 300}{   0}{   0}\x{ 300}{   0}{  20}
\y{ 320}{   0}{   0}\x{ 320}{   0}{  20}
\y{ 340}{   0}{   0}\x{ 340}{   0}{  20}
\y{ 360}{   0}{   1}\x{ 360}{   1}{  20}
\y{ 380}{   1}{   1}\x{ 380}{   2}{  20}
\y{ 400}{   2}{   1}\x{ 400}{   3}{  20}
\y{ 420}{   3}{   2}\x{ 420}{   5}{  20}
\y{ 440}{   5}{   5}\x{ 440}{  10}{  20}
\y{ 460}{  10}{  11}\x{ 460}{  21}{  20}
\y{ 480}{  21}{  13}\x{ 480}{  34}{  20}
\y{ 500}{  34}{  24}\x{ 500}{  58}{  20}
\y{ 520}{  58}{  26}\x{ 520}{  84}{  20}
\y{ 540}{  84}{  37}\x{ 540}{ 121}{  20}
\y{ 560}{ 121}{  34}\x{ 560}{ 155}{  20}
\y{ 580}{ 155}{  23}\x{ 580}{ 178}{  20}
\y{ 600}{ 178}{  20}\x{ 600}{ 198}{  20}
\y{ 620}{ 198}{  12}\x{ 620}{ 210}{  20}
\y{ 640}{ 210}{  10}\x{ 640}{ 220}{  20}
\y{ 660}{ 220}{   7}\x{ 660}{ 227}{  20}
\y{ 680}{ 227}{   1}\x{ 680}{ 228}{  20}
\y{ 700}{ 228}{   2}\x{ 700}{ 230}{  20}
\z{ 720}{ 230}{   1}\x{ 720}{ 229}{  20}
\z{ 740}{ 229}{   1}\x{ 740}{ 228}{  20}
\z{ 760}{ 228}{   2}\x{ 760}{ 226}{  20}
\y{ 780}{ 226}{   2}\x{ 780}{ 228}{  20}
\z{ 800}{ 228}{   3}\x{ 800}{ 225}{  20}
\z{ 820}{ 225}{  14}\x{ 820}{ 211}{  20}
\z{ 840}{ 211}{  10}\x{ 840}{ 201}{  20}
\z{ 860}{ 201}{  12}\x{ 860}{ 189}{  20}
\z{ 880}{ 189}{  13}\x{ 880}{ 176}{  20}
\z{ 900}{ 176}{  21}\x{ 900}{ 155}{  20}
\z{ 920}{ 155}{   4}\x{ 920}{ 151}{  20}
\y{ 940}{ 151}{  43}\x{ 940}{ 194}{  20}
\z{ 960}{ 194}{  58}\x{ 960}{ 136}{  20}
\z{ 980}{ 136}{ 110}\x{ 980}{  26}{  20}
\z{1000}{  26}{  22}\x{1000}{   4}{  20}
\z{1020}{   4}{   4}\x{1020}{   0}{  20}
\y{1040}{   0}{   0}\x{1040}{   0}{  20}
\y{1060}{   0}{   0}\x{1060}{   0}{  20}
\y{1080}{   0}{   0}\x{1080}{   0}{  20}
\y{1100}{   0}{   0}\x{1100}{   0}{  20}
\y{1120}{   0}{   0}\x{1120}{   0}{  20}
\y{1140}{   0}{   0}\x{1140}{   0}{  20}
\y{1160}{   0}{   0}\x{1160}{   0}{  20}
\y{1180}{   0}{   0}\x{1180}{   0}{  20}
\end{picture}} 
\put(300,250){\begin{picture}( 1200,1200)
\thicklines 
\newcommand{\x}[3]{\put(#1,#2){\line(1,0){#3}}}
\newcommand{\y}[3]{\put(#1,#2){\line(0,1){#3}}}
\newcommand{\z}[3]{\put(#1,#2){\line(0,-1){#3}}}
\newcommand{\e}[3]{\put(#1,#2){\line(0,1){#3}}}
\y{   0}{   0}{   0}\x{   0}{   0}{  20}
\y{  20}{   0}{   0}\x{  20}{   0}{  20}
\y{  40}{   0}{   0}\x{  40}{   0}{  20}
\y{  60}{   0}{   0}\x{  60}{   0}{  20}
\y{  80}{   0}{   0}\x{  80}{   0}{  20}
\y{ 100}{   0}{   0}\x{ 100}{   0}{  20}
\y{ 120}{   0}{   0}\x{ 120}{   0}{  20}
\y{ 140}{   0}{   0}\x{ 140}{   0}{  20}
\y{ 160}{   0}{   0}\x{ 160}{   0}{  20}
\y{ 180}{   0}{   0}\x{ 180}{   0}{  20}
\y{ 200}{   0}{   0}\x{ 200}{   0}{  20}
\y{ 220}{   0}{   0}\x{ 220}{   0}{  20}
\y{ 240}{   0}{   0}\x{ 240}{   0}{  20}
\y{ 260}{   0}{   0}\x{ 260}{   0}{  20}
\y{ 280}{   0}{   0}\x{ 280}{   0}{  20}
\y{ 300}{   0}{   0}\x{ 300}{   0}{  20}
\y{ 320}{   0}{   0}\x{ 320}{   0}{  20}
\y{ 340}{   0}{   0}\x{ 340}{   0}{  20}
\y{ 360}{   0}{   1}\x{ 360}{   1}{  20}
\y{ 380}{   1}{   0}\x{ 380}{   1}{  20}
\y{ 400}{   1}{   2}\x{ 400}{   3}{  20}
\y{ 420}{   3}{   3}\x{ 420}{   6}{  20}
\y{ 440}{   6}{   5}\x{ 440}{  11}{  20}
\y{ 460}{  11}{   8}\x{ 460}{  19}{  20}
\y{ 480}{  19}{  16}\x{ 480}{  35}{  20}
\y{ 500}{  35}{  24}\x{ 500}{  59}{  20}
\y{ 520}{  59}{  29}\x{ 520}{  88}{  20}
\y{ 540}{  88}{  31}\x{ 540}{ 119}{  20}
\y{ 560}{ 119}{  31}\x{ 560}{ 150}{  20}
\y{ 580}{ 150}{  26}\x{ 580}{ 176}{  20}
\y{ 600}{ 176}{  22}\x{ 600}{ 198}{  20}
\y{ 620}{ 198}{   9}\x{ 620}{ 207}{  20}
\y{ 640}{ 207}{  16}\x{ 640}{ 223}{  20}
\y{ 660}{ 223}{   1}\x{ 660}{ 224}{  20}
\y{ 680}{ 224}{   9}\x{ 680}{ 233}{  20}
\z{ 700}{ 233}{   1}\x{ 700}{ 232}{  20}
\y{ 720}{ 232}{   3}\x{ 720}{ 235}{  20}
\z{ 740}{ 235}{   6}\x{ 740}{ 229}{  20}
\y{ 760}{ 229}{   1}\x{ 760}{ 230}{  20}
\z{ 780}{ 230}{   1}\x{ 780}{ 229}{  20}
\z{ 800}{ 229}{  10}\x{ 800}{ 219}{  20}
\z{ 820}{ 219}{   8}\x{ 820}{ 211}{  20}
\z{ 840}{ 211}{  10}\x{ 840}{ 201}{  20}
\z{ 860}{ 201}{  11}\x{ 860}{ 190}{  20}
\z{ 880}{ 190}{  14}\x{ 880}{ 176}{  20}
\z{ 900}{ 176}{  18}\x{ 900}{ 158}{  20}
\z{ 920}{ 158}{   5}\x{ 920}{ 153}{  20}
\y{ 940}{ 153}{  39}\x{ 940}{ 192}{  20}
\z{ 960}{ 192}{  56}\x{ 960}{ 136}{  20}
\z{ 980}{ 136}{ 109}\x{ 980}{  27}{  20}
\z{1000}{  27}{  22}\x{1000}{   5}{  20}
\z{1020}{   5}{   5}\x{1020}{   0}{  20}
\y{1040}{   0}{   0}\x{1040}{   0}{  20}
\y{1060}{   0}{   0}\x{1060}{   0}{  20}
\y{1080}{   0}{   0}\x{1080}{   0}{  20}
\y{1100}{   0}{   0}\x{1100}{   0}{  20}
\y{1120}{   0}{   0}\x{1120}{   0}{  20}
\y{1140}{   0}{   0}\x{1140}{   0}{  20}
\y{1160}{   0}{   0}\x{1160}{   0}{  20}
\y{1180}{   0}{   0}\x{1180}{   0}{  20}
\end{picture}} 
\end{picture} 
\caption{\small
 The distribution of the  angle in the laboratory frame between  photon and the closer muon, 
as generated
from { \tt KKMC} and { \tt PHOTOS}; selection cuts are applied, see the text. The intermediate state of virtuality 
$M_Z+$6 GeV=97.187 GeV
with the transverse momentum $p_T=9$~GeV and pseudorapidity $\eta=2$ was created in the $u\bar u$ annihilation. 
 The samples of $40$M events were simulated,
ratio of the surfaces under distributions  is 0.9991. Relatively large (0.1\%) difference to unity is 
 typical for the  larger values of pseudorapidity.  For LO results 
(see web page \cite{webTestsPhotos} for extended results)
the ratio for the surfaces, is 
of the same order, for electrons  it is closer to 1.  
This spectrum feature plateau in rest frame of $Z/\gamma^*$ state, but boosted to laboratory frame  plateau is deformed. \label{LepAng}                     
}
\end{figure}

For all  plots of collection \cite{webTestsPhotos},
before selection cuts are applied, events   are boosted  to the laboratory frame
assuming fixed 4-vector of $Z$  defined by its $p_T^Z$ and  $\eta^Z$; the two dimensional grid in  $(p_T^Z,\;\eta^Z$) is constructed,
with 7 bins in  $p_T^Z$ 
spanning region 0-50 GeV and
 4 bins in $\eta^Z$ spanning from 0-2, for each point in the grid 40M events are simulated. Kinematical selection is applied
on lepton and photon 4-momenta and kinematical distributions are constructed from accepted events.
It is required that
 each lepton $p_T^l>20$~GeV and  $|\eta^\pm|<2.47$, with the gap  $1.37<|\eta^\pm|<1.52$  excluded. The gap in $\eta^\pm$ corresponds to
transition region between central and end-cap calorimeter  in ATLAS detector, and is used here to somewhat 
arbitrary enhance possible effect of exclusive selection.  Then, straightforward comparison between electron and muon cases is available; the only
difference originating from leptons mass. 
   For photon  $|\eta^\gamma|<2.37$ and region  $1.37<|\eta^\gamma|<1.52$ is excluded again,  $p_T^\gamma>15$~GeV  is required.
 In example  shown in Fig.~\ref{LepAng}, angle between photon and a closer lepton is shown for $p_T^Z=9$ GeV and $\eta^Z=2$.
Results from { \tt KKMC} and { \tt PHOTOS NLO}  are compared. 
In both cases multiphoton emission is generated in $Z/\gamma^*$ rest frame for
  $u \bar u \to Z/\gamma^* \to \mu^+\mu^- (n\gamma)$ production process, with  no 
initial state activity of any sort and  virtuality of intermediate state  $M_Z$+6~GeV=97.187~GeV, where $M_Z$ is the $Z$-boson mass.

 For each choice of $p_T^Z$ and $\eta^Z$ used to define $Z/\gamma^*$ state 4-momentum
a set of three observables: angle between photon and closer lepton, 
directions of leptons and an overall acceptance rate
is monitored in \cite{webTestsPhotos}. 
  A general agreement of 0.1\% can be concluded for all distributions. 
 The results for   LO restricted { \tt PHOTOS} are collected there as well.

For all cases 
one has to bear in mind that an overall normalization correction factors for cross section, like 
$(1+ \frac{3}{4}\frac{\alpha}{\pi})$ in case of $Z$ decay, have to be included when using {\tt PHOTOS} package.

\subsection{The $\phi^*_\eta$ observable}
Motivated by recent ATLAS publication \cite{Aad:2012wfa} on precise observable $\phi^*_\eta$, representing important improvement for
the measurement of $Z$ boson transverse momentum ($p^Z_T$) at LHC, we have decided to devote section of this paper
to discussion on the respective  QED FSR corrections.

The measurement of the $Z$ boson transverse momentum ($p^Z_T$ or $\phi^*_\eta$) offers a very 
sensitive way for studying dynamical effects of the strong interaction, 
complementary to the measurements of the associated production of bosons
with jets.   
The knowledge of the $p^Z_T$ distribution is crucial also to improve the modeling
of the $W$ boson production needed for a precise measurement of the $W$ mass 
\cite{Aaltonen:2012bp}, in particular in the low $p^Z_T$ region which dominates the 
cross section.  The study of the low $p^Z_T$  spectrum ($p^Z_T<  M_Z$), 
has also an important implications for the 
understanding of the Higgs signatures \cite{:2012gk} as well as for the New Physics searches 
at the LHC \cite{Collaboration:2011dca}.

The precision of the direct measurement of the spectrum at low $p^Z_T$  at the 
LHC and at the Tevatron using the $Z$ leptonic decay is limited by
 systematic uncertainties related to the knowledge and unfolding of the 
experiments resolution, in particular lepton energy scale \cite{Aad:2011gj,Chatrchyan:2011wt}.
 
In recent years, additional observables with better experimental resolution 
and less sensitive to experimental systematic uncertainties have been 
investigated \cite{Boonekamp:2010ik,Vesterinen:2008hx,Banfi:2010cf,Banfi:2011dx}. 
The optimal experimental observable to probe the low $p^Z_T$  domain of 
$Z/\gamma^*$ production at hadron colliders was found to be $\phi^*_\eta$ 
which is defined as
\begin{equation}
          \phi^*_\eta=  \tan(\phi_{acop}/2) \cdot  \sin(\Theta^*_\eta),
\end{equation}
where $\phi_{acop}$ is an azimuthal opening angle between the two leptons 
and the angle $\Theta^*_\eta$ is the scattering angle of the 
leptons with respect to the proton beam direction in the rest frame of 
the dilepton system. The $\Theta^*_\eta$ angle is defined as 
\begin{equation}
 \cos(\Theta^*_\eta) = \tanh\left(\frac{\eta^{-} - \eta^{+}}{2}\right),
\end{equation} 
where ${\eta}^-$  and $\eta^+$ are the pseudorapiditities of the negatively
 and  positively charged lepton, respectively. The variable $\phi^*_\eta$  
is correlated to the quantity $p^Z_T/m_{ll}$, where $m_{ll}$ is the invariant 
mass of the lepton pair. It therefore probes the same physics as 
the transverse momentum $p^Z_T$  and can be approximately related to it by 
$p^Z_T \simeq M_Z \cdot \phi^*_\eta$ . 
From the experimental point of view the  variable  $\phi^*_\eta$ relies 
entirely  on the angle reconstruction of the leptons in pair production, therefore on the tracking devices
of high precision.%

We have studied the theoretical error on $\phi^*_\eta$ distributions   due to photonic bremsstrahlung effects.
As in Section \ref{sub:KKMC-PHOTOS} comparison of results from { \tt PHOTOS} and { \tt KKMC} generators
was performed and stored on  web page \cite{webTestsPhotos}. Let us list 
the appropriate cuts and give example results.

We have requested that for both  leptons 
$p_T^l>$  20 GeV and   $|\eta^\pm|<2.4$~.
Distributions of $\frac{dN(Z\to l^+l^-)}{d\phi^*_\eta}$ from { \tt KKMC} 
and { \tt PHOTOS} were monitored. As before, the monochromatic samples of 
$q \bar q \to Z/\gamma^* \to l^+l^- (n\gamma)$ were generated for two virtualities: 
$M_Z+6$ GeV (97.187~GeV) and $M_Z-4$ GeV  (87.187~GeV), for incoming up and down quarks.
Generated events were boosted to the laboratory frame assuming fixed 
$p_T^Z$ and $\eta^Z$ of intermediate  $Z/\gamma^*$ state. Again the grid of 7 bins in  $p_T^Z$ 
spanning region 0-50 GeV and
 4 bins in $\eta^Z$ spanning from 0-2 was used.

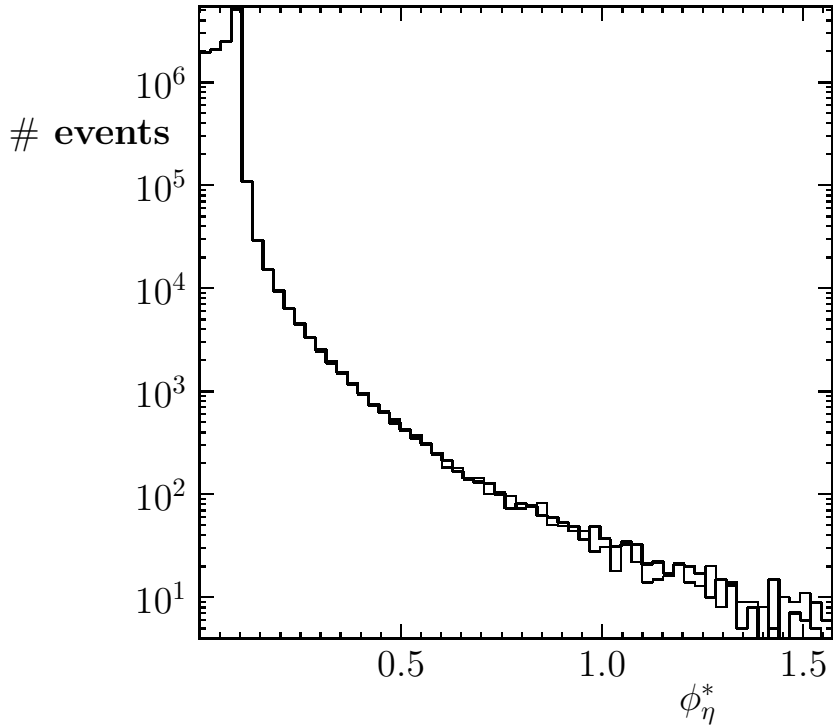
\begin{figure}[!ht]
\centering
\setlength{\unitlength}{0.07mm}
\begin{picture}(1600,1500)
\put(300,250){\begin{picture}( 1200,1200)
\put(0,0){\framebox( 1200,1200){ }}
\multiput(  381.97,0)(  381.97,0){   3}{\line(0,1){25}}
\multiput(    0.00,0)(   38.20,0){  32}{\line(0,1){10}}
\multiput(  381.97,1200)(  381.97,0){   3}{\line(0,-1){25}}
\multiput(    0.00,1200)(   38.20,0){  32}{\line(0,-1){10}}
\put( 382,-25){\makebox(0,0)[t]{\large $    0.5 $}}
\put( 764,-25){\makebox(0,0)[t]{\large $    1.0 $}}
\put(1146,-25){\makebox(0,0)[t]{\large $    1.5 $}}
\put( 946,-85){\makebox(0,0)[t]{\large $   \phi^*_\eta $}}
\multiput(0,   77.89)(0,  195.74){   6}{\line(1,0){25}}
\multiput(1200,   77.89)(0,  195.74){   6}{\line(-1,0){25}}
\multiput(0,  136.82)(0,  195.74){   6}{\line(1,0){10}}
\multiput(1200,  136.82)(0,  195.74){   6}{\line(-1,0){10}}
\multiput(0,  171.29)(0,  195.74){   6}{\line(1,0){10}}
\multiput(1200,  171.29)(0,  195.74){   6}{\line(-1,0){10}}
\multiput(0,    0.00)(0,  195.74){   7}{\line(1,0){10}}
\multiput(1200,    0.00)(0,  195.74){   7}{\line(-1,0){10}}
\multiput(0,   18.97)(0,  195.74){   7}{\line(1,0){10}}
\multiput(1200,   18.97)(0,  195.74){   7}{\line(-1,0){10}}
\multiput(0,   34.47)(0,  195.74){   6}{\line(1,0){10}}
\multiput(1200,   34.47)(0,  195.74){   6}{\line(-1,0){10}}
\multiput(0,   47.57)(0,  195.74){   6}{\line(1,0){10}}
\multiput(1200,   47.57)(0,  195.74){   6}{\line(-1,0){10}}
\multiput(0,   58.92)(0,  195.74){   6}{\line(1,0){10}}
\multiput(1200,   58.92)(0,  195.74){   6}{\line(-1,0){10}}
\multiput(0,   68.94)(0,  195.74){   6}{\line(1,0){10}}
\multiput(1200,   68.94)(0,  195.74){   6}{\line(-1,0){10}}
\put(-25,  78){\makebox(0,0)[r]{\large $ 10^{   1} $}}
\put(-25, 274){\makebox(0,0)[r]{\large $ 10^{   2} $}}
\put(-25, 469){\makebox(0,0)[r]{\large $ 10^{   3} $}}
\put(-25, 665){\makebox(0,0)[r]{\large $ 10^{   4} $}}
\put(-25, 861){\makebox(0,0)[r]{\large $ 10^{   5} $}}
\put(-25,1057){\makebox(0,0)[r]{\large $ 10^{   6} $}}
\put(-55, 957){\makebox(0,0)[r]{\large {\bf $ \# $ events}}}
\end{picture}}
\put(300,250){\begin{picture}( 1200,1200)
\thinlines 
\newcommand{\x}[3]{\put(#1,#2){\line(1,0){#3}}}
\newcommand{\y}[3]{\put(#1,#2){\line(0,1){#3}}}
\newcommand{\z}[3]{\put(#1,#2){\line(0,-1){#3}}}
\newcommand{\e}[3]{\put(#1,#2){\line(0,1){#3}}}
\y{   0}{   0}{1113}\x{   0}{1113}{  20}
\y{  20}{1113}{   6}\x{  20}{1119}{  20}
\y{  40}{1119}{  16}\x{  40}{1135}{  20}
\y{  60}{1135}{  61}\x{  60}{1196}{  20}
\z{  80}{1196}{ 329}\x{  80}{ 867}{  20}
\z{ 100}{ 867}{ 110}\x{ 100}{ 757}{  20}
\z{ 120}{ 757}{  56}\x{ 120}{ 701}{  20}
\z{ 140}{ 701}{  41}\x{ 140}{ 660}{  20}
\z{ 160}{ 660}{  33}\x{ 160}{ 627}{  20}
\z{ 180}{ 627}{  31}\x{ 180}{ 596}{  20}
\z{ 200}{ 596}{  24}\x{ 200}{ 572}{  20}
\z{ 220}{ 572}{  28}\x{ 220}{ 544}{  20}
\z{ 240}{ 544}{  22}\x{ 240}{ 522}{  20}
\z{ 260}{ 522}{  20}\x{ 260}{ 502}{  20}
\z{ 280}{ 502}{  17}\x{ 280}{ 485}{  20}
\z{ 300}{ 485}{  23}\x{ 300}{ 462}{  20}
\z{ 320}{ 462}{  20}\x{ 320}{ 442}{  20}
\z{ 340}{ 442}{  11}\x{ 340}{ 431}{  20}
\z{ 360}{ 431}{  15}\x{ 360}{ 416}{  20}
\z{ 380}{ 416}{  22}\x{ 380}{ 394}{  20}
\z{ 400}{ 394}{  14}\x{ 400}{ 380}{  20}
\z{ 420}{ 380}{  12}\x{ 420}{ 368}{  20}
\z{ 440}{ 368}{  20}\x{ 440}{ 348}{  20}
\z{ 460}{ 348}{  23}\x{ 460}{ 325}{  20}
\z{ 480}{ 325}{   1}\x{ 480}{ 324}{  20}
\z{ 500}{ 324}{  21}\x{ 500}{ 303}{  20}
\y{ 520}{ 303}{   1}\x{ 520}{ 304}{  20}
\z{ 540}{ 304}{  30}\x{ 540}{ 274}{  20}
\z{ 560}{ 274}{   2}\x{ 560}{ 272}{  20}
\z{ 580}{ 272}{   2}\x{ 580}{ 270}{  20}
\z{ 600}{ 270}{  23}\x{ 600}{ 247}{  20}
\y{ 620}{ 247}{   6}\x{ 620}{ 253}{  20}
\y{ 640}{ 253}{   4}\x{ 640}{ 257}{  20}
\z{ 660}{ 257}{  42}\x{ 660}{ 215}{  20}
\z{ 680}{ 215}{   2}\x{ 680}{ 213}{  20}
\z{ 700}{ 213}{   9}\x{ 700}{ 204}{  20}
\y{ 720}{ 204}{   0}\x{ 720}{ 204}{  20}
\z{ 740}{ 204}{  39}\x{ 740}{ 165}{  20}
\y{ 760}{ 165}{   9}\x{ 760}{ 174}{  20}
\z{ 780}{ 174}{  46}\x{ 780}{ 128}{  20}
\y{ 800}{ 128}{  56}\x{ 800}{ 184}{  20}
\z{ 820}{ 184}{  39}\x{ 820}{ 145}{  20}
\z{ 840}{ 145}{  39}\x{ 840}{ 106}{  20}
\y{ 860}{ 106}{   6}\x{ 860}{ 112}{  20}
\y{ 880}{ 112}{   6}\x{ 880}{ 118}{  20}
\y{ 900}{ 118}{  23}\x{ 900}{ 141}{  20}
\z{ 920}{ 141}{  35}\x{ 920}{ 106}{  20}
\z{ 940}{ 106}{   6}\x{ 940}{ 100}{  20}
\y{ 960}{ 100}{  37}\x{ 960}{ 137}{  20}
\z{ 980}{ 137}{  78}\x{ 980}{  59}{  20}
\y{1000}{  59}{  47}\x{1000}{ 106}{  20}
\z{1020}{ 106}{  37}\x{1020}{  69}{  20}
\y{1040}{  69}{   0}\x{1040}{  69}{  20}
\z{1060}{  69}{  10}\x{1060}{  59}{  20}
\z{1080}{  59}{  40}\x{1080}{  19}{  20}
\y{1100}{  19}{  59}\x{1100}{  78}{  20}
\z{1120}{  78}{   9}\x{1120}{  69}{  20}
\y{1140}{  69}{  17}\x{1140}{  86}{  20}
\z{1160}{  86}{  67}\x{1160}{  19}{  20}
\z{1180}{  19}{  19}\x{1180}{   0}{  20}
\end{picture}} 
\put(300,250){\begin{picture}( 1200,1200)
\thicklines 
\newcommand{\x}[3]{\put(#1,#2){\line(1,0){#3}}}
\newcommand{\y}[3]{\put(#1,#2){\line(0,1){#3}}}
\newcommand{\z}[3]{\put(#1,#2){\line(0,-1){#3}}}
\newcommand{\e}[3]{\put(#1,#2){\line(0,1){#3}}}
\y{   0}{   0}{1113}\x{   0}{1113}{  20}
\y{  20}{1113}{   6}\x{  20}{1119}{  20}
\y{  40}{1119}{  16}\x{  40}{1135}{  20}
\y{  60}{1135}{  61}\x{  60}{1196}{  20}
\z{  80}{1196}{ 328}\x{  80}{ 868}{  20}
\z{ 100}{ 868}{ 112}\x{ 100}{ 756}{  20}
\z{ 120}{ 756}{  55}\x{ 120}{ 701}{  20}
\z{ 140}{ 701}{  41}\x{ 140}{ 660}{  20}
\z{ 160}{ 660}{  33}\x{ 160}{ 627}{  20}
\z{ 180}{ 627}{  30}\x{ 180}{ 597}{  20}
\z{ 200}{ 597}{  25}\x{ 200}{ 572}{  20}
\z{ 220}{ 572}{  24}\x{ 220}{ 548}{  20}
\z{ 240}{ 548}{  23}\x{ 240}{ 525}{  20}
\z{ 260}{ 525}{  19}\x{ 260}{ 506}{  20}
\z{ 280}{ 506}{  23}\x{ 280}{ 483}{  20}
\z{ 300}{ 483}{  17}\x{ 300}{ 466}{  20}
\z{ 320}{ 466}{  21}\x{ 320}{ 445}{  20}
\z{ 340}{ 445}{  15}\x{ 340}{ 430}{  20}
\z{ 360}{ 430}{  21}\x{ 360}{ 409}{  20}
\z{ 380}{ 409}{  13}\x{ 380}{ 396}{  20}
\z{ 400}{ 396}{  10}\x{ 400}{ 386}{  20}
\z{ 420}{ 386}{  16}\x{ 420}{ 370}{  20}
\z{ 440}{ 370}{  18}\x{ 440}{ 352}{  20}
\z{ 460}{ 352}{  14}\x{ 460}{ 338}{  20}
\z{ 480}{ 338}{  21}\x{ 480}{ 317}{  20}
\z{ 500}{ 317}{  14}\x{ 500}{ 303}{  20}
\z{ 520}{ 303}{   6}\x{ 520}{ 297}{  20}
\z{ 540}{ 297}{   3}\x{ 540}{ 294}{  20}
\z{ 560}{ 294}{  18}\x{ 560}{ 276}{  20}
\z{ 580}{ 276}{  29}\x{ 580}{ 247}{  20}
\y{ 600}{ 247}{  10}\x{ 600}{ 257}{  20}
\z{ 620}{ 257}{   6}\x{ 620}{ 251}{  20}
\z{ 640}{ 251}{  18}\x{ 640}{ 233}{  20}
\z{ 660}{ 233}{   3}\x{ 660}{ 230}{  20}
\z{ 680}{ 230}{  10}\x{ 680}{ 220}{  20}
\z{ 700}{ 220}{   7}\x{ 700}{ 213}{  20}
\z{ 720}{ 213}{  26}\x{ 720}{ 187}{  20}
\y{ 740}{ 187}{  26}\x{ 740}{ 213}{  20}
\z{ 760}{ 213}{  24}\x{ 760}{ 189}{  20}
\z{ 780}{ 189}{  15}\x{ 780}{ 174}{  20}
\y{ 800}{ 174}{   5}\x{ 800}{ 179}{  20}
\y{ 820}{ 179}{   0}\x{ 820}{ 179}{  20}
\z{ 840}{ 179}{  38}\x{ 840}{ 141}{  20}
\y{ 860}{ 141}{   4}\x{ 860}{ 145}{  20}
\z{ 880}{ 145}{  22}\x{ 880}{ 123}{  20}
\y{ 900}{ 123}{  18}\x{ 900}{ 141}{  20}
\z{ 920}{ 141}{   4}\x{ 920}{ 137}{  20}
\z{ 940}{ 137}{  14}\x{ 940}{ 123}{  20}
\z{ 960}{ 123}{  45}\x{ 960}{  78}{  20}
\y{ 980}{  78}{  34}\x{ 980}{ 112}{  20}
\z{1000}{ 112}{  12}\x{1000}{ 100}{  20}
\z{1020}{ 100}{  81}\x{1020}{  19}{  20}
\y{1040}{  19}{  40}\x{1040}{  59}{  20}
\z{1060}{  59}{  59}\x{1060}{   0}{  20}
\y{1080}{   0}{ 112}\x{1080}{ 112}{  20}
\z{1100}{ 112}{ 112}\x{1100}{   0}{  20}
\y{1120}{   0}{  48}\x{1120}{  48}{  20}
\z{1140}{  48}{  14}\x{1140}{  34}{  20}
\y{1160}{  34}{  35}\x{1160}{  69}{  20}
\z{1180}{  69}{  35}\x{1180}{  34}{  20}
\end{picture}} 
\end{picture} 
\caption{\small The distribution of the $\phi^*_\eta$  in $Z/\gamma^* \to \mu^+\mu^- (n\gamma)$ as generated
from { \tt KKMC} and { \tt PHOTOS}; selection cuts are applied, see the text. 
The intermediate state of virtuality 97.187 GeV
with the transverse momentum $p_T=9$~GeV and pseudorapidity $\eta=2$ decaying into muon pair was created in the $u\bar u$ annihilation. 
 The  samples of $40$M events were used, 
ratio of the surfaces under distributions  is 0.9991. Relatively large (0.1\%) difference to unity 
is typical for the  larger value of pseudorapidity.  For LO results 
(see web page \cite{webTestsPhotos} for extended results)
the ratio for the surfaces, is 
of the same order. For electrons, both in LO and NLO cases,  this ratio is closer to 1.  
\label{PhiStar}                   
} 
\end{figure}

In Fig.~\ref{PhiStar} an example for one point of $(p_T^Z,\eta^Z)$ grid is chosen and the  $\phi^*_\eta$ distribution is shown. 
An agreement  significantly better than 0.1\% between {\tt KKMC} and {\tt PHOTOS} is observed.
For other choices of flavour of incoming quarks, $Z/\gamma^*$ momentum and 
virtuality agreement is equally good  \cite{webTestsPhotos}.

\subsection{ Case of universal kernel} 

So far, in all numerical results  { \tt  PHOTOS} with first order matrix 
elements as available in {\tt C++} version were 
used both  in $Z$ and $W$ decays. If only the universal kernel was used, 
as available in public { \tt FORTRAN PHOTOS} version 2.14 or higher, the loss of precision would be 
noticeable, but the uncertainty calculated with respect of the total rate would remain
at 0.2\% level, 
for photonic bremsstrahlung corrections to the shapes of distributions \cite{webTestsPhotos}. As in the NLO case,
overall normalization factor has to be  corrected for separately.  

\section{Non photonic Final State Bremsstrahlung}

In this Section we concentrate on those effects which go beyond multiple 
photon emissions. They can be divided into three groups. 
Emission of additional pairs, the effect which certainly belong to final state 
emissions, effect of interference of initial-final state QED effects and finally
all effects which are not directly related to final state radiation, but nonetheless may affect their 
matrix element  calculations.

\subsection{Emission of pairs}

Emission of light fermion pairs should be included starting from the second order of QED, {\it i.e.} 
from the ${\mathcal O}(\alpha^2)$ corrections. 
There are two classes of diagrams which need to be taken into account.
Emission of real pairs (Fig.~\ref{fig:realp}) and the corresponding correction to the vertex
(Fig.~\ref{fig:virtp}). These two effects cancel each other to a large degree due 
to the Kinoshita-Lee-Nauenberg theorem.
The generic size of the effect can be expected to be of the order of higher order photonic bremsstrahlung
corrections discussed so far. Moreover, it is well known from direct calculations in
particular cases that the pair corrections are typically several times smaller than the photonic bremsstrahlung ones
in the same order in $\alpha$. Let us recall that
careful studies of pair radiation effects were performed at LEP times~\cite{Kobel:2000aw,Arbuzov:2001rt}.

\begin{figure}[h]
\label{fig:pair-bouble}
\vspace*{-8mm}
\begin{eqnarray*}
\begin{array}{ccc}
\vcenter{\hbox{
  \begin{picture}(165,100)(0,0)
    \ArrowLine(25,25)(50,50)        \Text(10,25)[lc]{$q$}
    \ArrowLine(50,50)(25,75)        \Text(10,75)[lc]{$\bar q$}
    \Photon(50,50)(75,50){2}{7}     \Text(62.5,55)[cb]{$\gamma,Z$}
    \ArrowLine(95,30)(75,50)        \Text(75,40)[lt]{$e^+$}
    \ArrowLine(115,10)(95,30)       \Text(130,10)[rc]{$e^+$}
    \ArrowLine(75,50)(115,90)       \Text(130,90)[rc]{$e^-$}
    \Photon(95,30)(125,30){2}{7}    \Text(110,35)[cb]{$\gamma$}
    \ArrowLine(150,15)(125,30)      \Text(165,15)[rc]{$\bar{f}_{2}$}
    \ArrowLine(125,30)(150,45)      \Text(165,45)[rc]{$f_{2}$}
    \Vertex(50,50){2.5}
    \Vertex(75,50){2.5}
    \Vertex(95,30){2.5}
    \Vertex(125,30){2.5}
  \end{picture}}}
&\quad+&
\vcenter{\hbox{
  \begin{picture}(165,100)(0,0)
    \ArrowLine(25,25)(50,50)        \Text(10,25)[lc]{$ q$}
    \ArrowLine(50,50)(25,75)        \Text(10,75)[lc]{$\bar q$}
    \Photon(50,50)(75,50){2}{7}     \Text(62.5,55)[bc]{$\gamma,Z$}
    \ArrowLine(115,10)(75,50)       \Text(130,10)[rc]{$e^+$}
    \ArrowLine(75,50)(95,70)        \Text(77,62)[lb]{$e^-$}
    \ArrowLine(95,70)(115,90)       \Text(130,90)[rc]{$e^-$}
    \Photon(95,70)(125,70){2}{7}    \Text(110,65)[ct]{$\gamma$}
    \ArrowLine(150,55)(125,70)      \Text(165,55)[rc]{$\bar{f}_{2}$}
    \ArrowLine(125,70)(150,85)      \Text(165,95)[rc]{$f_{2}$}
    \Vertex(50,50){2.5}
    \Vertex(75,50){2.5}
    \Vertex(95,70){2.5}
    \Vertex(125,70){2.5}
  \end{picture}}}
\qquad\mbox{\large{\bf }}
\end{array}
\end{eqnarray*}
\vspace*{-5mm}
\caption[]{A typical example of real pair correction.
\label{fig:realp}}
\vspace*{-1mm}

\vspace{-4mm}
\begin{eqnarray*}
  \vcenter{\hbox{
  \begin{picture}(160,100)(0,0)
  \Photon(50,50)(100,50){2}{10}
  \ArrowLine(0,0)(50,50)
  \ArrowLine(50,50)(0,100)
  \ArrowLine(150,100)(125,75)
  \ArrowLine(125,75)(100,50)
  \ArrowLine(100,50)(125,25)
  \ArrowLine(125,25)(150,0)
  \Photon(125,75)(125,60){2}{5}
  \Photon(125,25)(125,40){2}{5}
  \ArrowArc(125,50)(10,90,270)
  \ArrowArc(125,50)(10,270,90)
  \Text(75,55)[bc]{$\gamma,Z$}
  \Text(-15,0)[lc]{$q$}
  \Text(-15,100)[lc]{$\bar q$}
  \Text(165,0)[rc]{$e^-$}
  \Text(165,100)[rc]{$e^{+}$}
  \end{picture}}}
\end{eqnarray*}
\vspace{-4mm}
\caption[]{A typical example of virtual pair correction.
\label{fig:virtp}}
\end{figure}

In the context of this paper we will estimate theoretical uncertainty due to 
neglection of pair effects in the Drell-Yan observables.

The {\tt SANC} integrators allow to perform a quick calculation of the pair corrections
within the leading logarithmic approximation. We apply here the formalism of electron
structure (fragmentation) functions~\cite{Kuraev:1985hb,Skrzypek:1992vk,Arbuzov:1999cq} 
which describe radiation in the approximation of collinear kinematics.

The leading logarithmic approximation (LLA) was applied to take into account
the corrections of the orders $\order{\alpha^nL^n}$, $n=2,3$. Let us
remind that in the first order $\order{\alpha}$ SANC has the complete calculation.
The large logarithm $L=\ln(\mu^2/m_l^2)$ depends on the lepton mass $m_l$ 
and on the factorization scale $\mu$. The latter is taken to be equal
to the cms incoming parton energy (other choices are also possible). 

The pure photonic contribution to the non-singlet electron fragmentation function 
in the collinear leading logarithmic approximation reads:
\begin{eqnarray}
{\mathcal D}^{\gamma}_{ee}(y,L) &=& \delta(1-y) + \frac{\alpha}{2\pi}(L-1)P^{(1)}(y)
+ \frac{1}{2}\biggl(\frac{\alpha}{2\pi}(L-1)\biggr)^2P^{(2)}(y)
\nonumber \\
&+& \frac{1}{6}\biggl(\frac{\alpha}{2\pi}(L-1)\biggr)^2P^{(3)}(y) 
+ {\mathcal O}(\alpha^4L^4).
\end{eqnarray}
Analytic expressions for the relevant higher order splitting functions 
can be found in Refs.~\cite{Skrzypek:1992vk,Arbuzov:1999cq}.

For numerical evaluations of the splitting functions regularized by the plus prescription, 
we applied the phase space slicing as follows:
\begin{eqnarray}
P^{(1)}(y) = \left[\frac{1+y^2}{1-y}\right]_+ = \lim_{\Delta\to 0}\left\{
\delta(1-y)\biggl(2\ln\Delta+\frac{3}{2}\biggr) 
+ \Theta(1-y-\Delta)\frac{1+y^2}{1-y}\right\}, 
\end{eqnarray}
where an auxiliary small parameter $\Delta$ is introduced. 
In actual computations we used $\Delta=10^{-4}$ and verified
the independence of the numerical results from the variations of
this parameter, definition of $P^i(y), \; i>1$ is given in Ref.~\cite{Skrzypek:1992vk}.

The effect due to emission of real and virtual electron-positron pairs 
can be estimated using the non-singlet and singlet pair contributions
to the LLA electron fragmentation functions:
\begin{eqnarray}
{\mathcal D}^{\mathrm{pair}}_{ee}(y,L) &=& 
\biggl(\frac{\alpha}{2\pi}(L-1)\biggr)^2
\biggl[\frac{1}{3}P^{(1)}(y) + \frac{1}{2}R^s(y)\biggr]
+ \biggl(\frac{\alpha}{2\pi}(L-1)\biggr)^3
\nonumber \\
&\times& \biggl[\frac{1}{3}P^{(2)}(y) + \frac{4}{27}P^{(1)}(y)
+ \frac{1}{3}R^{s}\otimes P^{(1)}(y) - \frac{1}{9}R^s(y)\biggr]
+ {\mathcal O}(\alpha^4L^4).
\end{eqnarray}
Expressions for the relevant singlet splitting functions $R^s$
and $R^{s}\otimes P^{(1)}$ can be found in Ref.~\cite{Skrzypek:1992vk}.

The differential cross section of the neutral current Drell-Yan process
with FSR leading logarithmic corrections takes the factorized form
\begin{eqnarray}
&& d^5\sigma^{\mathrm{FSR}}_{\mathrm{LLA}}(p+p\to X+ l^+(y_1p^+)+l^-(y_2p^-))
= d^3\sigma^{\mathrm{Born}}(p+p\to X+ l^+(p^+)+l^-(p^-))
\nonumber \\
&& \quad \times d y_1 \left({\mathcal D}^{\gamma}(y_1,L) 
+{\mathcal D}^{\mathrm{pair}}(y_1,L) \right)
d y_2 \left({\mathcal D}^{\gamma}(y_2,L) 
+{\mathcal D}^{\mathrm{pair}}(y_2,L) \right).
\end{eqnarray}
For consistency we expand the product of the fragmentation functions
in $\alpha$ and take in  only terms of the order $\order{\alpha^2L^2}$
and $\order{\alpha^3L^3}$. Terms coming from the product of the photonic
and pair parts of the fragmentation functions are treated as a part
of pair corrections.

\begin{figure}[htp!]
\begin{tabular}{ccc}
\subfigure{
\includegraphics[%
  width=0.40\columnwidth]{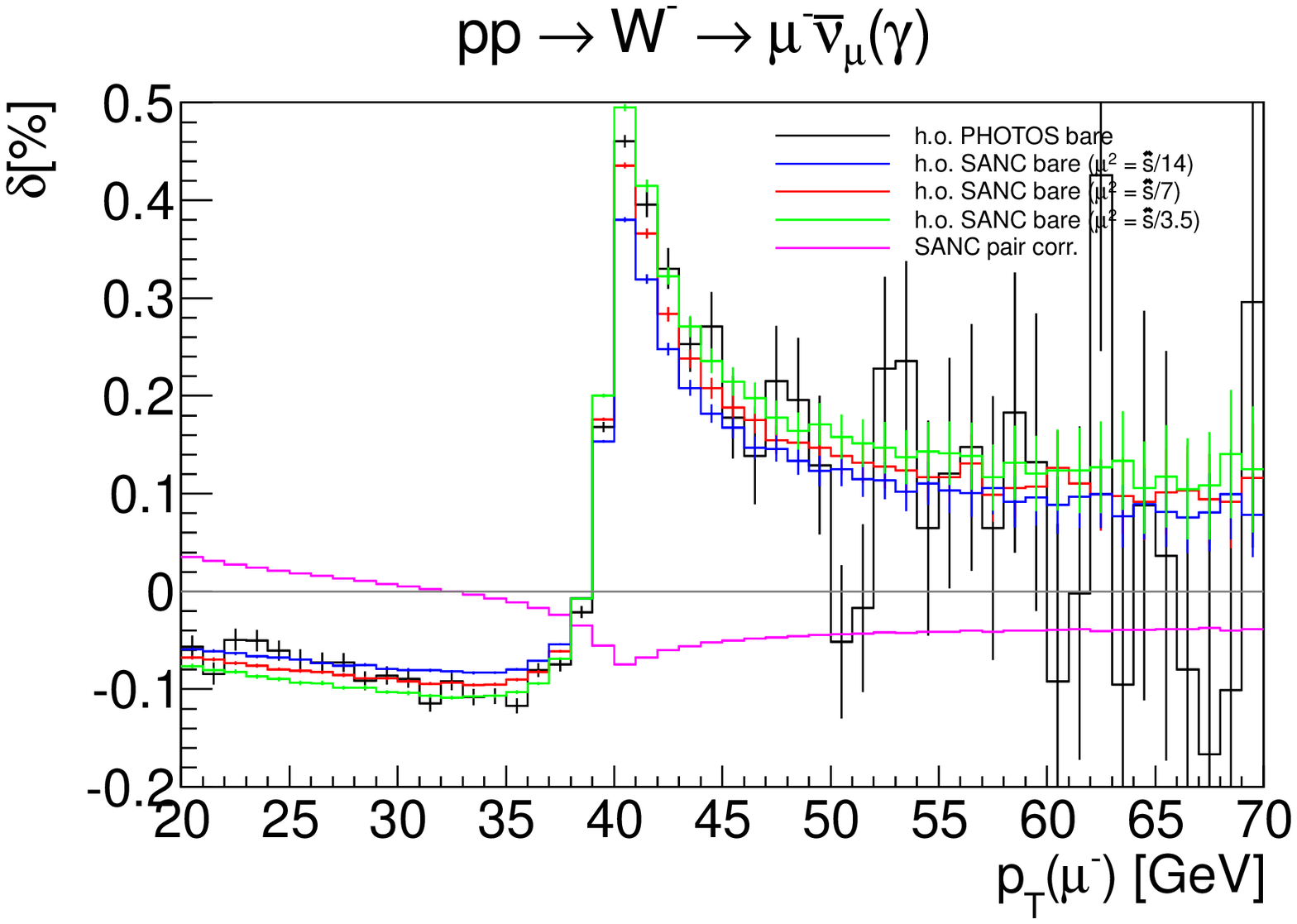}} & \subfigure{\includegraphics[%
  width=0.40\columnwidth]{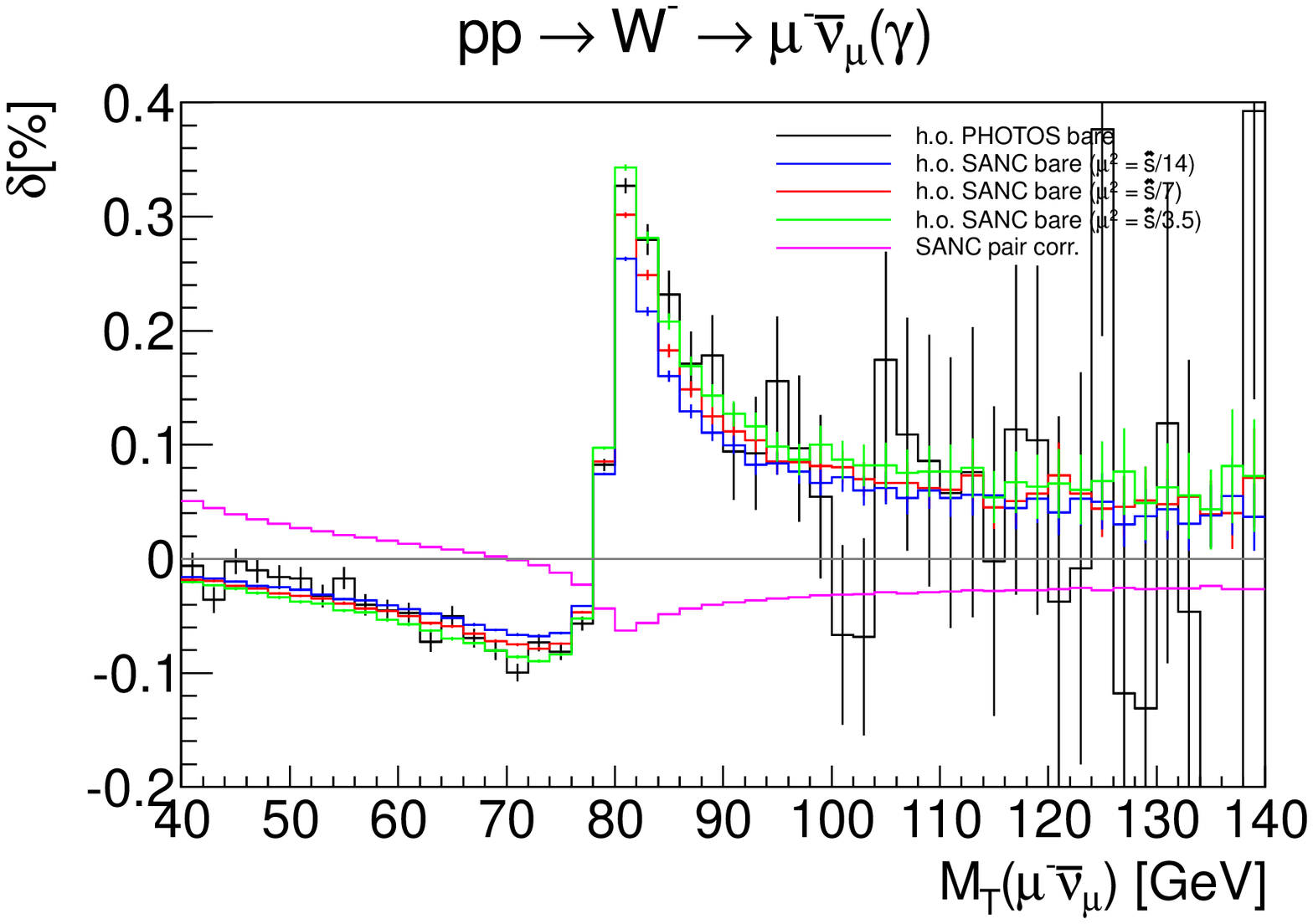}} \\
\subfigure{  \includegraphics[%
  width=0.40\columnwidth]{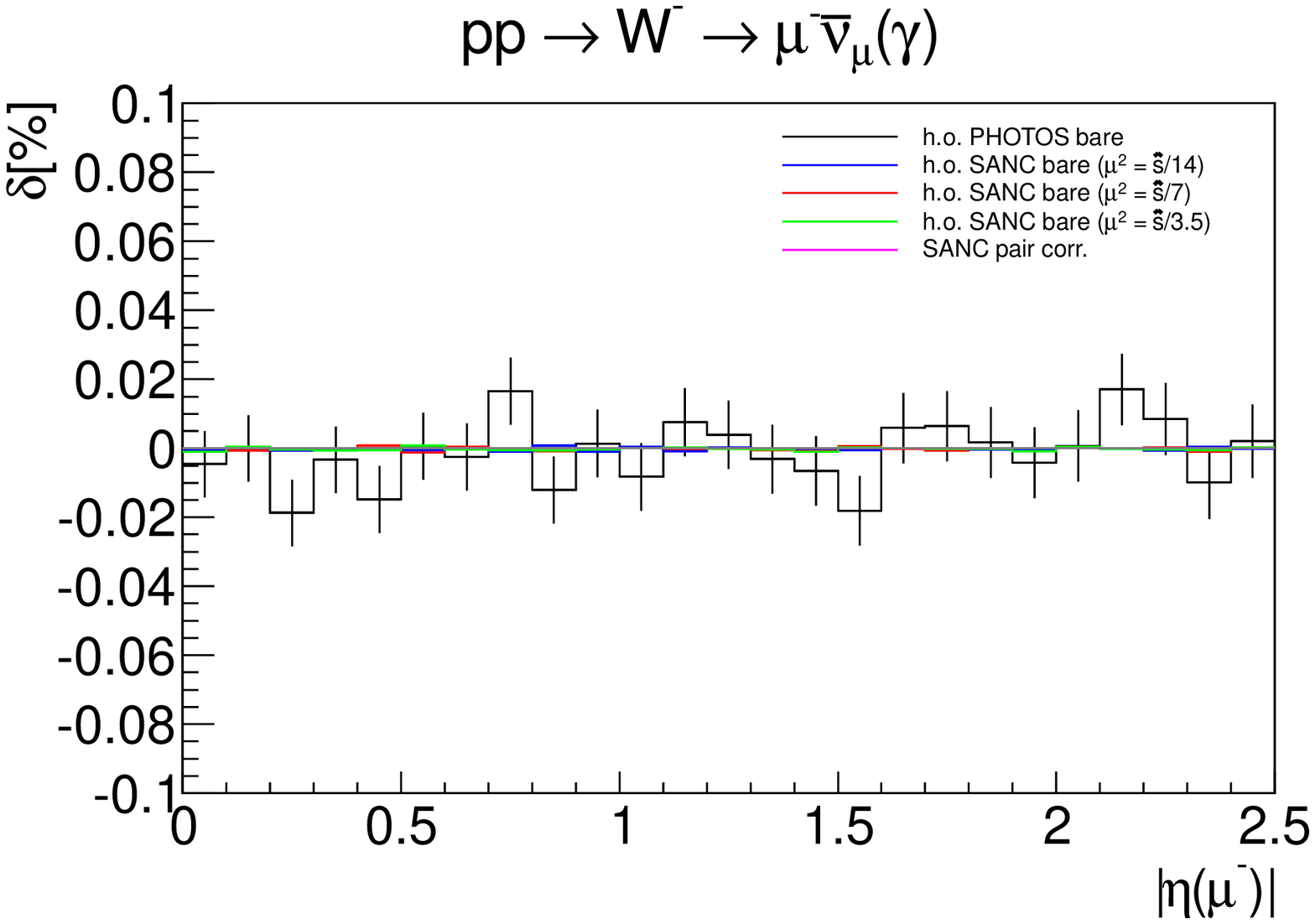}}
\end{tabular}
\caption{Higher order photonic and pair corrections ($\delta$ in \%) for  basic  distributions from { \tt PYTHIA+PHOTOS} and { \tt SANC} in $W^-\to \mu^- \bar \nu$ decay.  
 \label{WpairMUbare}}
\end{figure}

\begin{figure}[htp!]
\begin{tabular}{ccc}
\subfigure{
\includegraphics[%
  width=0.40\columnwidth]{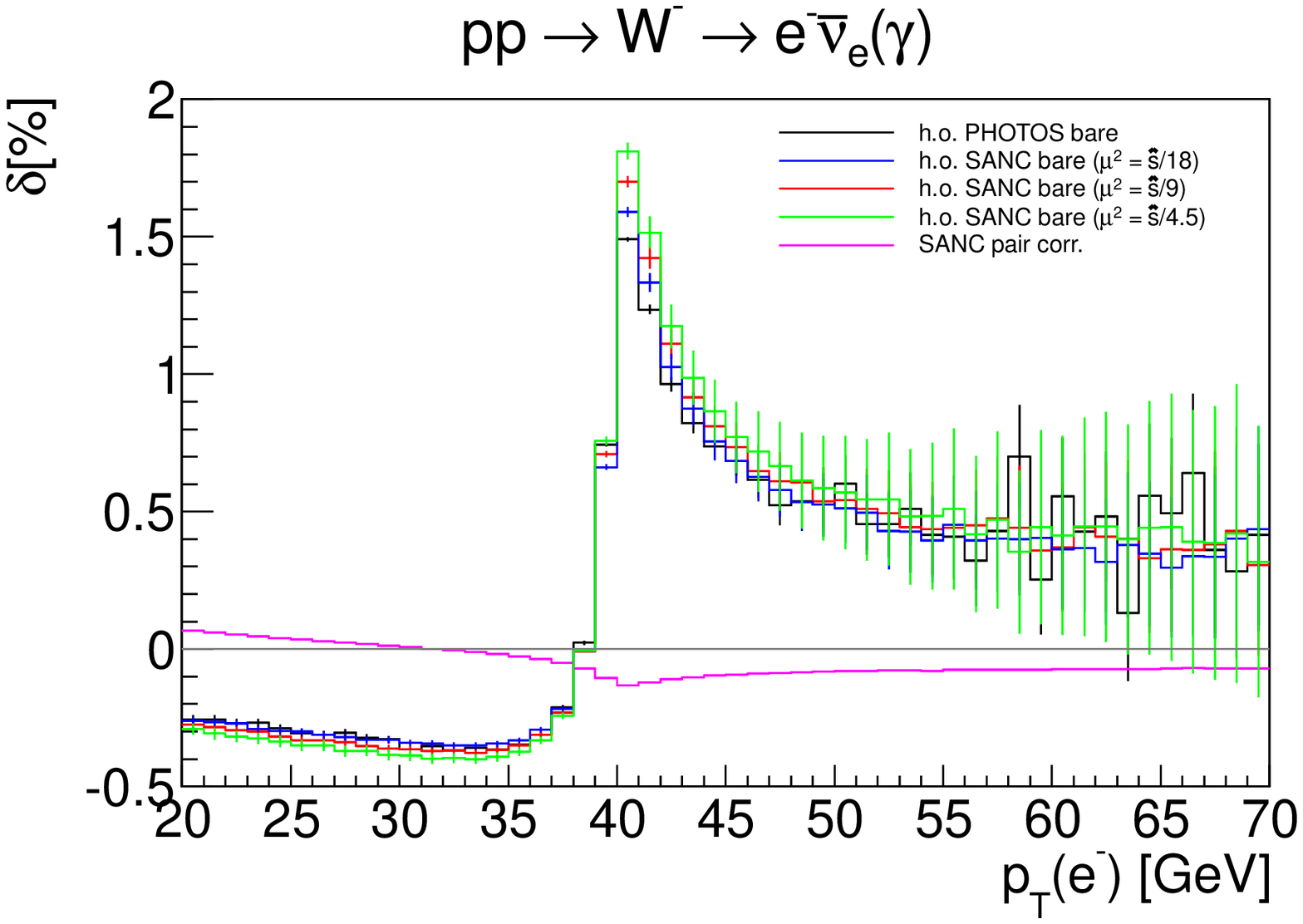}} & \subfigure{\includegraphics[%
  width=0.40\columnwidth]{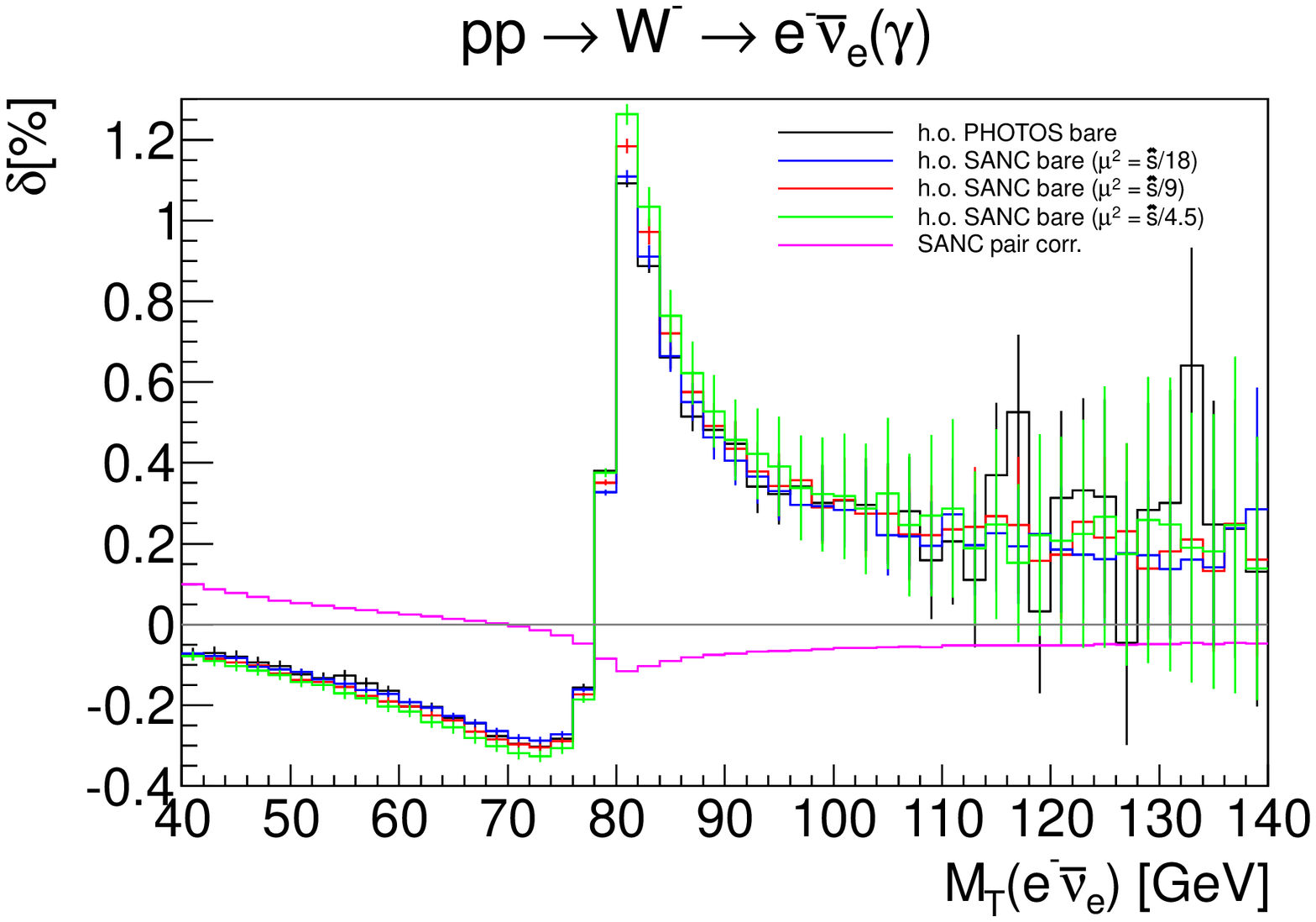}} \\
\subfigure{  \includegraphics[%
  width=0.40\columnwidth]{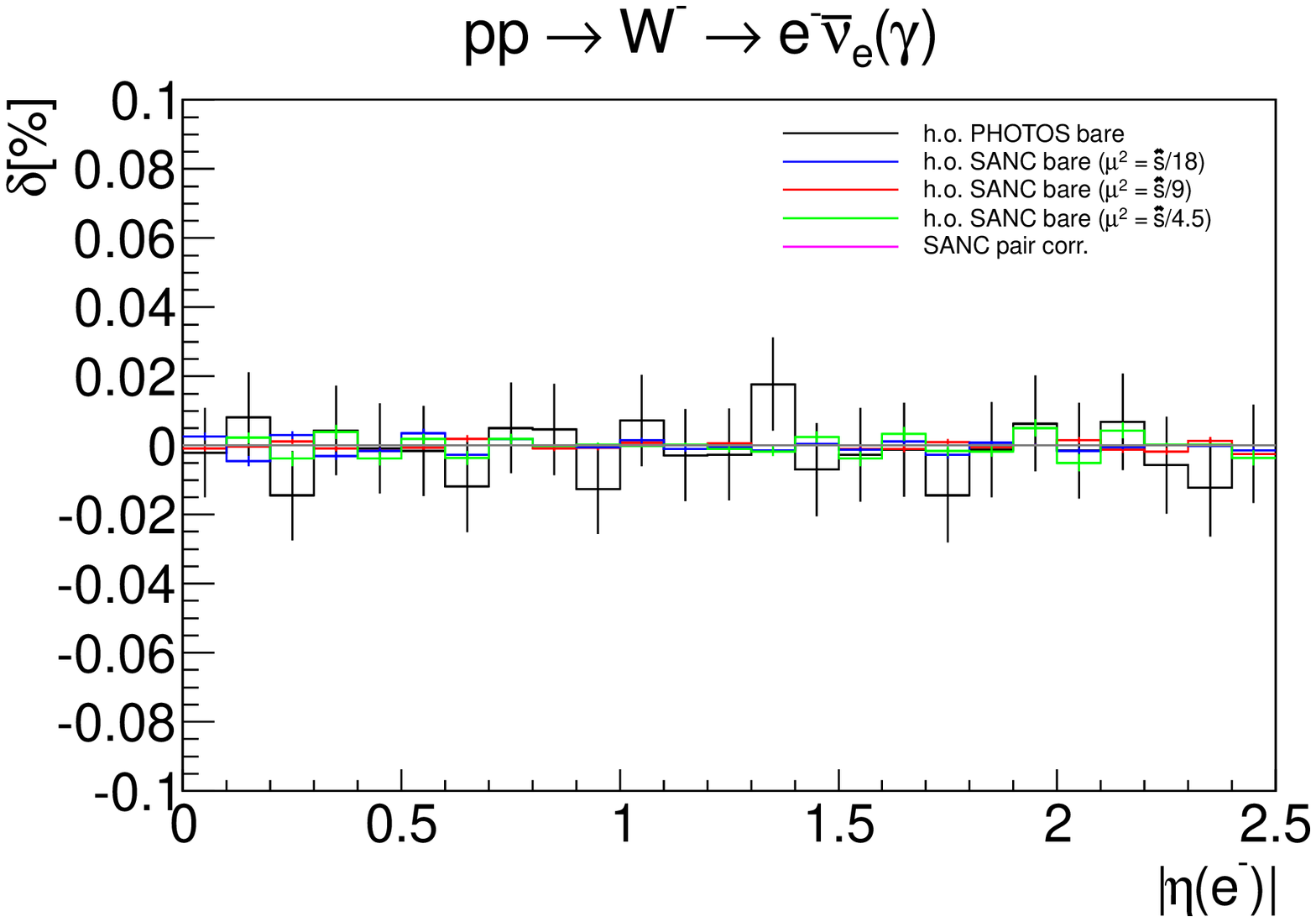}}
\end{tabular}
\caption{Higher order photonic and pair corrections ($\delta$ in \%) for  basic  distributions from { \tt PYTHIA+PHOTOS} and { \tt SANC} in $W^- \to e^- \bar \nu$ decay.
 \label{WpairELbare}}
\end{figure}

Numerical results for the contribution of pair corrections 
are presented in Figs.~\ref{WpairMUbare}-\ref{WpairELbare}, for $W\to l \nu$ decays. 

 If further improvement in precision would be required, 
pair emission can be implemented {\it e.g.} into C++ { \tt PHOTOS} generator~\cite{Davidson:2010ew}.

We assume that in the experimental analysis, there is no specific implicit  cut rejecting $Z \to l^+l^- f \bar f $ events.
Our cut on $p_T^l$  may reject some  
$Z \to l^+l^- f \bar f $ events,   since with the real $f \bar f $ pair 
emission, 
leptons $l$ will have  a  somewhat softer energy spectrum. But this require relatively 
high energy to be carried out by the 
  $f \bar f $ pair, the dominant triple logarithmic term will thus cancel 
between contributions from diagrams 
of Figs. \ref{fig:realp} and \ref{fig:virtp}. Let us stress that special care 
is needed in case of selection cuts sensitive to  the soft, small virtuality and collinear to primary lepton pairs.
Only this region of phase space may contribute significantly to pair corrections. 
Cuts affecting additional  leptons of larger energies 
are thus of no concern.

 The direction of the leptons originating from Z decays (and therefore our $\phi^*_\eta$ observable) may be affected by partial
reconstruction of  two leptons (or two hadronized quarks) of the extra pair. Resulting phenomena may be important only if the pair 
$f \bar f $ is not collinear to any of the primary leptons $l$. Again this represents a non dominant effect, thus substantially below 
 required precision goal of a permille level. 

Summarizing, details of pair corrections are still not important for  precision tag at the level  of 
0.1-0.2\%.

\subsection{Initial-Final state bremsstrahlung interference}

The effect of QED-type interference between spin amplitudes for emission from  the initial and final states 
represents important, even if numerically small, class of corrections.
Even if separation of the initial and final state bremsstrahlung at the spin amplitude
level is clear, large interference effects may make this separation of limited practical convenience.
However in certain approaches interferences can be combined with  QED final state effects in a convenient way. 

Before we will present numerical results,
let us recall first some details of the discussion 
from LEP times, see Ref.~\cite{Kobel:2000aw}.  The discussion was devoted to energies higher than resonance peak, 
that is why interference cancellations as explained {\it e.g.} in \cite{Jadach:1988zp,Jadach:1999gz} did not apply.
In the case of discussed in this paper LHC observables, oriented on production of $W$ and $Z$ resonances, 
such suppression is nonetheless expected. 
The reason is of a physical origin;  time separation between boson production and decay.
However, as a consequence of the uncertainty principle this suppression can be broken with strong event
selection cuts if these cuts would constrain the final state energies.

 In the studies of Drell-Yan processes at LHC
one can restrict discussion of the interference to the first order only. 
On the technical level control of the QED $\order{\alpha}$ interference contribution is realized
in the {\tt SANC} Mote Carlo integrator rather simply. The corresponding effect is computed
by switching a respective flag in the code. 

For the $W$ decays similar arguments  related to intermediate state
life-time apply. In this case however, size of corrections is calculation scheme dependent, {\it i.e.} depend on 
the way how diagrams of photon emission off $W$ line are treated. 
Studies with {\tt SANC} demonstrated that the interference is below 0.1\% for LHC applications. They 
were completed not only for $W$ but for $Z$ as well, see {\it e.g.} \cite{Aad:2011dm}.

\begin{figure}[htp!]
\includegraphics[width=0.5\textwidth]{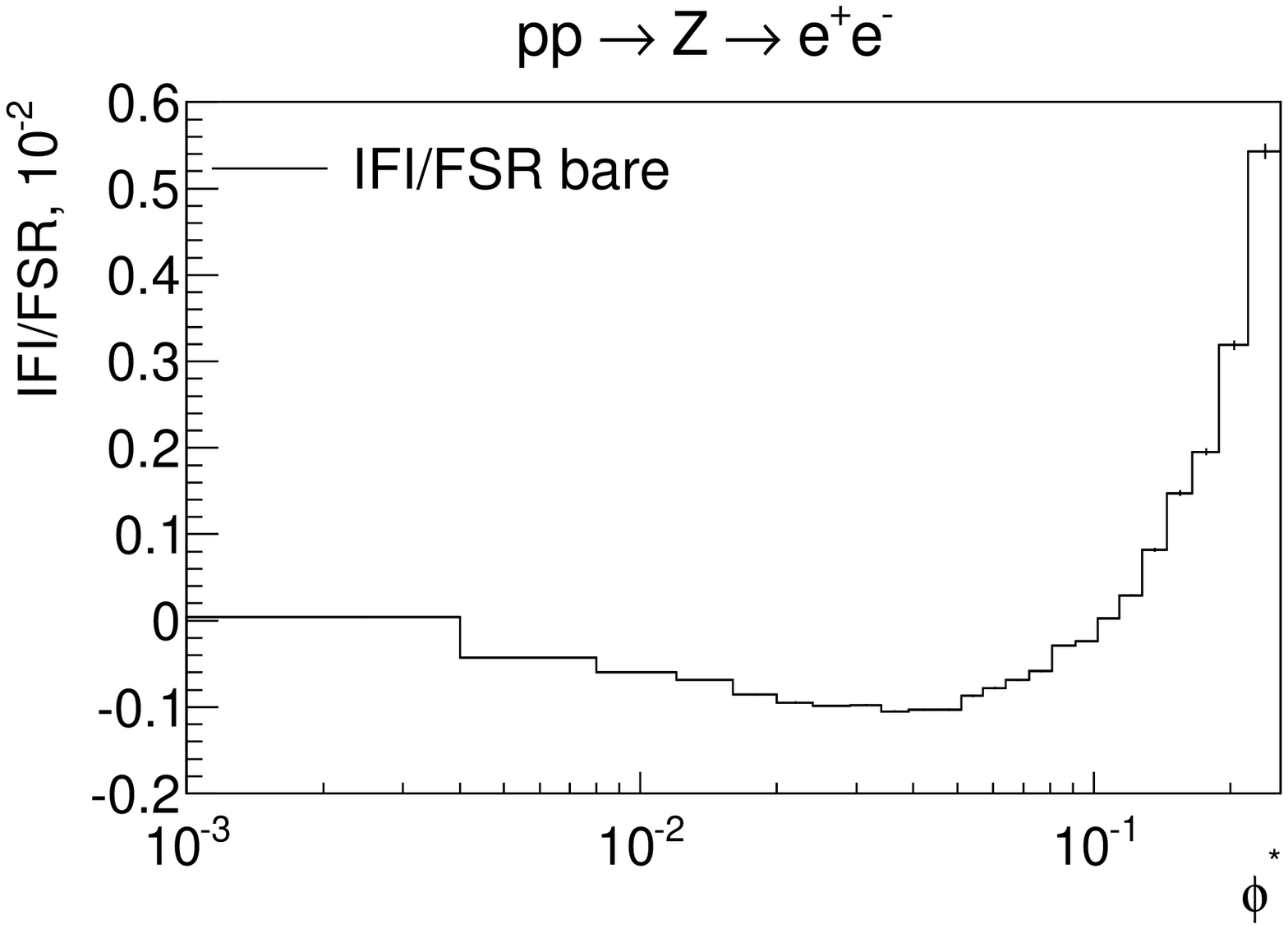}
\includegraphics[width=0.5\textwidth]{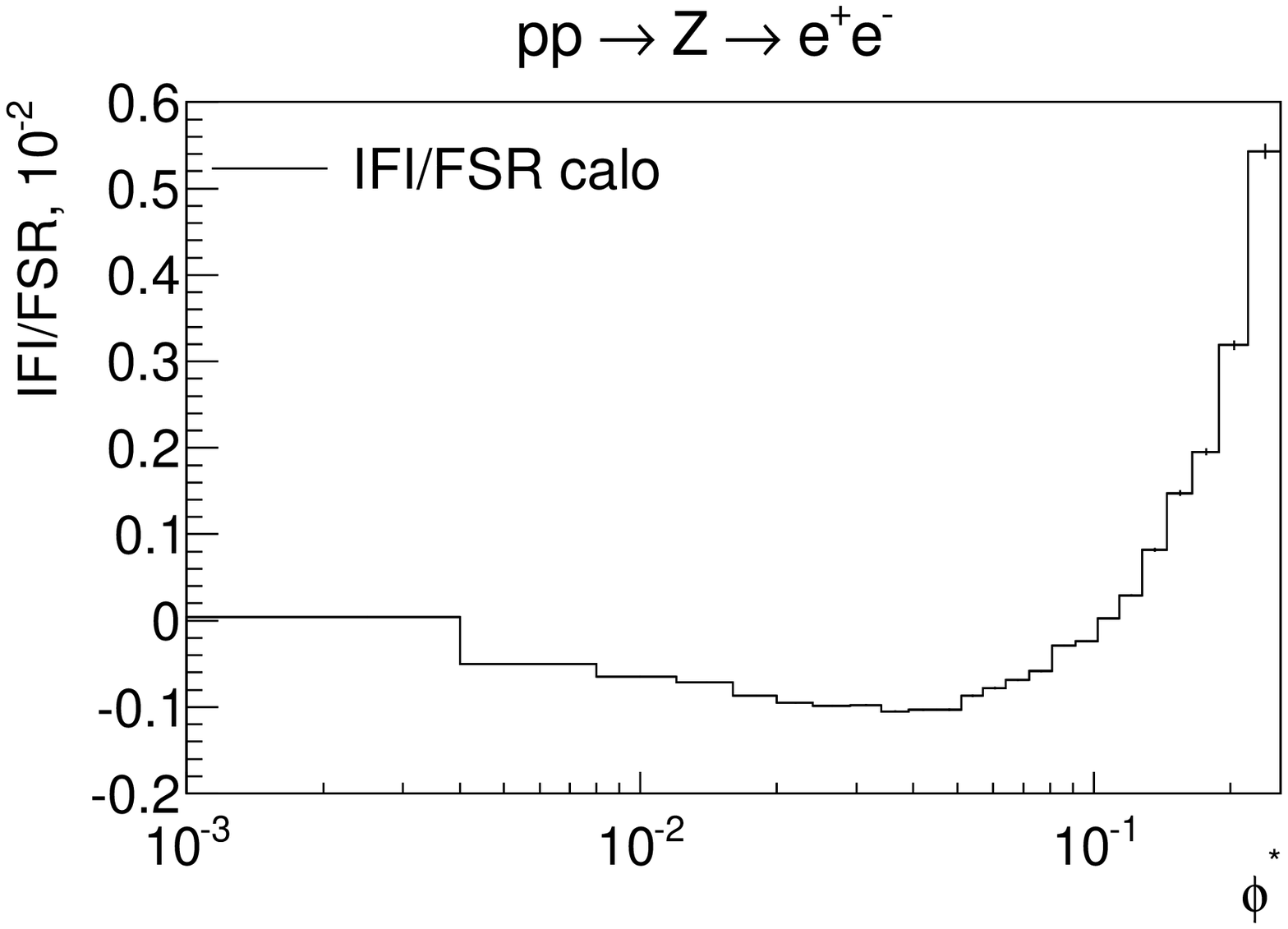}
\caption{IFI/FSR ratio in $Z$ decay for $\phi^\ast$ distribution. For  $\phi^\ast_\eta>0.2$ interference 
effects become sizable. }
\label{IFI}
\end{figure}

Let us show in Fig.~\ref{IFI}, as an example,
corrections from  interference to the $\phi^*_\eta$ observable discussed previously.
The effect is small, below  0.1\% for $\phi^*_\eta<0.15$ and  rises to 0.5\% for $\phi^*_\eta \simeq 0.3$. 
Recent measurement from
ATLAS collaboration \cite{Aad:2012wfa} of  $\phi^*_\eta $ observable extends to 1.3 but with large so far, statistical errors 
in range $\phi^*_\eta>0.2 $.
If precision requirements would become 
more demanding, the effect
 should  be included together with  the FSR corrections. At present, size of the interference effect can be used
to  estimate the size of the corresponding  theoretical uncertainty due to its omission.

We can conclude that the initial-final state interference does not represent a problem
 for separating final state photonic bremsstrahlung
from the remaining electroweak corrections in processes of $W$ and $Z$ production at LHC.
This conclusion  is justified for the precision of  0.1\%, but it will have to be studied in more detail for 
more exclusive configurations, like {\it e.g.} larger values of $\phi^\ast_\eta$ distribution or for observables
defined for off peak regions of lepton pair invariant masses.

\subsection{Relations with other electroweak and hadronic interactions}

In calculation of final state radiation matrix element, dependence on 
the direction of incoming quarks is present. However
one can see from \cite{Golonka:2006tw} that such effects due to {\it e.g.} initial state interactions,  
affect numerical results for   
final state bremsstrahlung in a minor way,
through the term  which is in itself at the permille level, thus well
below present precision requirement  of 0.1\%. 

For decays of 
 narrow width states  or  when gauge symmetry can be used to separate phenomena 
from other parts of the interactions, there are no major  difficulties to identify QED effects at the spin amplitude level.  
In the general case,
 QED FSR can be defined and its systematic error can be discussed as well. However, if {\it e.g. }
contribution of diagrams featuring t-channel exchange of bosons complicate the separation,  discussion 
of systematic errors of other parts of calculations may become scheme dependent.
Our discussion on QED FSR corrections only will nonetheless be  still useful for experimental
applications.

\section {Summary}

We have addressed question of theoretical error for predictions of 
QED final state bremsstrahlung in decays of $W$ and $Z$ bosons, used in 
 precision measurements at LHC experiments.

Tests and comparisons of { \tt PHOTOS} versus { \tt SANC} programs 
for final state photonic  bremsstrahlung were  performed in realistic conditions.
Hard processes and initial state hadronic interactions were simulated 
with the help of {\tt PYTHIA8} \cite{Sjostrand:2007gs} Monte Carlo program,
for results with {\tt PHOTOS}. For {\tt SANC} its own set-up was used.
Related differences in electroweak Born-level processes required careful tuning until agrement was 
established. 

We have started our discussion with technical tests and results obtained at the first order.
Separation into 
QED FSR and remaining electroweak corrections have been studied and verified at the spin amplitude level.
We have checked that in {\tt PHOTOS} and {\tt SANC} 
numerically compatible, down to 0.01\% precision level, schemes of such separation 
were defined. This agreement confirms  proper installation of 
matrix elements and numerical stability of { \tt SANC} and { \tt PHOTOS}  
as well. Then the comparison was repeated after allowing multi-photon emission in both programs. 
An agreement necessary to estimate systematic error 
in implementation of QED FSR photonic bremsstrahlung for 
the precision level of 0.1\% was found. This conclusion holds  for the decays of intermediate states,
produced from annihilation of light quarks, predominantly close to 
the $W$ and $Z$ resonance peaks but with  tails of the distributions taken into account.
The conclusion holds  if { \tt NLO} kernel is active
in  { \tt PHOTOS} and for  { \tt SANC} multiphoton option.
For {\tt PHOTOS} with the {\tt LO} kernel theoretical precision  is estimated to be 0.2\%. 
This  conclusion is limited  to effects resulting from shapes 
of distributions and for the selection cuts  discussed in the paper. In principle, whenever
new type of cuts is applied such comparison needs to be repeated.
Effects on normalization have to be taken into account independently, either as part of 
genuine electroweak corrections (thanks to the proper choice of $\mu_{PW}$ in $W$ decays), or as an simple overall factor, 
like $(1+ \frac{3}{4}\frac{\alpha}{\pi})$ in case of the 
Z decay.

We have estimated the size of the  
higher orders QED photonic bremsstrahlung corrections using other programs.  The { \tt KKMC} \cite{Jadach:1999vf,Jadach:2000ir} Monte Carlo 
program of LEP era,  featuring exclusive exponentiation and second order matrix element
 for final state photonic bremsstrahlung was used for reference results. With the help of this program  
monochromatic intermediate $Z/\gamma^*$ states of fixed virtuality were produced from
annihilation of light quarks. This provided interesting test while grid of predefined values
of $(p_T,\eta)$ was populated, in particular for $\phi^*_\eta$ observable. The differences
versus {\tt NLO PHOTOS} was found below 0.1\% (0.2\% for {\tt PHOTOS} kernel restricted to {\tt LO} only).

 Interferences of QED FSR with QED ISR was found to be below 0.1\% 
 for selected  $W$ and $Z$ observables, 
as expected from the physics arguments.
 Separation of FSR radiation from the remaining  electroweak effects 
is of a practical importance as it facilitate  phenomenological work. Our calculation schemes
are  convenient from that point of view. Interference effect was found to be below required 
precision level.

We estimate  precision level of photonic final state corrections at 0.1\%.
With such precision tag  separation of QED FSR from the rest of the process can be used 
for the sake of detector studies on final state leptons. Such detector studies represent also a well
defined  segment in comparison of theoretical predictions with the measured data.
 One exception is $\phi^*_\eta$ distribution in region of large  $\phi^*_\eta> 0.15$. Already at 
$\phi^*_\eta \simeq 0.3$ interference reach 0.5\%. In this region of phase space spin amplitudes for bremsstrahlung in initial and final state become gradually of comparable size.
Emission of additional pairs was discussed as well and a size of effect was estimated at 0.1\% level.

We  estimate an overall systematic error for FSR implementation
in {\tt PHOTOS} and {\tt SANC}
 at 0.2\% (0.3\% for {\tt PHOTOS} with {\tt LO} kernel). 
Further improvement of precision is possible, but  requires more detailed discussion. 
Details of experimental acceptance have to be taken into account.

At a margin of the discussion we entered investigation of dependence on scheme specific parameters such
 as electromagnetic factorization scale $\mu^2$ or photon enegy threshold $k_0=\epsilon$ used in fixed order
simulations. This may be of some interest for further studies of
 uncertainties resulting from some choices of matching of FSR with hard process and/or initial state 
interactions and/or hard emission matrix elements.

\newpage
\centerline{\bf Acknowledgments}
\vskip 3 mm

Part of this work devoted to observable of $\phi^*_\eta$ angle 
has been inspired by the discussion with Lucia di Ciaccio Elzbieta Richter-Was and other members 
of ATLAS LAPP-Annecy group, 
 continuous encouragements and comments on intermediate steps of the work are  
acknowledged. Work was performed in frame of IN2P3 collaboration between Krakow and Annecy.
Useful discussion with Ashuthos Kotwal on 
distributions of importance for benchmarking algorithm in context of its 
reliability for the simulations of importance for CDF $W$ mass measurements
is to be mentioned as well.


\providecommand{\href}[2]{#2}\end{document}